\documentclass[12pt]{article}
\usepackage[T1]{fontenc}
\usepackage{amsmath,amssymb,mathrsfs}
\usepackage{paralist}
\usepackage{slashed}
\usepackage{graphicx}
\usepackage{subfigure}
\usepackage{color}
\usepackage{fullpage}
\usepackage{multirow}
\usepackage{hyperref}
\usepackage{amscd}
\usepackage{xcolor}
\usepackage{braket}
\usepackage{amsfonts}
\usepackage{tikz}
\usepackage{pgfplots} 
\usepackage{cite}
\usepackage{booktabs}

\catcode`\@=11 
\def\numberbysection{\@addtoreset{equation}{section} 
        \def\theequation{\thesection.\arabic{equation}}} 

\def\be{\begin{equation}} 
\def\ee{\end{equation}} 
\def\ba{\begin{eqnarray}} 
\def\ea{\end{eqnarray}} 
\def\bali{\begin{align}}
\def\eali{\end{align}}
 
\def\ov{\overline}

\def\C{\mathbb{C}}

\def\nl{\nonumber \\}

\def\wt{\widetilde} 
\def\wh{\widehat}

\def\CR{\color{red}}
\def\CG{\color{green}}
\def\CC{\color{cyan}}
\def\Bell{\ensuremath{\boldsymbol\ell}}

\def\a{\alpha} 
 
\def\g{\gamma} 
 
\def\D{\Delta} 
 
\def\e{\epsilon} 
\def\eps{\varepsilon}

\def\l{\lambda}

\def\s{\sigma}

\def\f{\phi}

\def\c{\chi} 
\def\w{\omega}

\begin{document} 
 
\begin{titlepage} 
\begin{center} 
\vskip .6 in 
{\LARGE Critical Ising Model in Varying Dimension}\\
\medskip
{\LARGE by Conformal Bootstrap} 
\vskip 0.2in 
Andrea CAPPELLI${}^{(a)}$, 
Lorenzo MAFFI${}^{(a,b)}$ and Satoshi OKUDA${}^{(c)}$
 \medskip

{\em ${}^{(a)}$INFN, Sezione di Firenze\\
Via G. Sansone 1, 50019 Sesto Fiorentino - Firenze, Italy}\\ 
{\em ${}^{(b)}$Dipartimento di Fisica, Universit\`a di Firenze\\ 
Via G. Sansone 1, 50019 Sesto Fiorentino - Firenze, Italy} \\
{\em ${}^{(c)}$Department of Physics, Rikkyo University\\ 
Toshima, Tokyo 171-8501, Japan}
\end{center} 
\vskip .2 in 

\begin{abstract} 
The single-correlator conformal bootstrap is solved numerically for
several values of dimension $4>d>2$ using the available SDPB and Extremal
Functional methods. Critical exponents and other conformal data of
low-lying states are obtained over the entire range of dimensions with
up to four-decimal precision and then compared with several existing
results.
The conformal dimensions of leading-twist fields are also determined
up to high spin, and their $d$-dependence shows how the conformal
states rearrange themselves around $d=2.2$ for matching 
the Virasoro conformal blocks in the $d=2$ limit.
The decoupling of states at the Ising point is studied for $3>d>2$ and 
the vanishing of one structure constant at $d=3$ is found to
persist till $d=2$ where it corresponds to 
a Virasoro null-vector condition.
\end{abstract} 
 
\vfill 
\end{titlepage} 
\pagenumbering{arabic} 
\numberbysection

%-1--------------------------------------------- 
 
\section{Introduction} 

The study of conformal invariance in two dimensions, following the
fundamental paper \cite{BPZ}, has led to a deep understanding of
non-perturbative phenomena in massless quantum field theories and the
exact solution of many models with countless
physical applications \cite{cft}.

Conformal invariance above two dimensions has long been considered of
limited help in solving non-perturbative physics due to the lack of an
analog of the infinite-dimensional Virasoro algebra. This implies among
other things the existence of infinite conserved local currents,
and the integrability of the theory (in principle, at least). 
A new perspective has recently emerged due to the success of the
conformal bootstrap approach above two dimensions \cite{slava-I}, that
provided many remarkable non-perturbative results, both numerical 
and analytic, on critical exponents and other conformal data,
most notably for the Ising model in three-dimensions \cite{slava-rev}.

In view of these developments, the interplay between conformal
invariance in two and higher dimensions requires some better understanding
and a convenient approach is that of studying the dependence of theories
on the continuous dimension $d\ge2$. For example, the decomposition
of correlators in `conformal partial waves' \cite{dolan}, a basic ingredient
for setting up the bootstrap, depends smoothly on $d$.
On the contrary, the algebraic structures related to representations
of the Virasoro algebra and the existence of minimal models with reduced sets
of states are very specific of $d=2$. 
Thus, a rather general question is whether any of these
structures  can be generalized to $d>2$ (and how). More specifically,
we might ask:
\begin{itemize}
\item 
How the reduced sets of states of the $d=2$ Ising minimal model 
extends when the dimension is deformed to $d>2$?
\item
How the vast degeneracy of scaling dimensions of $d=2$ 
conformal fields breaks down for $d>2$?
\item
Are the $d>2$ conformal theories sitting on the boundary of the unitary region
`minimal' in some sense?
\item
In particular, do analogs of Virasoro null vectors exist for $d>2$
in the form of projection of states?
\end{itemize}

In this paper, we describe the numerical solution of the simplest
one-correlator conformal bootstrap for thirteen values of $4>d>2$,
directly improving the earlier works
\cite{slava-d}\cite{behan}. For each $d$ value, we use the SDPB
program \cite{SDPB} for finding the boundary of the unitary region and
then run the Extremal Functional Method \cite{extr} to solve the
truncated bootstrap equations along this boundary.
We first reproduce the $d=3$ data and patterns of the 2014 paper by El-Showk et
al. \cite{slava-IM} and then extend to other dimensions with similar
(high) precision. As in that paper, the Ising criticality is identified
by the point with smallest value of central charge along the boundary, 
where also a kink is present.
The use of the Extremal Functional Method greatly reduces
the numerical work and allows for a better resolution of subleading fields,
which are then compared with the advanced three-correlator bootstrap results
of Ref. \cite{sd} \cite{sd-old} \cite{slava-rev}.

In Section two, we present these checks at $d=3$ and then proceed
to describe the conformal dimensions of six best identified
low-lying fields ${\cal O}$, 
respectively $\s,\e,\e'$ for spin $\ell=0$, $T'$ for $\ell=2$ and $C,C'$ for
$\ell=4$, and the relative structure constants $f_{\s\s\cal O}$,
for the values: 
\be
d=3.75,\ 3.5,\ 3.25,\ 2.75,\ 2.5,\ 2.25,\ 2.2,\ 2.15,\ 2.1,\ 2.05,
\ 2.01,\ 2.00001.
\label{d-val}
\ee
These conformal data are then expressed as polynomials in $y=4-d$
obtained by least chi-square fits and then compared with earlier bootstrap
results in non-integer dimension \cite{slava-d}\cite{behan}, the
epsilon expansion of $\l \phi^4$ theory
\cite{zinn} \cite{panzer}, and Monte Carlo simulations \cite{mc}.
We thus obtain a description of Ising criticality in continuous dimension
$4> d \ge 2$ that can be useful for many applications, such as, for example,
in developing some conjectures on universality classes
\cite{slava-long}\cite{tromb} and as a benchmark for testing
resummations of the epsilon expansion \cite{zinn}\cite{panzer}
\cite{serone}.

We remark that the six low-lying states depends smoothy on the dimension 
$d$ and their behaviour does not allow to answer
any of the previous questions on the interplay between $d=2$ and $d>2$. 
As is well known, degeneracies, null vectors and other features of
Virasoro representations involve the higher part of the $d=2$
conformal spectrum: unfortunately, this is not described 
accurately enough in our numerical setting.

Nonetheless, some hints of the $d$-dependent changes can be observed.
In Section three, we describe the spectrum of least-dimensional
fields for each spin value $\ell$, the so called leading-twists. Their
dimension $\D_\ell=d-2+\ell +\g_\ell$ include a large classical part
and a small anomalous dimension $\g_\ell$, whose asymptotic behaviour
in $\ell$ is given by:
\be
\lim_{\ell\to\infty}\g_\ell=2\g_\s,  \qquad \quad \D_\ell=d-2+\ell +\g_\ell,
\qquad \D_\s=\frac{d-2}{2}+\g_\s,
\label{asymp}
\ee
where $\D_\s$ is the dimension of the Ising spin.  The numerical
bootstrap gives values of $\D_\ell$ that are unstable, oscillating
between $\g_\ell=0$ and $\g_\ell\neq0$ within the range of $\D_\s$
values identifying the Ising point - a known feature of the Extremal
Functional Method \cite{sd}.  Yet, the non-vanishing $\g_\ell$ values
match rather well the precise $d=3$ values found by the
three-correlator bootstrap \cite{sd}, and give us confidence for
analyzing $\g_\ell(d)$ for $3>d>2$ and $4 \le \ell \le 20$. For any dimension
$d>2.2$, the $\g_\ell(d)$ obey the asymptotic limit (\ref{asymp}) and
also satisfy the Nachtmann theorem \cite{nacht}, i.e. they form a
curve in $\ell$ that is monotonically increasing and convex.

However, at $d\le 2.2$ such a behavior is lost and all anomalous
dimensions converge to the values $\g_\ell=0$, that pertain to 
higher-spin conserved currents fitting the Virasoro tower of the identity 
field $I$ in the $d=2$ Ising model. The subleading twist fields analogously
converge to $\g_\ell=1$ for entering the tower of the energy field $\e$.
This result establishes that $d\sim 2.2$ is the dimension at which 
conformal theories actually acquire the $d>2$ structure, and that 
the transition between the $d>2$ and $d=2$ regimes takes place in the
region $2.2>d>2$.

In Section four, we analyze the fate of the Virasoro null vectors as
the dimension is increased above two. We start in $d=2$ by recalling
the Zamolodchikov counting of quasi-primary fields in the minimal
models \cite{zam}, and use it to formulate a necessary condition for
the occurrence of null vectors. This indicates when a bootstrap
channel decouples as the Ising point is approached from $\D_\s\ge1/8$
along the unitarity boundary, corresponding to a proper null vector of
the Ising model.

Among the low-lying states numerically accessible at $d=2$, 
we find that the simplest $\ell=2$ null vector of the Ising energy field
$\e=\phi_{1,3}\sim\phi_{2,1}$ (in the Kac table \cite{BPZ}) is clearly seen, 
while the higher ones are blurred.
Following the evolution of this state in the bootstrap spectrum for
$d>2$ and $\D_\s\ge \D_\s^{\rm Ising} $, we find that it
keeps decoupling at the Ising point till $d=3$ where it had been
observed before \cite{slava-IM}.
This is a very interesting and encouraging result w.r.t. the 
questions raised at the beginning, suggesting that $d>2$ Ising conformal
theory could be characterized by specific decouplings of states.
However, some words of caution are necessary, because 
a single occurrence does not prove the existence of a pattern.

Finally, in Section five we draw our conclusions and in appendix A
we discuss  the numerical procedures employed in this work.

%-2--------------------------------------------- 
 
\section{Precise critical exponents and structure constants as functions
of dimension} 

\subsection{Method and $d=3$ checks}

The first step of our analysis is to reproduce the results of the
one-correlator bootstrap at $d=3$ by El-Showk et al. \cite{slava-IM}
\cite{slava-IM-old}. To this effect, we consider precisely the same
setting, truncating the functional space of the bootstrap equation to
153 and 190 components.  We use the SDPB algorithm \cite{SDPB} to find
the unitarity boundary in the $(\D_\s,\D_\e)$ plane and solve the
bootstrap equations on this boundary by the Extremal Functional Method
\cite{extr}. We obtain the plots of conformal dimensions as a function
of $\D_\s$ that indeed reproduce the patterns shown in the Figures 7
to 16 of Section 3 in Ref. \cite{slava-IM}.  The numerical methods
employed here are freely available and have been adapted to varying
dimension; the details of their implementation are discussed in
Appendix A. A key feature of our approach is the gain of computational
speed provided by the Extremal Functional Method that allowed us to
study several values of $d$ with reasonable effort.

\begin{figure}
\centering
\subfigure[\label{fig1a}]{\includegraphics[scale=0.35]{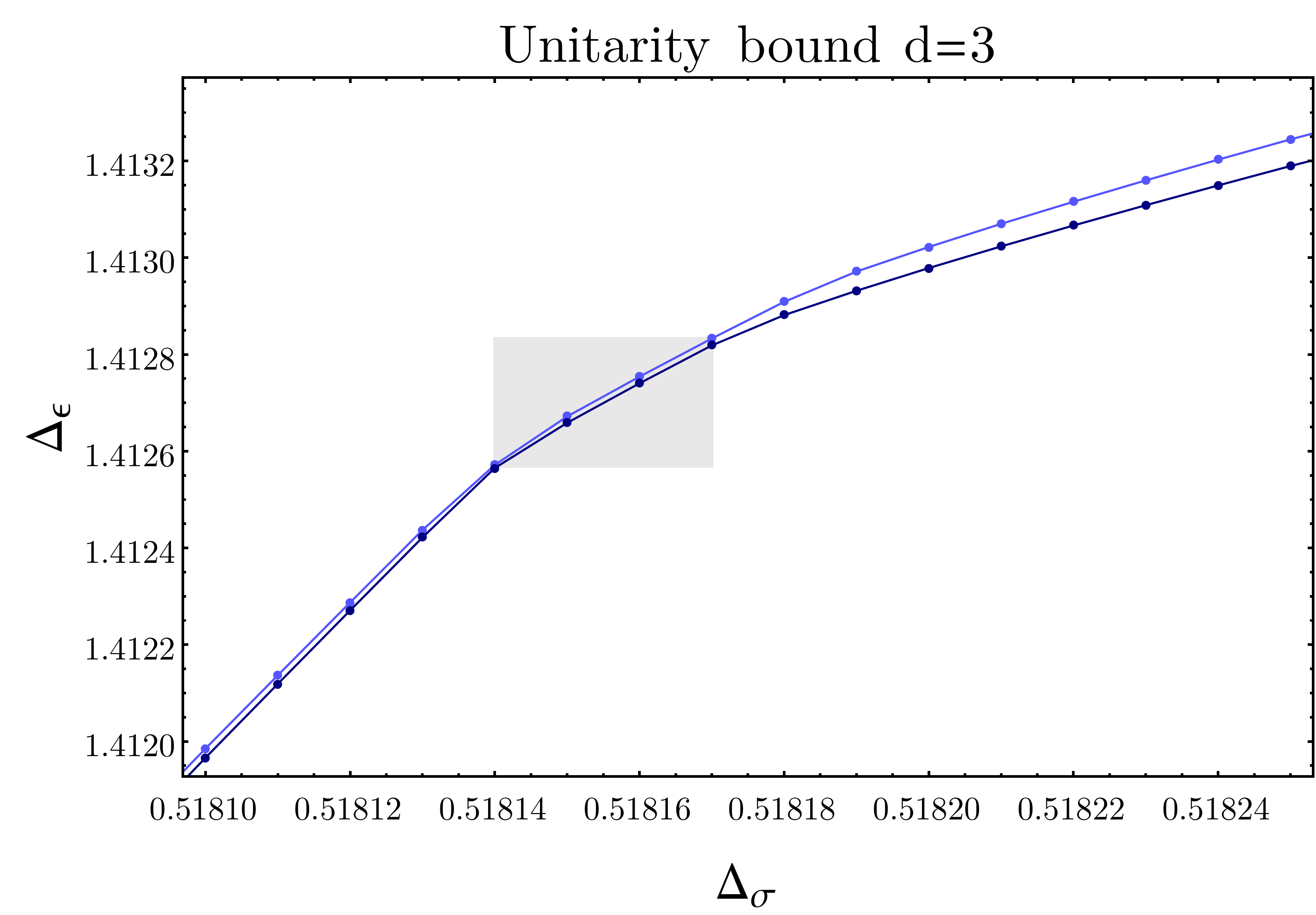}}
\subfigure[\label{fig1b}]{\includegraphics[scale=0.35]{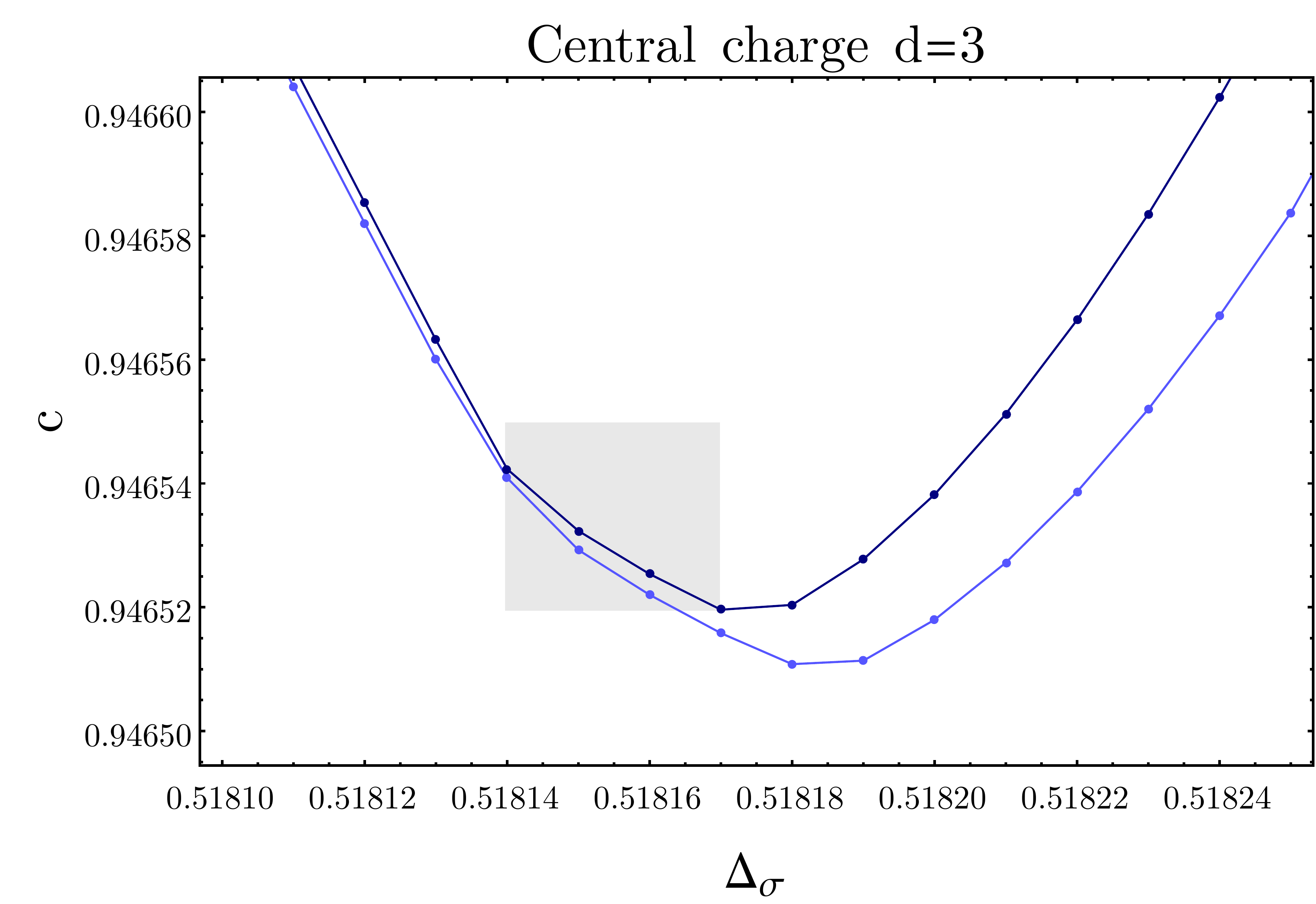}}
\caption{Determination of the Ising critical point for $d=3$ 
from: (a) kink on unitarity boundary; (b) minimum of central charge.
The light (dark) blue lines correspond to bootstrap equations with 153 (190)
components. The grey area indicates the estimated error.}
\label{fig1}
\end{figure}

A crucial point is the determination of the value of $\D_\s$ on the unitarity
boundary corresponding to the Ising model, which then fixes the other 
conformal data. In earlier works, this has been identified by the
point where the boundary in the $(\D_\s,\D_\e)$ plane has a kink
and by the minimum of the central charge $c$. 
These two features are shown in Fig. \ref{fig1a} and \ref{fig1b},
respectively. Let us discuss this issue in detail.

The analysis of Ref. \cite{slava-IM} identified the Ising point by 
the minimum of $c$ in Fig. \ref{fig1b}: the authors noted that
by improving the bootstrap precision from 153 (light blue points) 
to 190 (dark blue points) and 231 (not shown) components,
the minimum moves to the left and eventually stays within
the range $\D_\s= 0.51814 - 0.51817$, from which they extracted the 
precise value $\D_\s= 0.518154(15)$ and the estimated error.

In this work, we adopt a slightly different approach that is
convenient for the other dimensions $d$ as well: we consider both the
$c$ minimum and the kink position. In Fig. \ref{fig1b}, the former is
placed to the right of $\D_\s= 0.518170$, as said; in Fig. \ref{fig1a},
the latter is identified by the crossing of the left and right
tangents to the boundary, around $\D_\s= 0.518150$.  As expected, the
kink and the minimum are not exactly at the same point, thus we should
choose a value in between, and use the mismatch as an estimate of the
error. We thus take an error range of four data points as indicated by the 
gray area in Fig. \ref{fig1a}.
The values of $\D_\e$ and $c$ and their errors are found by reading 
the corresponding ranges on the $y$ axes of  Fig. \ref{fig1a} and
\ref{fig1b}. 

In summary, our identification of the Ising point gives the values
$(d=3)$:
\ba
\D_\s&=& 0.518155(15), 
\label{ds}\\
\D_\e&=&  1.41270(15),
\label{de}\\
c&= & 0.946535(15).
\label{cc}
\ea 
These results are very similar to those of Ref. \cite{slava-IM},
(see Tables 1 and 2 for a comparison); our errors are slightly worse
owing to the conventions just explained. The enhanced $c$ error 
in Fig. \ref{fig1b} will be clarified later.
It is very interesting and reassuring that we find consistent 
results by using different numerical routines. 

Let us remark that the value of $c$ is determined by the structure constant
$f_{\s\s T}$ of the stress tensor $T$ using the formula:
\be
f_{\s\s T}^2=\frac{d\D_\s^2}{4(d-1) c}.
\ee
The central charge is normalized to $c=1$ for the free bosonic theory
and the structure constants are defined within the conventions 
of the 3-correlator study \cite{sd}, differing from those of
the 1-correlator work \cite{slava-IM} by the factor:
\be
\left(f_{\s\s{\cal O}_\ell} \right)^2_{\rm 3-corr} =
\left(f_{\s\s{\cal O}_\ell} \right)^2_{\rm 1-corr} 
\frac{\left(\frac{d-2}{2}\right)_\ell}{\left(d-2\right)_\ell},
\label{convent}
\ee
where $\ell$ is the spin of the field ${\cal O}_\ell$ and
 $(x)_\ell$ is the Pochhammer symbol.

The solution of the bootstrap equations on the unitarity boundary
yields a spectrum of conformal dimensions $\D_{\ell,i}$ as functions
of the varying parameter $\D_\s$, divided into sectors of spin
$\ell=0,2,4,\dots$ and numbered by increasing size, $i=1,2,\dots$.
As said, our results for these curves are
almost identical to those of Ref. \cite{slava-IM} and are not redrawn
here. The data for the Ising model are obtained
by reading the values of the curves at the point $\D_\s$ given in (\ref{ds});
the corresponding errors are given by the maximal curve variation within the  
$\D_\s$ error range, as in the case of $\D_\e$ in Fig. \ref{fig1a}. 
The subleading fields, $\D_{i,\ell}$, $i=2,\dots$, 
show larger fluctuations within this range,
because the spectrum changes considerably 
around the Ising point and the errors become larger and larger. 
Of course, we report the best data obtained by the bootstrap truncated
to 190 components; the comparison with the 153-components data is sometimes
used as an estimate of systematic errors.

Our results for first few low-lying dimensions are shown in
Tab. \ref{tab1}: they are slightly better than those of Ref. \cite{slava-IM},
owing to the improved numerical algorithms, and  
are then compared with the three-correlator bootstrap
results \cite{sd}, basically one order of magnitude more
precise.  The analogous comparison of structure constants is given in
Tab. \ref{tab2}.

\begin{table}
\centering
\be 
\begin{array}{|c|l|l|l|l|}  
\hline   
\D(d=3)& \ell &\ \ {\rm this\ work} &{\rm 3-correlator} & 
{\rm 1-correlator}\\  
\hline
\D_\s & 0 & 0.518155(15) & 0.5181489(10) & 0.518154(15)\\
\D_\e  & 0 & 1.41270(15) & 1.412625(10) & 1.41267(13) \\  
\D_{\e '} & 0 & 3.8305(15) & 3.82968(23) & 3.8303(18) \\ 
\D_{\e ''} & 0  & 7.01(5) & 6.8956(43) &  \\ 
\hline
\D_{T'} & 2 & 5.505(10) & 5.50915(44)& 5.500(15) \\ 
\D_{T''} & 2 & 7.25(55) & 7.0758(58) & \approx 7\\ 
\hline
\D_C  & 4 & 5.026(4) & 5.022665(28) & \\ 
\D_{C'} & 4 & 6.67(23) & 6.42065(64) &\\ 
\D_{C''} & 4 & 7.45(5) & 7.38568(28) &\\ 
\hline
\end{array}
\nonumber
\ee
\caption{Comparison of $d=3$ conformal 
dimensions of low-lying fields with earlier results
of one-correlator \cite{slava-IM} and three-correlator \cite{sd} bootstrap.}
\label{tab1}
\end{table}

\begin{table}
\centering
\be 
\begin{array}{|c|l|l|l|}
\hline 
f_{\s\s\cal O} & {\rm\ this\ work} & {\rm 3-correlator} & 
{\rm 1-correlator}\\ 
\hline 
f_{\s\s\epsilon} & 1.051835(35) & 1.0518537(41) & 1.05184(4) \\ 
f_{\sigma\sigma\epsilon '} & 0.05300(5) & 0.053012(55) & 0.05301(5) \\ 
f_{\sigma\sigma T} & 0.326142(7) & 0.32613776(45) & 0.326142(8)  \\ 
c & 0.946535(15) & 0.9463385(60) & 0.946534(11)\\
f_{\sigma\sigma T'} & 0.010575(15) & 0.0105745(42) & 0.01055(4)  \\ 
f_{\sigma\sigma C} & 0.065(5) & 0.069076(43)  &\\ 
f_{\sigma\sigma C'} & 0.0020(5) & 0.0019552(12)  &\\
\hline
\end{array}
\nonumber
\ee
\caption{Comparison of $d=3$ structure constants of six low-lying
  fields with earlier results of one-correlator \cite{slava-IM} and
  three-correlator \cite{sd} bootstrap (the values of $f_{\s\s T}$ and
  $c$ are related).}
\label{tab2}
\end{table}

The analysis of the data in the two tables let us to conclude that there are
six fields, ${\cal O}=\s,\e,\e',T',C,C'$ of spin $\ell=0,2,4$, 
whose dimensions $\D_{\cal O}$ and structure constants $f_{\s\s\cal O}$
are sufficiently precise to be worth analyzing in other dimensions
$4> d> 2$. Note that the structure constants $f_{\s\s\cal O}$ 
are sometimes better determined than the corresponding dimensions $\D_{\cal O}$.

\begin{figure}
\centering
\subfigure[\label{fig2a}]{\includegraphics[scale=0.21]{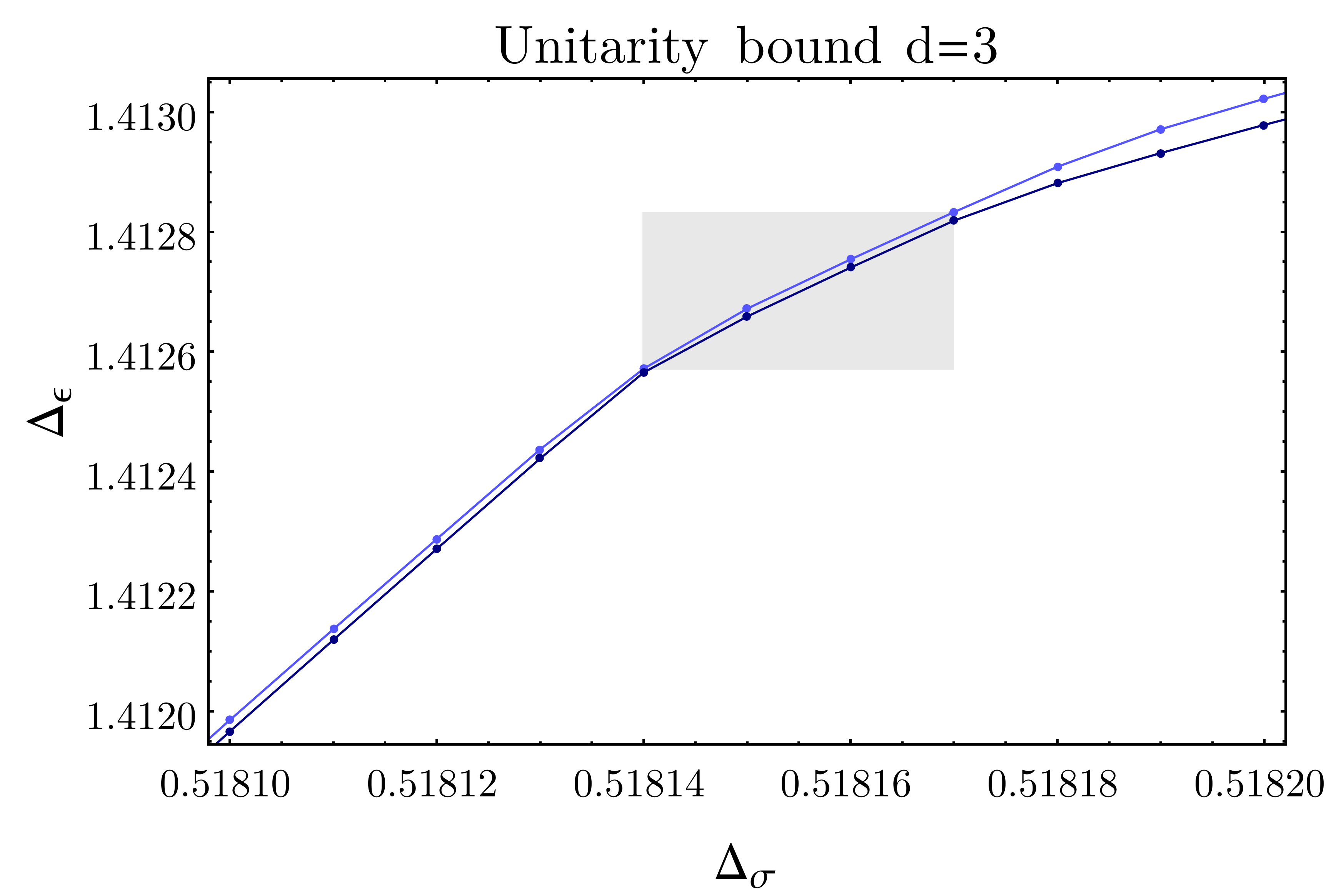}} 
\hspace{.1cm}
\subfigure[\label{fig2b}]{\includegraphics[scale=0.21]{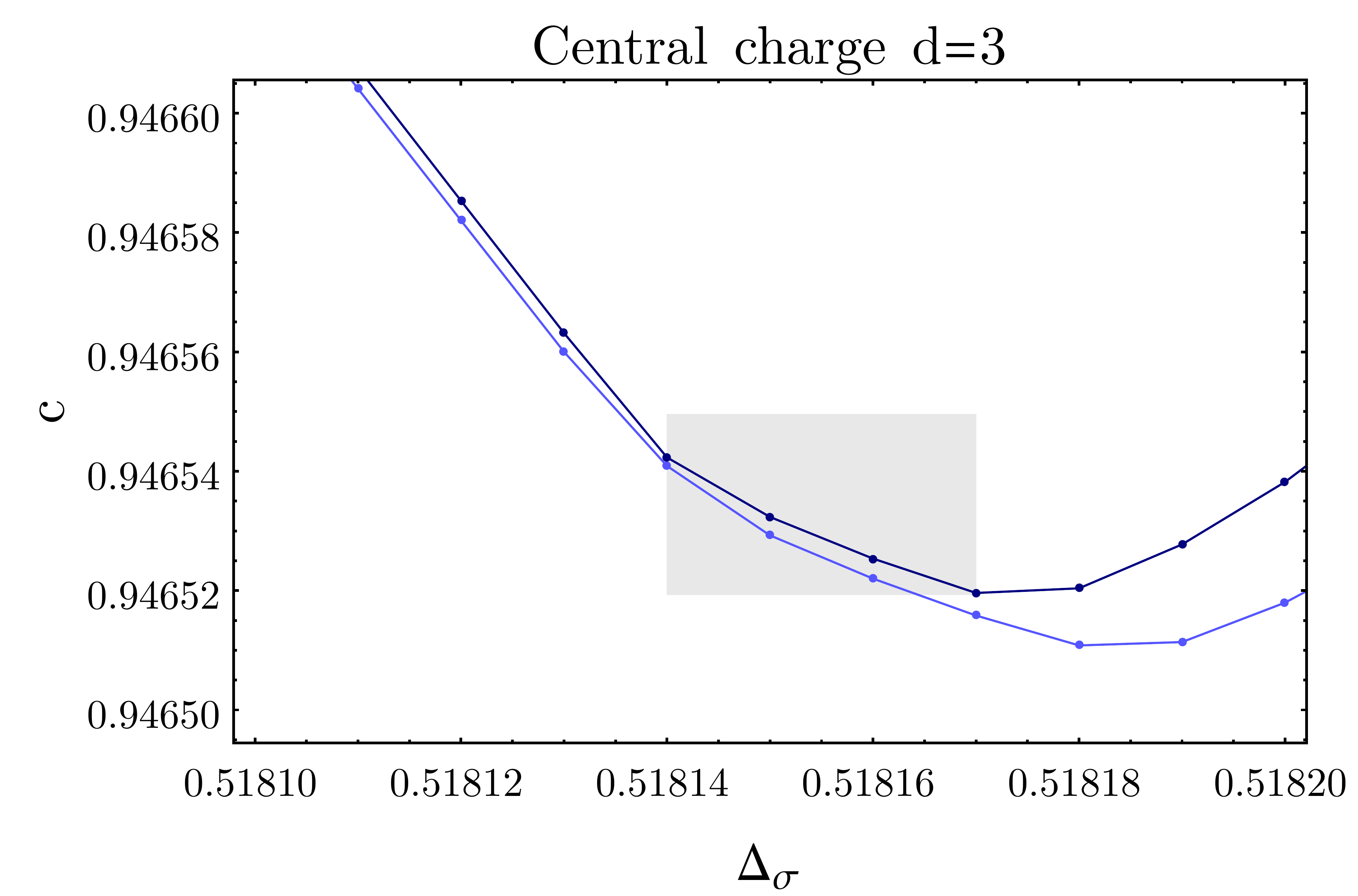}} 
\subfigure[\label{fig2c}]{\includegraphics[scale=0.21]{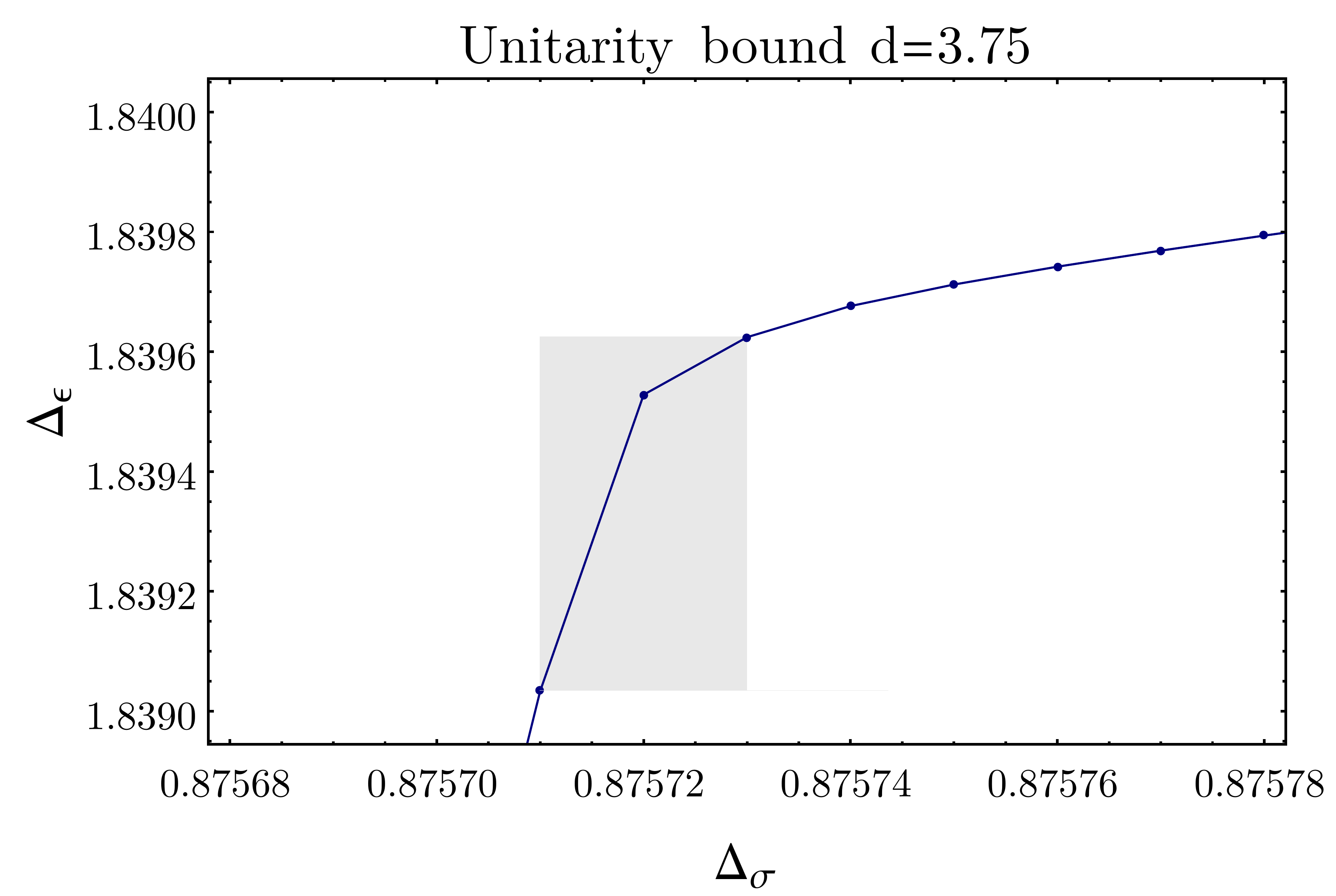}} 
\hspace{.1cm}
\subfigure[\label{fig2d}]{\includegraphics[scale=0.21]{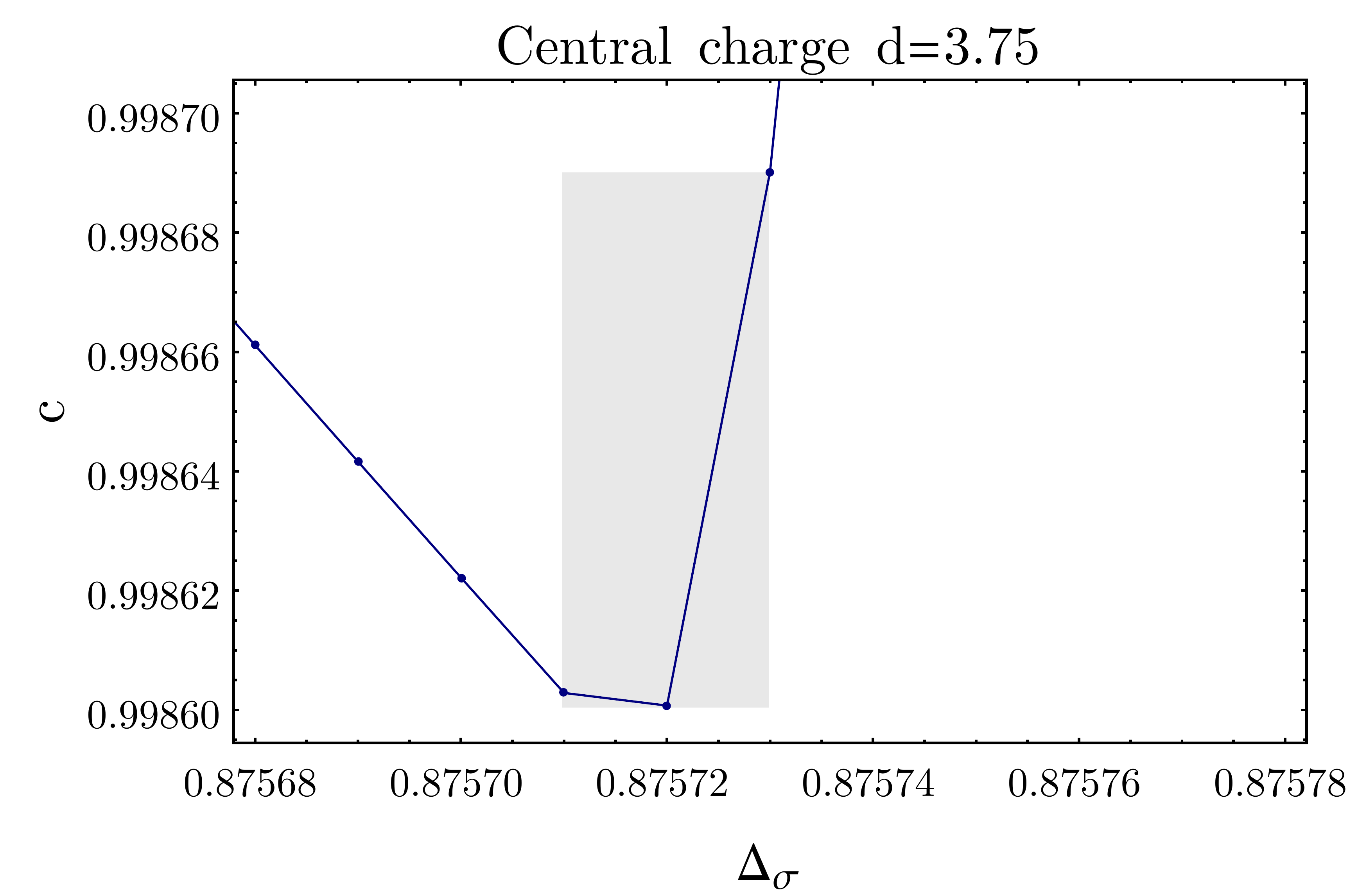}} 
\subfigure[\label{fig2e}]{\includegraphics[scale=0.205]{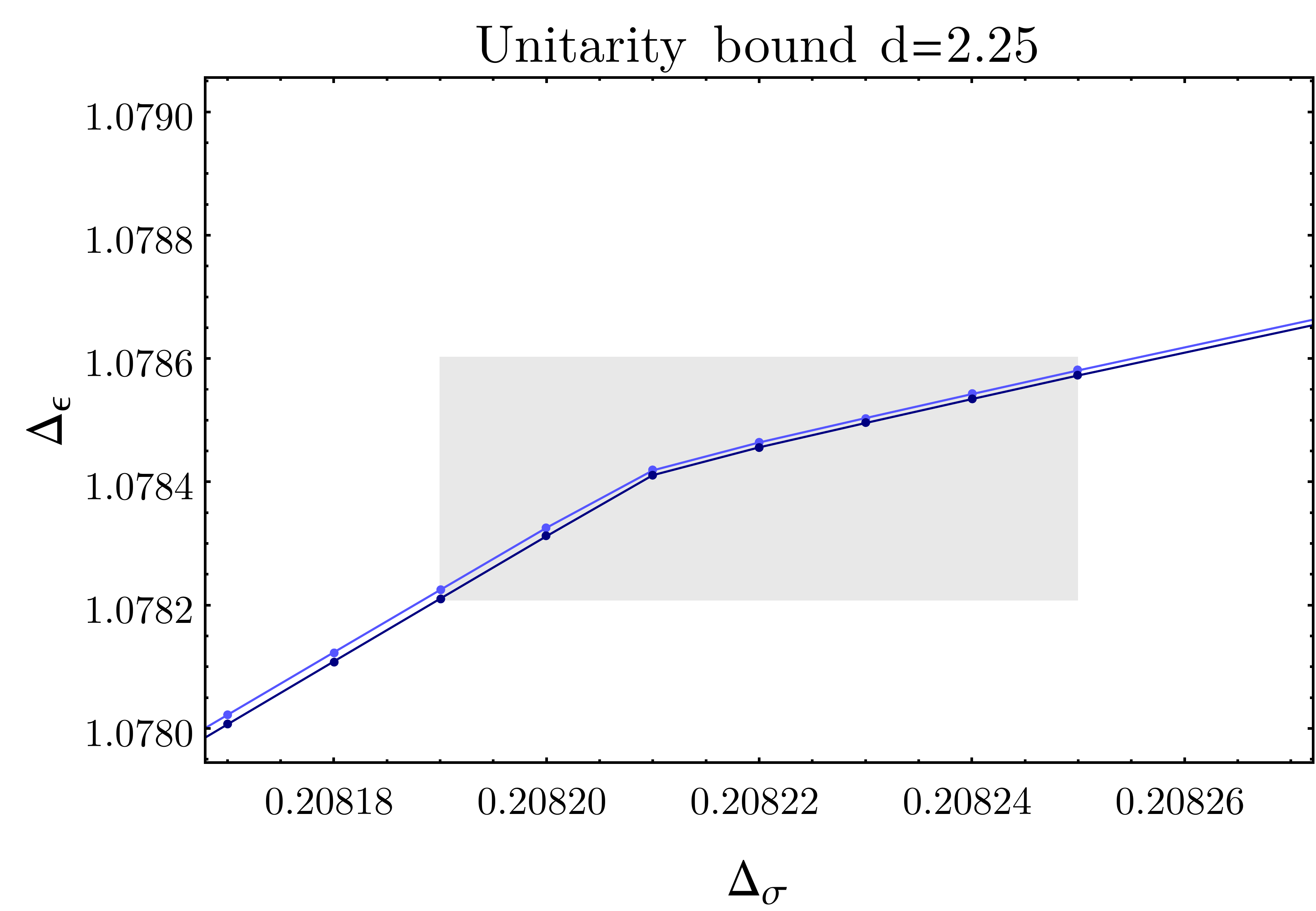}} 
\hspace{.3cm}
\subfigure[\label{fig2f}]{\includegraphics[scale=0.205]{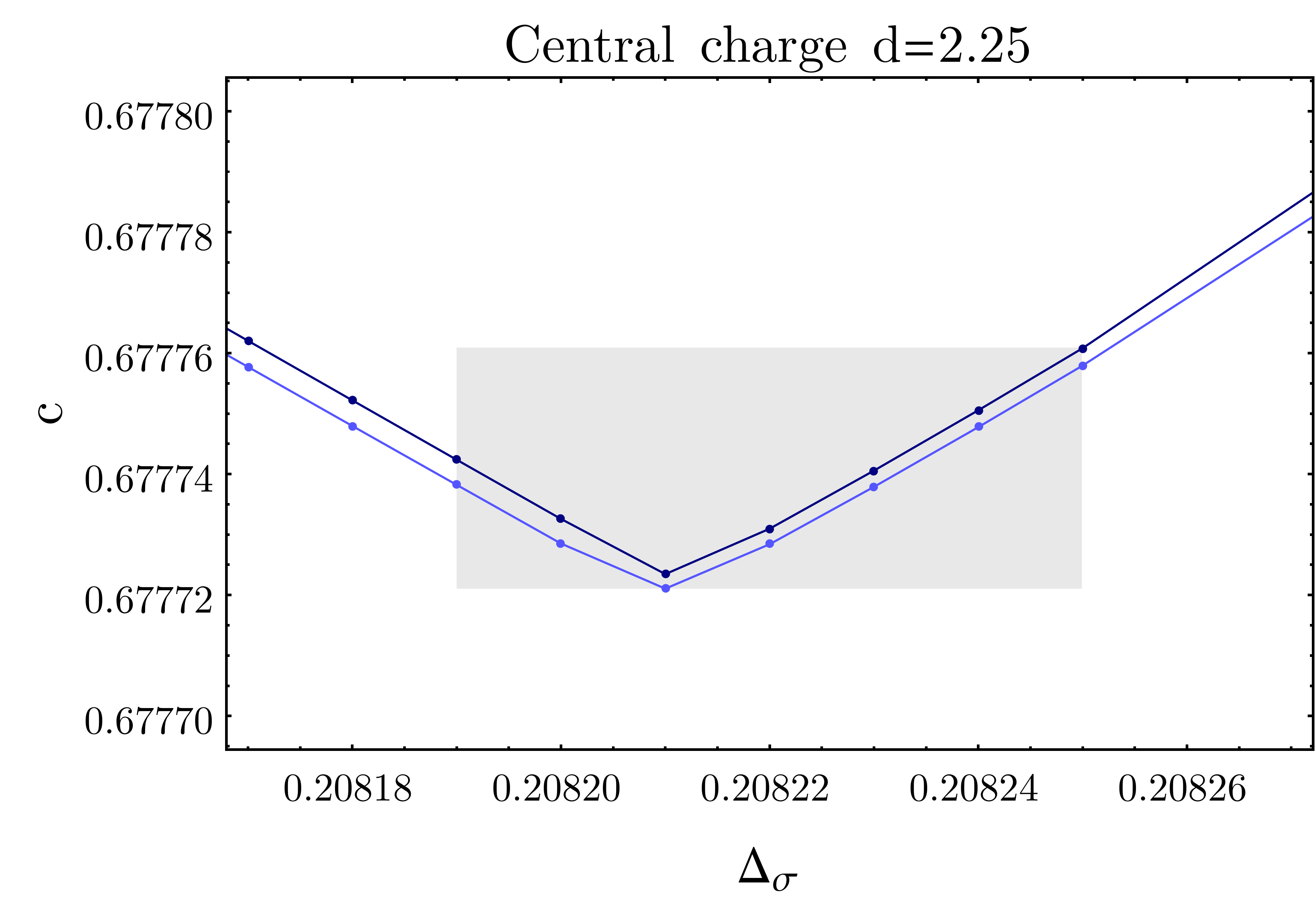}} 
\caption{Determination of the Ising critical point for $d=3$ 
(rescaled view of Fig. \ref{fig1}), $d=3.75$ and $d=2.25$.
As in Fig. \ref{fig1}, the light (dark) blue lines correspond 
to bootstrap equations with 153 (190)
components and the grey area is the error.}
\label{fig2}
\end{figure}

\begin{figure}[t]
\centering
\subfigure[\label{fig3a}]{\includegraphics[scale=0.21]{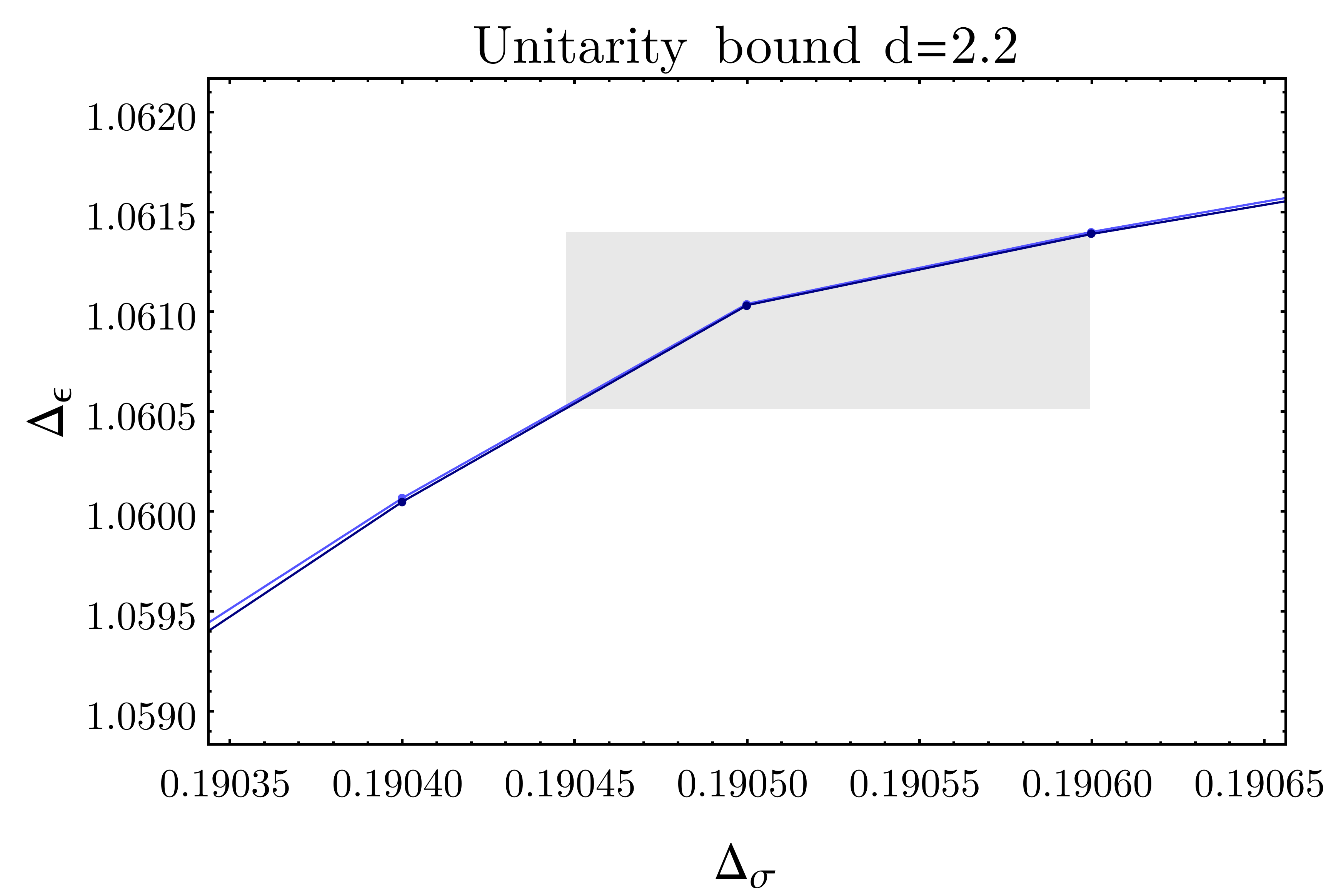}} 
\hspace{.1cm}
\subfigure[\label{fig3b}]{\includegraphics[scale=0.21]{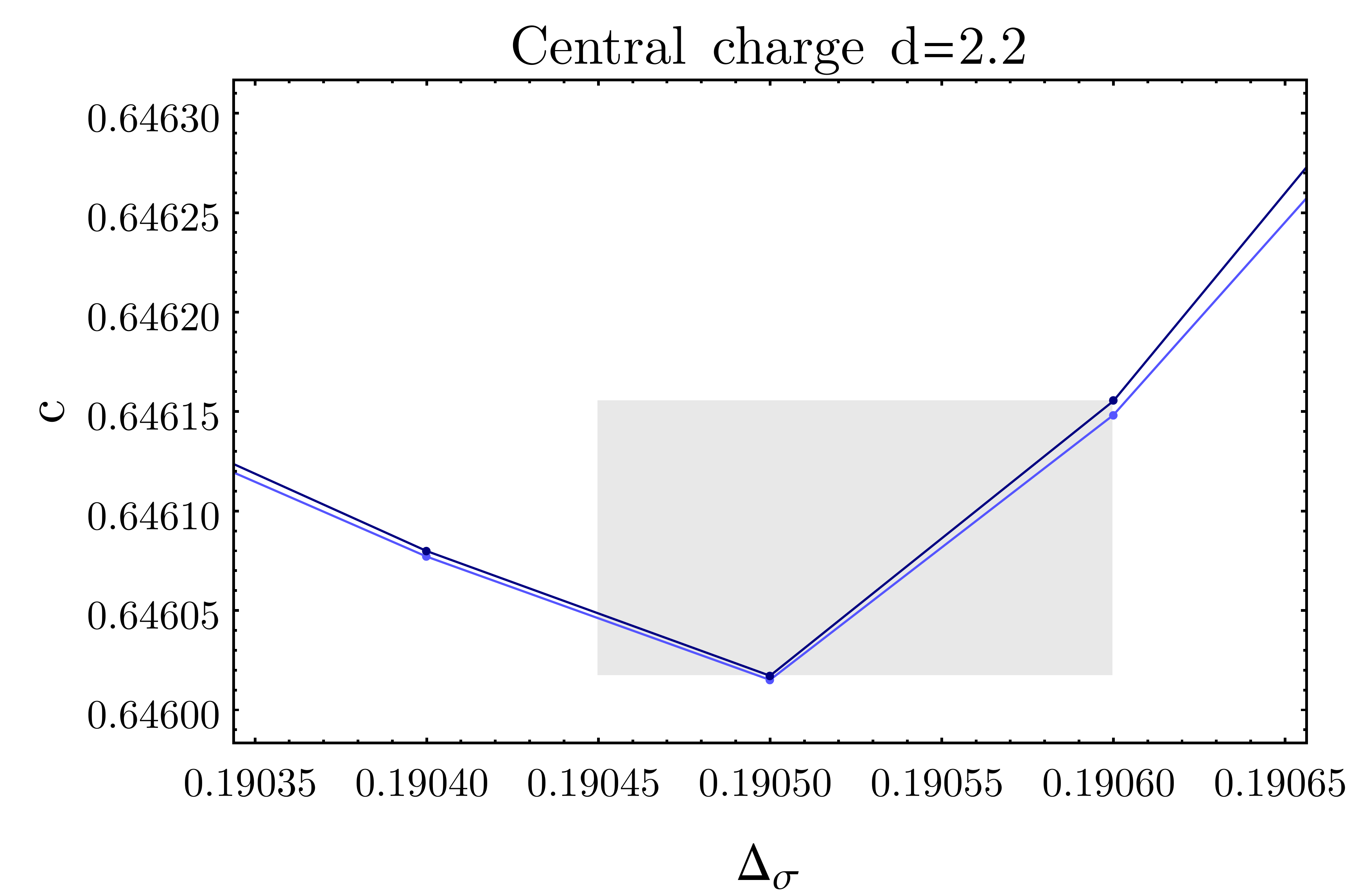}} 
\subfigure[\label{fig3c}]{\includegraphics[scale=0.2]{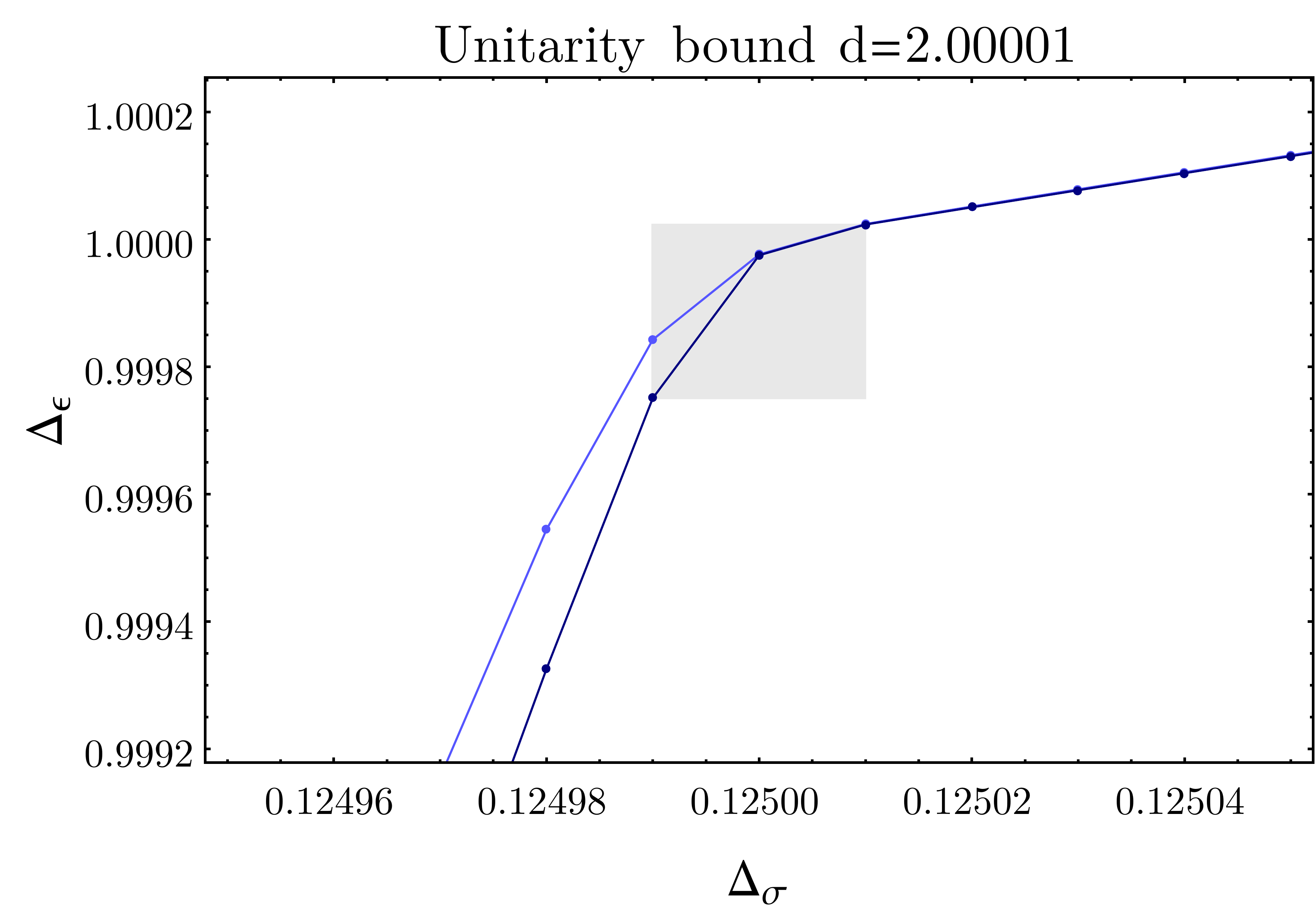}} 
\hspace{.3cm}
\subfigure[\label{fig3d}]{\includegraphics[scale=0.2]{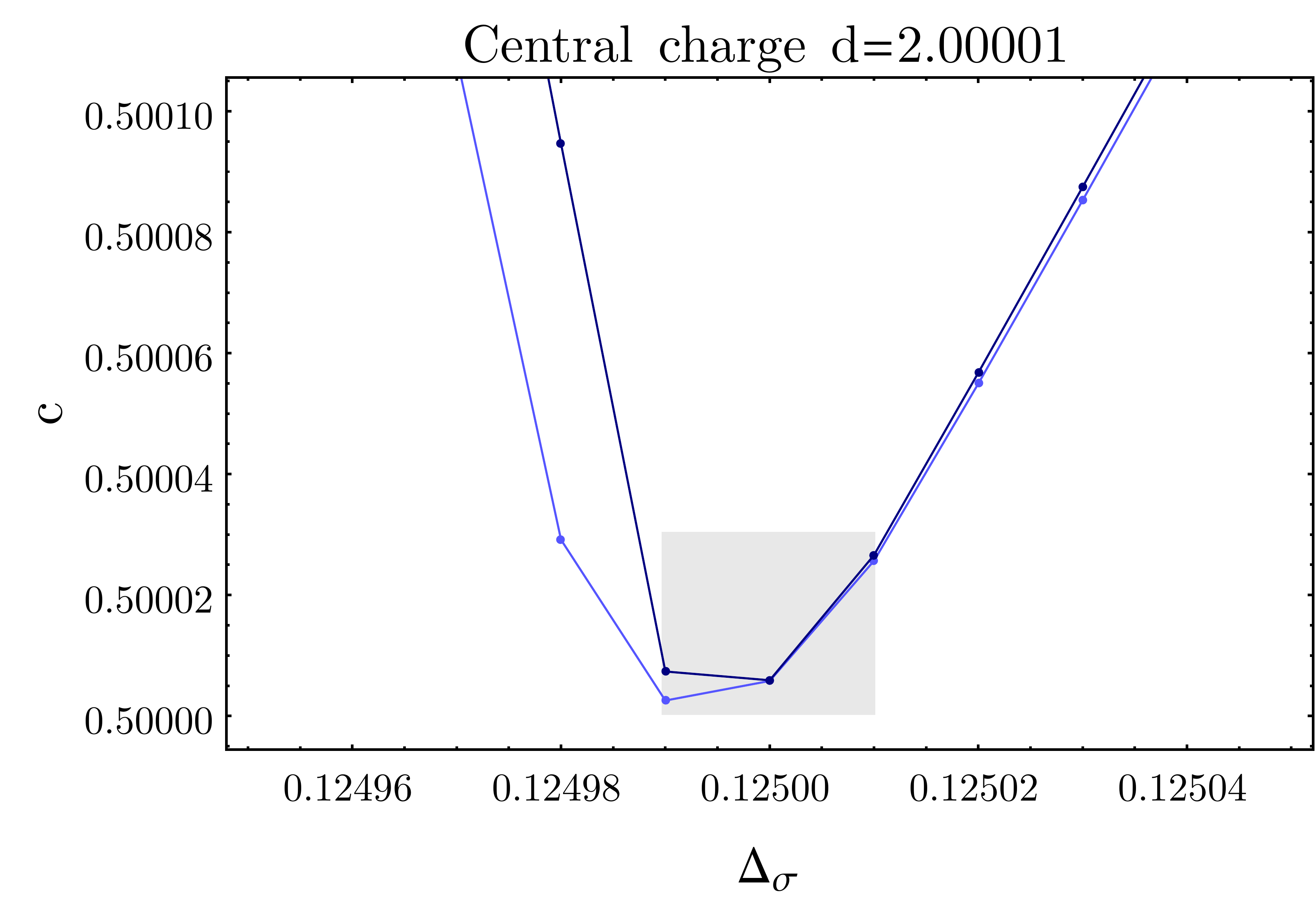}} 
\caption{Determination of the Ising critical point for $d=2.2$ 
(smaller scale) and $d=2.00001$.}
\label{fig3}
\end{figure}

%-2.2--------------------------------------------- 

\subsection{Polynomial fits of $4>d> 2$ data and comparison with 
other approaches}

The determination of the Ising critical point on the unitarity
boundary for dimensions $d=3$, $3.75,$ $2.25,$ $2.2,$ $2.00001$ is shown in
Fig. \ref{fig2}, \ref{fig3}: the plots are drawn on equal intervals
$(0.0001,0.001, 0.0001)$ of $(\D_\s,\D_\e,c)$, respectively, for
comparison, with the exception of the $d=2.2$ case, whose ranges are
three times larger.  We note that the minimum and kink are 
of comparable shape for $3\ge d\ge 2.2$ and become sharper at
$3.75$ and at $d=2.00001$ again.  
The data shows that it is not always clear where the $c$ minimum
moves as the precision of the bootstrap is improved (from 153 to 190
components) and that
the kink position helps in the identification of the Ising point.
The plots at $d=2.2$ with reduced scale is shown as an example of error
determination in the region $2.2 \ge d\ge 2.01$,
where our sampling of the unitary boundary is coarser.
In all cases the error range is chosen to be no less than $2-3$ data points.

\subsubsection{Conformal dimensions}

Tables \ref{tab3} summarize the determination of dimensions for the
six low-lying fields at twelve values of dimension. One sees that the
precision attained at $d=3$ is roughly maintained for $4>d\ge 3$, it
decreases in going from $d=3$ to $d=2.01$ and then is good again 
at $d=2.00001$.
The lower quality around $d=2.25$ is due to the reordering of 
the conformal towers of states for approaching the $d=2$ theory 
that will be discussed in Section three.
For $2.2\ge d\ge 2.01$, the reduced data sampling also affects the results.

\begin{table}[h]
\centering
\be 
\resizebox{\hsize}{!}{$
\begin{array}{|c|l|l|l|l|l|l|}
\hline 
d & \Delta_\sigma & \Delta_\epsilon  & \Delta_ {\e'} & \Delta_{ T'} 
& \Delta_ C  & \Delta_ {C'} \\ 
\hline 
\mathbf{4} & \mathbf{1} & \mathbf{2} & \mathbf{4}  & 
\mathbf{6} & \mathbf{6} & \mathbf{8} \\ 
3.75 & 0.87572(1) & 1.83932(30) & 3.958(23) & 5.8622(14) & 5.750995(25) 
& 7.805(15) \\ 
3.5 & 0.753395(15) & 1.68851(31) & 3.921(11) & 5.734(6) & 5.50465(15) 
& 7.48(6) \\ 
3.25 & 0.633885(15) & 1.54638(18) & 3.8770(25) & 5.613(8) & 5.2625(15) 
& 7.145(35)  \\ 
3 & 0.518155(15) & 1.41270(15) &  3.8305(15) &  5.505(10) &  5.026(4)
&  6.67(23)\\ 
2.75 & 0.407465(35) & 1.2887(2) & 3.800(2) & 5.445(15) & 4.790(5) & 6.3(2) 
\\ 
2.5 & 0.30341(1) & 1.17625(15) & 3.797(1) & 5.455(25) & 4.574(9) & 5.78(13)
 \\
2.25 & 0.20822(3) & 1.0784(2) & 3.847(1) & 5.575(45) & 4.344(14) & 5.36(6) 
\\
2.2 & 0.19053(8) & 1.06095(45) & 3.864(4) & 5.685(35) & 4.325(15) & 5.29(4)
 \\
2.15 & 0.17333(8) & 1.04435(35) & 3.891(6) & 5.64(13) & 4.275(25) & 5.19(1)
 \\
2.1 & 0.14663(8) & 1.02855(45) & 3.9215(5) & 5.82(1) & 4.165(35) & 
5.115(35) \\
2.05 & 0.14048(8) & 1.0134(7) & 3.9565(5) & 5.905(1) & 4.13(6) & 5.065(15)
 \\
2.01 & 0.12803(8) & 1.0011(17) & 3.990(1) & 5.9815(5) & 4.0144(1)& 
5.0115(15)\\
2.00001 & 0.12500(1) & 0.99989(14) & 4.00015(20) & 6.0006(2) & 
4.000055(10) & 5.00048(8) \\
\mathbf{2} & \mathbf{0.125} & \mathbf{1} & \mathbf{4}  & \mathbf{6} 
& \mathbf{4} & \mathbf{5} \\ 
\hline
\end{array}
\nonumber
$}
\ee
\caption{Dimensions of six low-lying states for $4>d>2$. The exact values
for $d=2,4$ are given in bold.}
\label{tab3}
\end{table}

\begin{figure}[h]
\centering
\includegraphics[scale=0.35]{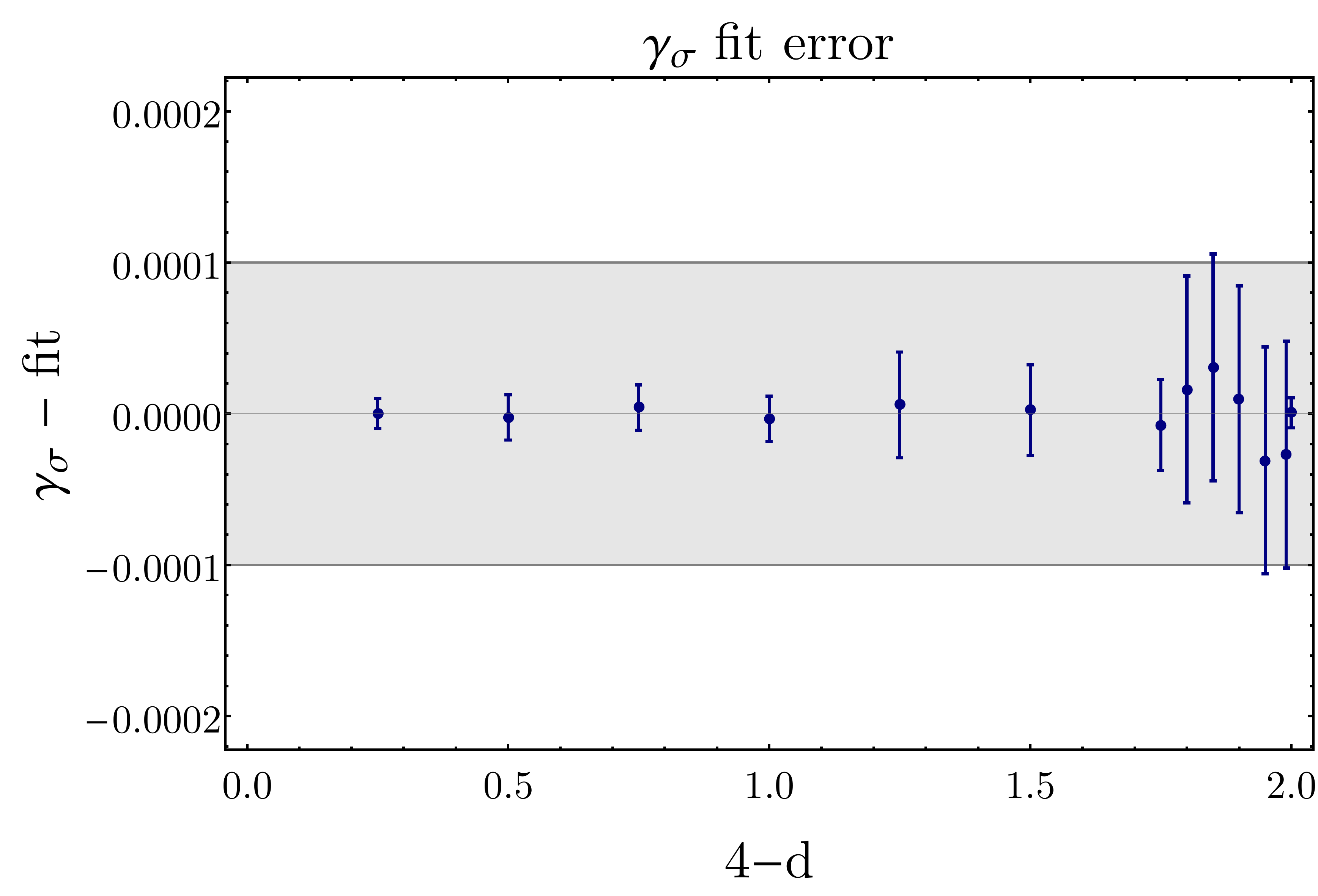}
\caption{Quality of the polynomial fit for $\g_\s(4-d)$ bootstrap data
in the range of dimensions $4> d\ge 2$. The maximal overall error of 
the fit is indicated by a grey band.}
\label{fig4}
\end{figure}

\begin{figure}[h]
\centering
\includegraphics[scale=0.35]{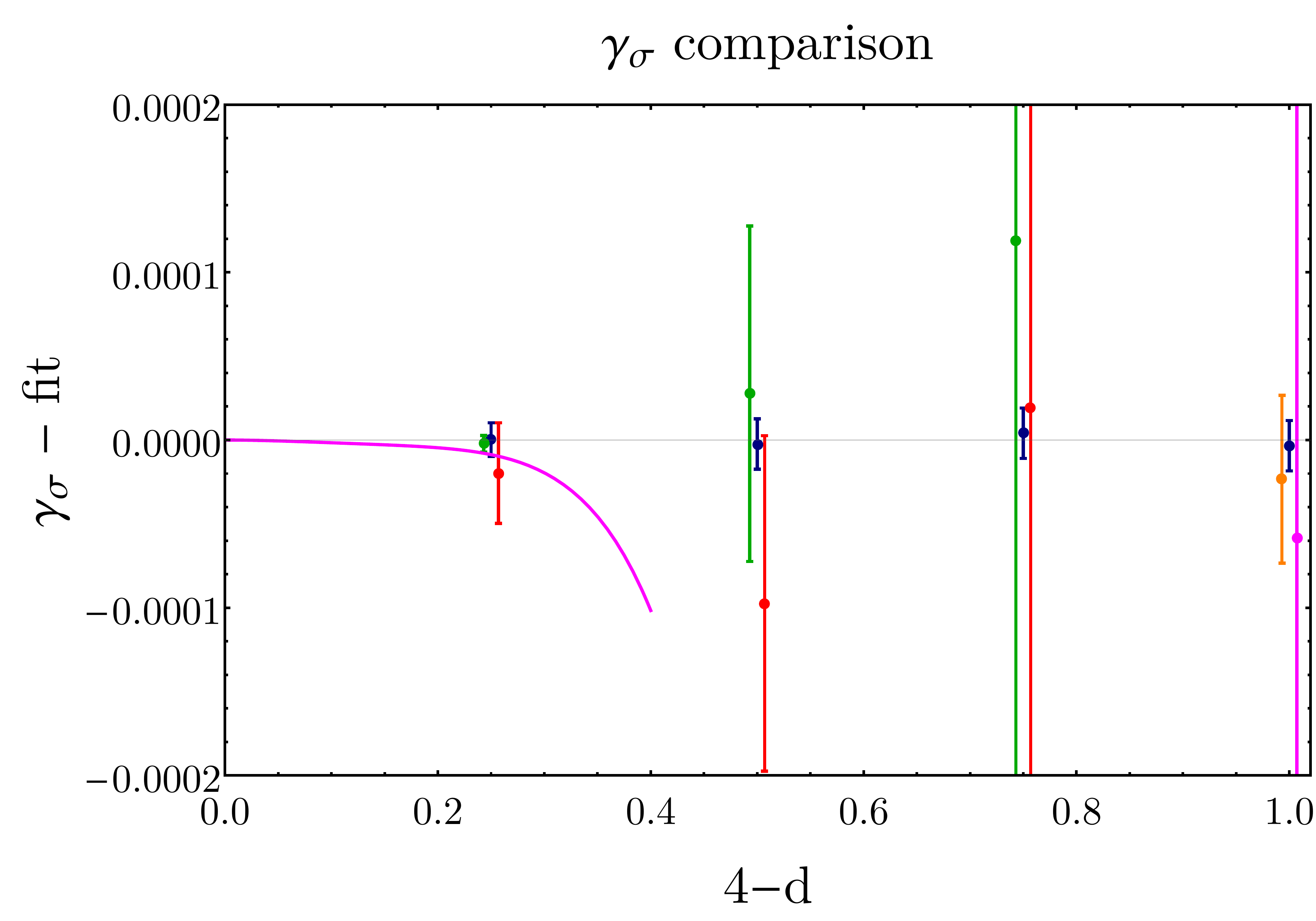}
\caption{Comparison of $\g_\s$ bootstrap data (blue) with: 
i) Borel-resummed epsilon expansion  (green) \cite{zinn} and ii)
3-correlator bootstrap data (red) \cite{behan} for $d=3.75,3,5,3.25$ 
(the $d=3$ coarse data are omitted); 
iii) unresummed high-order epsilon expansion and resummed estimate
at $d=3$ (magenta) \cite{panzer} 
iv) Monte Carlo data at $d=3$ (orange) \cite{mc} (Note that data points 
are slightly displaced around the same $d$ value for improving readability).}
\label{fig5}
\end{figure}

We now proceed to analyze the data in each of the columns of Table 3
and study their dependence on spacetime dimension, starting with
the case of $\D_\s$. We perform a least chi-square fit of the
anomalous dimension $\g_\s=\D_\s-(d-2)/2$ with a polynomial in
$y=4-d$, and obtain the result:
\ba\label{s-fit}
&\g_\s(y) =& 0.00955001 y^2 + 0.00764826 y^3 + 0.00091284 y^4 
- 0.00024948 y^5 
\nl
&& +  0.000296768 y^6, \qquad\qquad\qquad\qquad
{\rm Err} \left(\g_\s\right) < 0.0001 .
\ea
The polynomial is of $(2-6)$-th order, i.e. it is supposed to start
quadratically at $d=4$, as suggested by the epsilon
expansion of $\l\f^4$ theory (a linear term would not improve the fit). 
Moreover, it includes a sufficient number
of terms for obtaining a reasonably small chi-square 
value\footnote{
Some digits in the coefficients of the polynomial (\ref{s-fit}) might
be redundant for the achieved precision but are kept anyway.}.
Since the fit is very precise, we do not plot the curve $\g_\s(y)$
but its difference with respect to the data points (see Fig. \ref{fig4}): 
one sees that  $\g_\s(y)$ is determined 
with an overall error that is less than $10^{-4}$, although some individual
points are far more precise. Since $\g_\s =O(10^{-1})$, the relative error
is less than $10^{-3}$. Note that the very good matching of the $d=2$ 
value is not imposed but obtained by the fit.

Equation (\ref{s-fit}) is one of the most interesting results of this work,
expressing the conformal dimension $\D_\s$ as a function of $d$ with
high precision, a long-sought goal of quantum field theory
since the seventies \cite{wf}.
Let us compare with the results of other methods.
Fig. \ref{fig5} shows again the difference between
data and fit (limited to the range $4>d\ge 3$), together
with the following inputs:
\begin{itemize}
\item 
Green data points -- Borel-resummed epsilon expansion of 
$\l\f^4$ theory obtained in the
nineties \cite{zinn} at $d=3.75,3.5,3.25$ (the $d=3$ value is less precise
and is omitted); note the very good match at $d=3.75$.
\item
Magenta curve and data point -- unresummed
high-order epsilon expansion series \cite{panzer}:
\ba
\!\!\!\!
\g_\s(y)&=&0.00925925 y^2 + 0.009345 y^3 - 0.00416438 y^4 + 0.0128283 y^5 
\nl
&&- 0.0406359 y^6 ,
\qquad\qquad\qquad \qquad\qquad {\rm (Epsilon\ expansion)},
\label{s-ee}
\ea
and $d=3$ value obtained by Borel resummation.
\item
Orange data point -- Monte Carlo result at $d=3$ \cite{mc}.
\item
Red data points -- three-correlator bootstrap at $d=3.75,3.5,3.25$
obtained in Ref. \cite{behan}\footnote{
We thank C. Behan for providing his data to us.}.  
In this approach,  the Ising point is uniquely identified 
by a small unitarity island in the $(\D_\s,\D_\e)$ plane; thus,
these data provide a check for our kink-minimum criteria discussed
before.
\end{itemize}
Finally, our results have been checked against the early bootstrap
data\footnote{
Nicely sent us by S. Rychkov.} of Ref. \cite{slava-d},
finding agreement within their uncertainties.  Let us also
quote the works \cite{mc-bad} where other numerical and
perturbative results for the Ising critical exponents have been discussed.

The comparison in Fig. \ref{fig5} shows  very
good consistency of our $\g_\s(y)$ formula (\ref{s-fit})
with all other methods. Note incidentally that the unresummed
epsilon expansion series (\ref{s-ee}) is extremely good up to $y=0.2$.
Let us consider the following naive argument: if this series
were asymptotic as 
the $\l\f^4$ perturbative expansion at $d=4$ (no proof of this),
the $n$-th term would grow as $O(n!)$ and the sixth-order
unresummed series (\ref{s-ee}) would start diverging at about $y\sim 1/6$. 
The data in Fig. \ref{fig5} are consistent with this estimate
but actually show a milder behaviour.

The best $d=3$ results from bootstrap, epsilon expansion and Monte Carlo
are summarized in Table \ref{tab4}. In this table, some data are obtained
from the critical exponents $\g,\nu, \w$, by assuming the scaling relations 
$\g_\e=2 -1/\nu$, $\g_\s=1-\g/(2\nu)$ and $\D_{\e'}=d+\w$, and simple
independent error propagation.

\begin{table}[h]
\centering
\be 
\begin{array}{|c|l|l|l|l|}  
\hline   
d=3 & {\rm this\ work} &  {\rm 3-correlator} & 
 {\rm eps-expansion} &  {\rm Monte\ Carlo} \\
\hline
\D_\s & 0.518155(15) &  0.5181489(10) & 0.5181(3)& 0.518135(50) \\
\D_\e & 1.41270(15)  &  1.412625(10) & 1.4107(13) & 1.41275(25) \\
\D_{\e'} & 3.8305(15)  &  3.82968(23) & 3.820(7) & 3.832(6) \\
\hline
\end{array}
\nonumber
\ee
\caption{Comparison of $d=3$ results for low-lying
fields from bootstrap data of this work and the three-correlator analysis
\cite{sd}, resummed epsilon expansion \cite{panzer} and Monte Carlo \cite{mc}.}
\label{tab4}
\end{table}

\begin{figure}
\centering
\subfigure[\label{fig6a}]{\includegraphics[scale=0.35]{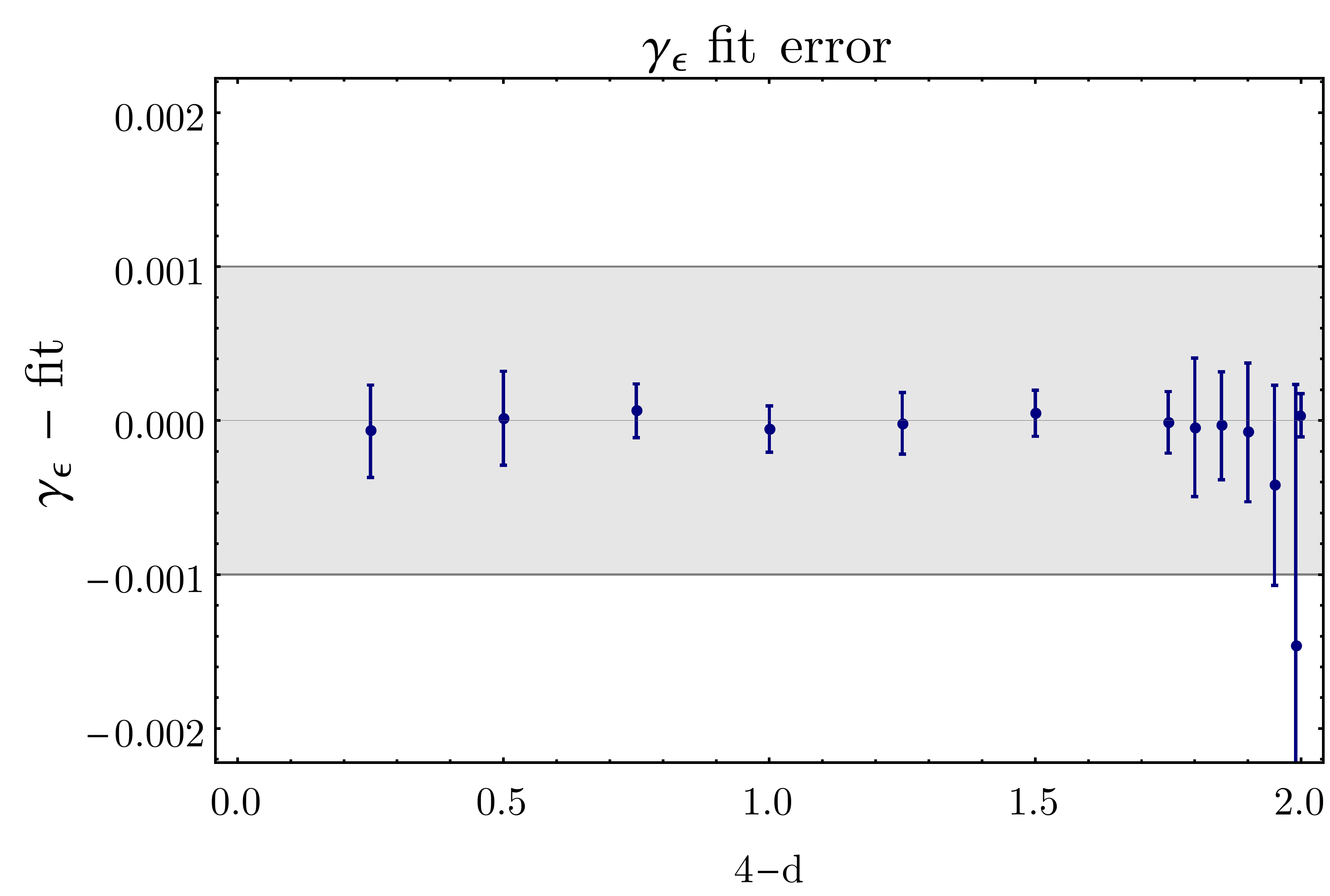}} 
\subfigure[\label{fig6b}]{\includegraphics[scale=0.35]{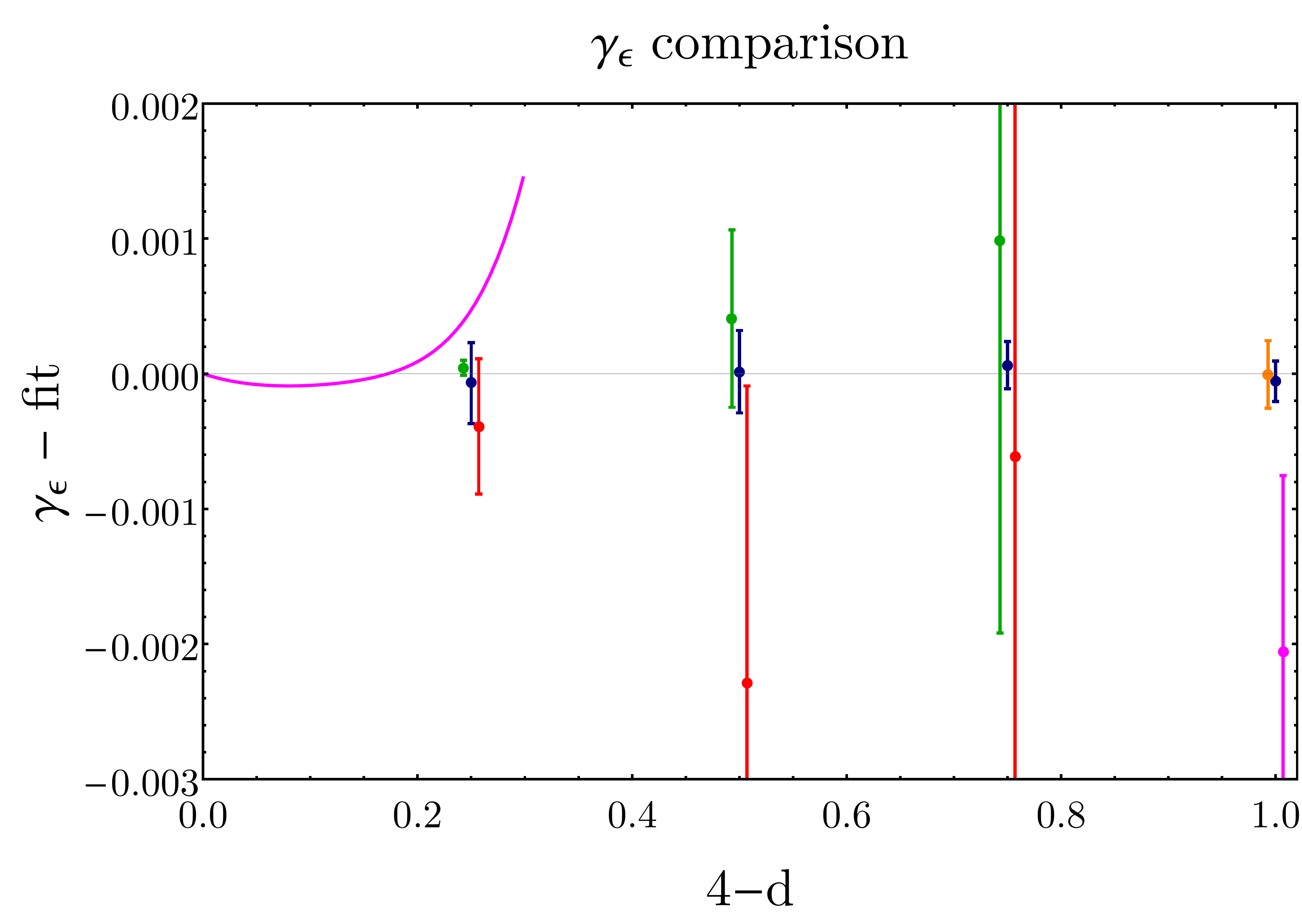}} 
\caption{(a) Polynomial fit of $\g_\e$ bootstrap data with overall error.
(b) Comparison of bootstrap data (blue) with: resummed epsilon expansion 
(green); 3-correlator bootstrap (red);
high-order epsilon expansion and resummed estimate
at $d=3$ (magenta); Monte Carlo (orange) (references as in Fig. \ref{fig5}).}
\label{fig6}
\end{figure}

The previous analysis of the conformal dimension as a function of $d$
is now repeated for the $\e$ field, 
leading to the results in Fig. \ref{fig6}. The part $(a)$ shows the
difference between the data for the
anomalous dimension $\g_\e=\D_\e-d+2$ and the fitting polynomial:
\ba
\g_\e(y) &=& 0.336000 y + 0.0914812 y^2 - 0.0229152 y^3 + 0.00729869 y^4 
\nl
&& + 0.000890045 y^5 ,
\qquad\qquad\qquad\qquad
{\rm Err}\left(\g_\e \right) < 0.001 .
\label{e-fit}
\ea
The fit assumes vanishing $\g_\e$ at $d=4$ and involves five terms
(the sixth one would have negligible coefficient). Note again 
the good match of the exact $d=2$ value $\g_\e=1$. 
The overall error is less than $10^{-3}$ and 
the relative error is also less than $10^{-3}$ as in the case of $\g_\s$.

Figure \ref{fig6b} shows the comparison with other results, that are
plotted using the same color code of Fig. \ref{fig5}. 
The unresummed epsilon expansion is \cite{panzer}:
\ba
\g_\e(y) &=& 0.333335 y + 0.117285 y^2 - 0.124528 y^3 + 0.30685 y^4 - 
 0.95125 y^5 
\nl
&& + 3.57266 y^6,
\qquad\qquad \qquad\qquad {\rm (Epsilon\ expansion)};
\ea
this is also accurate for $y<0.2$. The consistency of all methods is
again rather surprising. 

\begin{figure}
\centering
\includegraphics[scale=0.35]{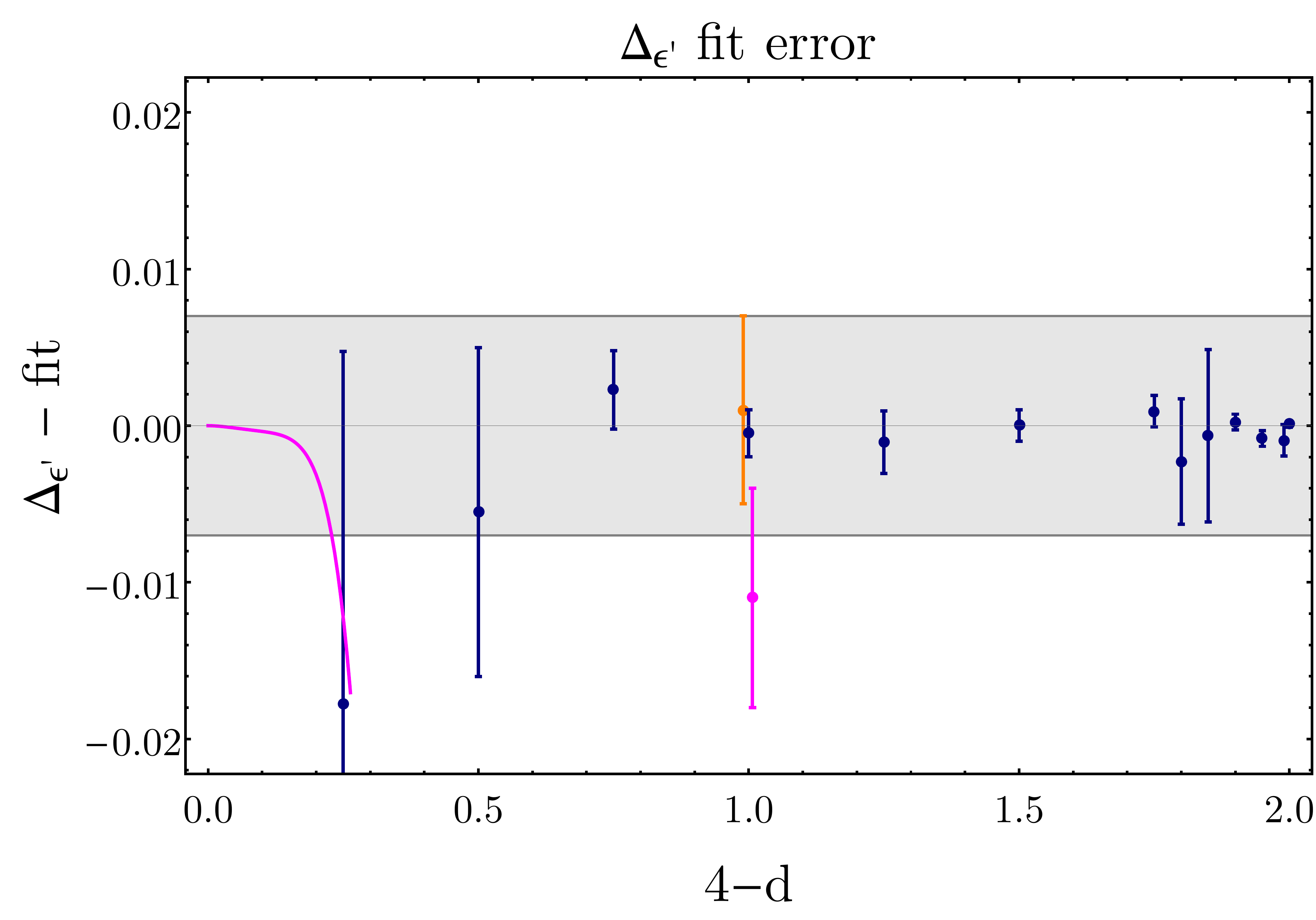}
\caption{Fit of $\D_{\e'}$ bootstrap data (blue) with estimated error,
  high-order epsilon expansion with resummed estimate at $d=3$
  (magenta) and Monte Carlo (orange) (references as in
  Fig. \ref{fig5}).}
\label{fig7}
\end{figure}

The next quantity to analyze is the subleading scalar field $\e'$,
that is related to the scaling exponent $\w=\D_{\e'}-d$, widely analyzed
in the literature. The least chi-square fit of our data gives:
\ba
\D_{\e'}(y) &=& 4 - 0.530509 y^2 + 0.616593 y^3 - 0.334523 y^4 
+ 0.0794284 y^5,
\nl
&&\qquad \qquad\qquad \qquad\qquad \qquad \qquad 
{\rm Err\ }\left(\D_{\e'}\right)  <0.007.
\label{ep-fit}
\ea
It starts quadratically as the epsilon expansion (a linear term would
worsen the fit). Figure \ref{fig7} shows the difference between the data
and the fit, together with the Monte Carlo results \cite{mc} and 
the epsilon expansion series \cite{panzer}:
\ba
\D_{\e'}(y) &=& 4 - 0.629625 y^2 + 1.61825 y^3 - 5.23513 y^4 + 20.7497 y^5 
- 93.1109 y^6,
\nl
&&\qquad\qquad\qquad \qquad\qquad {\rm (Epsilon\ expansion)}.
\ea
The two large errors in our $d=3.75,3.5$ data are probably due to the
coarse sampling of the unitarity boundary, since the spectrum changes 
rapidly at these dimensions, as shown in Figs. \ref{fig2c},\ref{fig2d}. 
Apart from this, the bootstrap
results are rather good and consistent with those of the other methods.
The overall error of the fit  for $\D_{\e'}$ is estimated
to be less than $0.007$ (grey band). The $d=3$ results of different
methods are again summarized in Table \ref{tab4}.

\begin{figure}[h]
\centering
\subfigure[\label{fig8a}]{\includegraphics[scale=0.21]{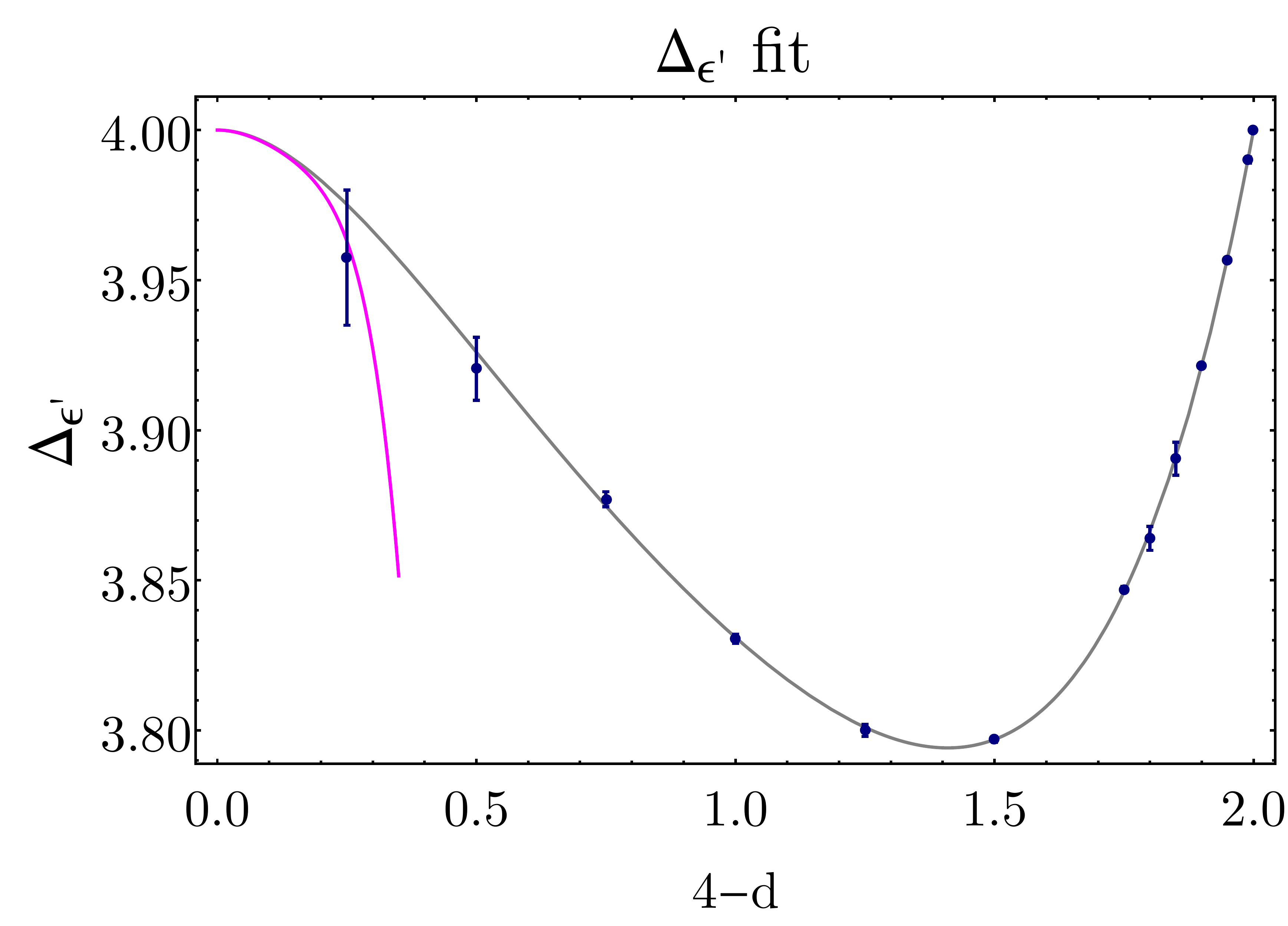}}
\hspace{.1cm}
\subfigure[\label{fig8b}]{\includegraphics[scale=0.21]{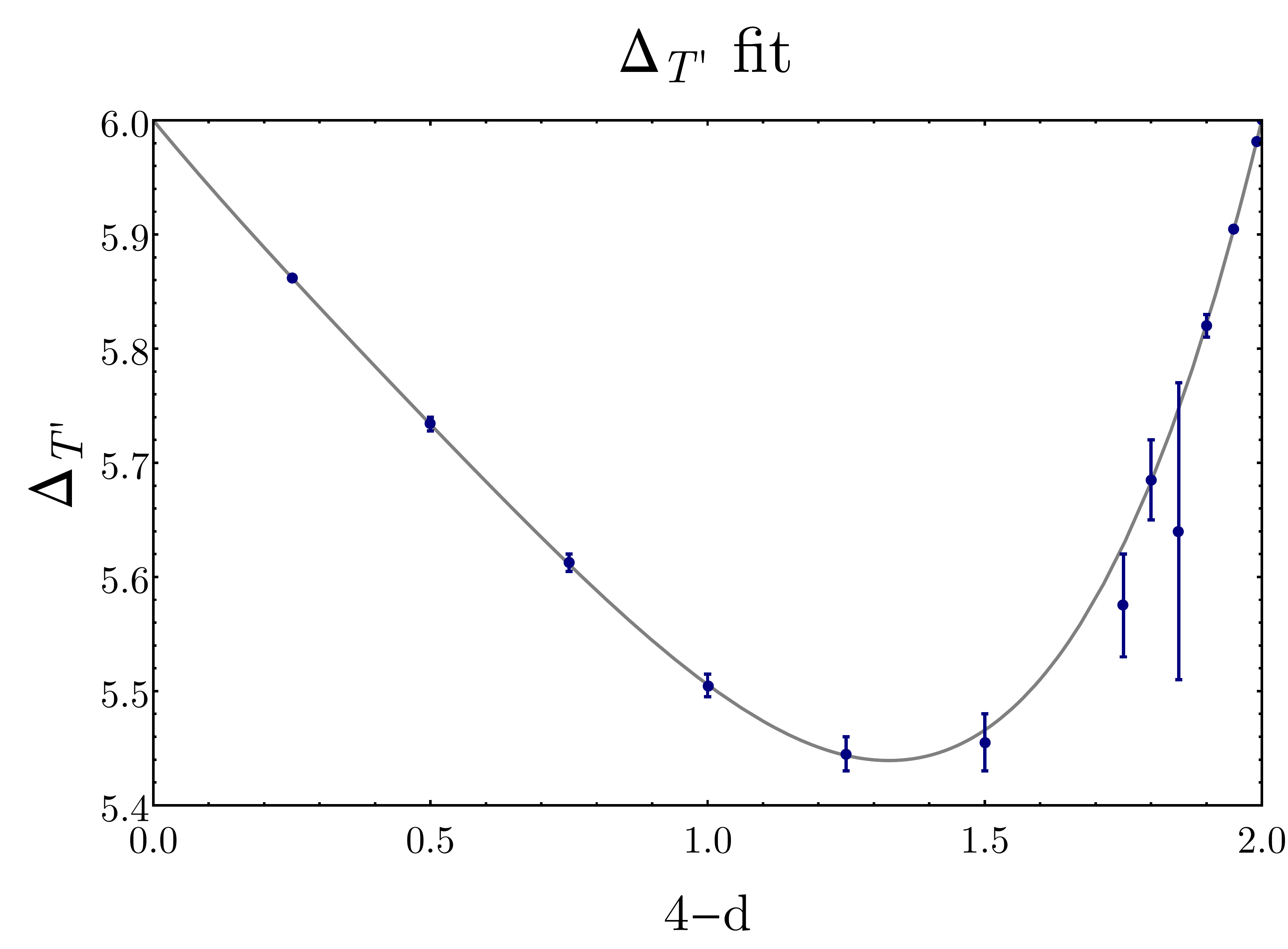}}
\subfigure[\label{fig8c}]{\includegraphics[scale=0.21]{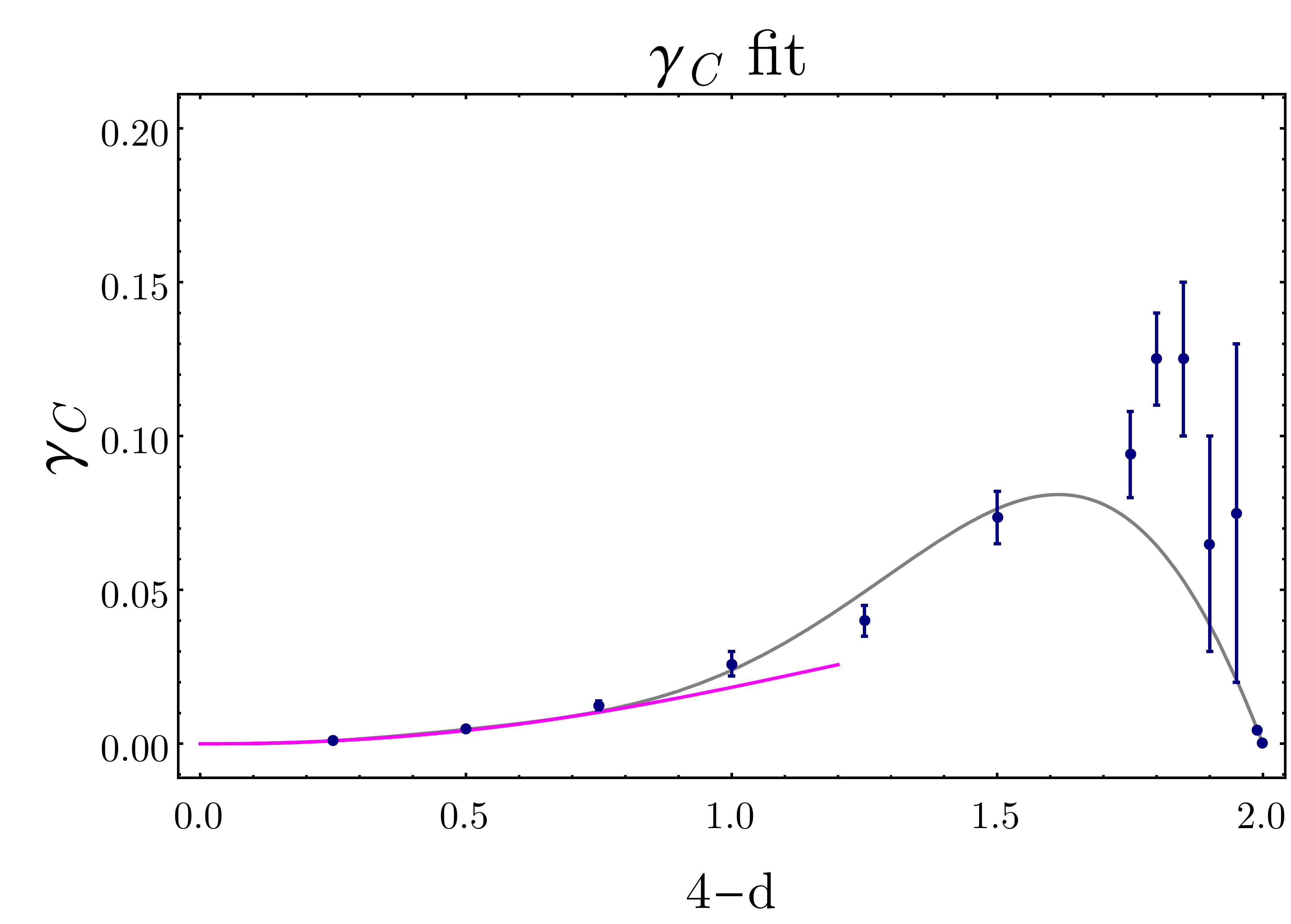}}
\hspace{.1cm}
\subfigure[\label{fig8d}]{\includegraphics[scale=0.21]{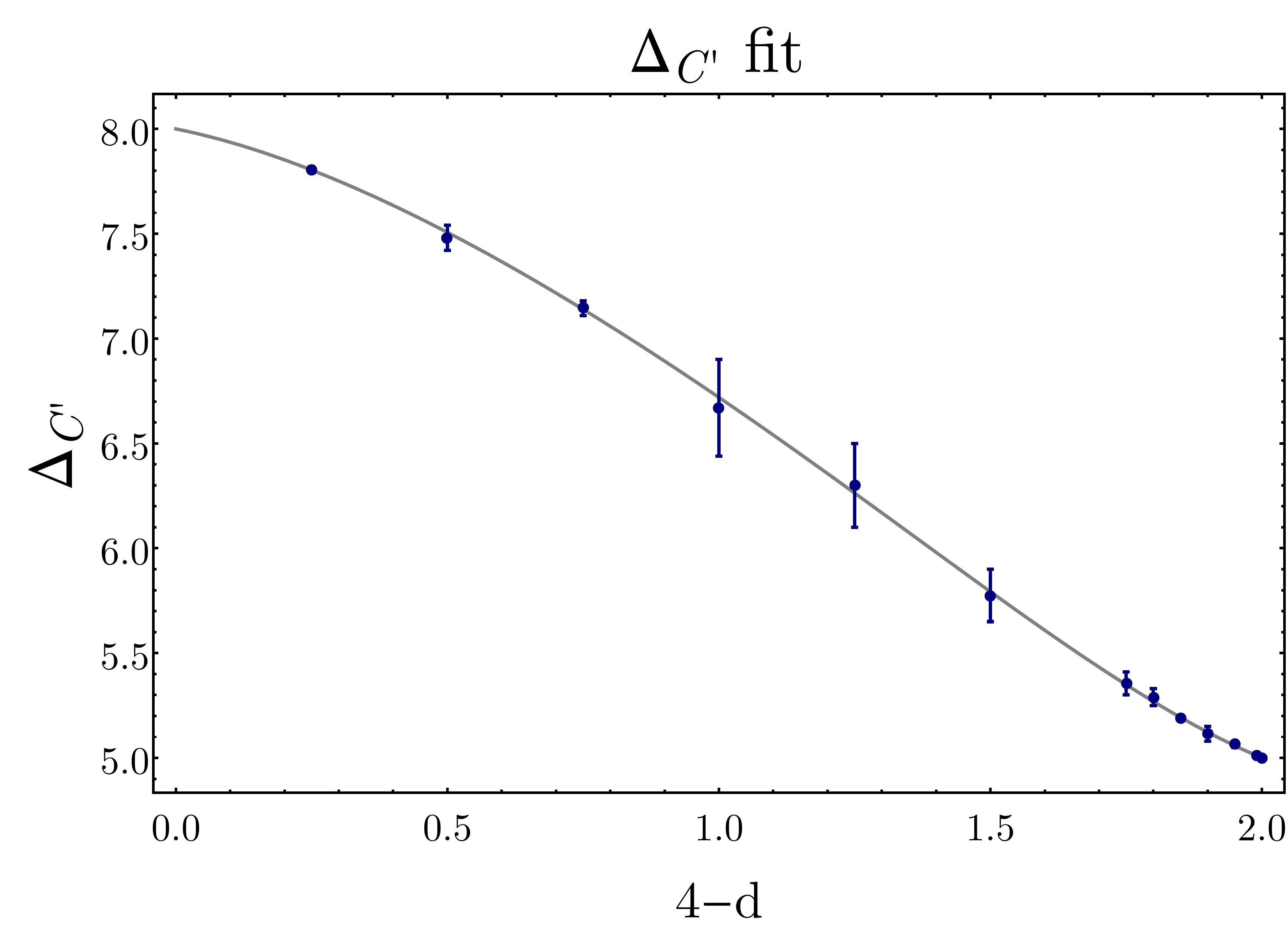}}
\caption{Polynomial fit of $\D_{\e'}, \D_{T'},\g_{C},\D_{C'},$
as functions of $y=4-d$. Figs. (a) and (c) include the known epsilon
expansion results.}
\label{fig8}
\end{figure}

The analysis of the other fields $T',C,C'$ can be done along the same 
lines. The resulting polynomials are as follows:
\ba
\D_{T'}(y) &=& 6 - 0.589135 y + 0.210294 y^2 - 0.273167 y^3 
\nl
&& + 0.157697 y^4,
\\
\D_C (y)&=& 6 - y +0.0148373 y^2 + 0.233465 y^3 - 0.584595 y^4 + 0.642985 y^5 
\nl
&& -  0.302706 y^6 + 0.0495537 y^7,
\\
\D_{C'}(y) &=& 8 - 0.518833 y - 1.22242 y^2 + 0.729856 y^3 - 0.356502 y^4 
\nl
&& + 0.0872808 y^5,
\ea
and the fits are shown in Fig. \ref{fig8}.
The errors are now bigger and vary considerably with $d$, thus it
is not possible to give a $d$-independent bound for the error of 
the fit. We find:

\be
\begin{array}{|c|c|c|}
\hline
& 4>d>2.3 & 2.3\ge d \ge 2 \\
\hline
{\rm Err}\left(\D_{T'} \right) & < 0.03 &\sim 0.15\\
\hline
{\rm Err}\left(\D_{C} \right) & < 0.01 &\sim 0.10  \\
\hline
\end{array}
\ee

\be
\begin{array}{|c|c|c|c|}
\hline
& 4>d>3 & 3 \ge d \ge 2.5 & 2.5>d \ge 2\\
\hline
{\rm Err}\left(\D_{C'} \right) & < 0.1 &\sim 0.2 & < 0.1 \\
\hline
\end{array}
\ee

Comparison with the epsilon expansion series is possible in the case
of the $\ell=4$ leading-twist $C$ \cite{gopa}\cite{4th}\footnote{
The results of Refs. \cite{gopa}\cite{4th} are obtained by a combination of
epsilon expansion and analytic bootstrap approaches and might differ
from the fully perturbative calculations of Ref. \cite{panzer}, beyond the 
leading term. As such they might be closer to the numerical data 
for $y=O(1)$.}:
\be
\D_C(y)=6 - y + 0.01296296 y^2 + 0.01198731 y^3 - 0.006591585 y^4 ,
\qquad {\rm (Epsilon\ expansion)}.
\ee
This field corresponds to a conserved current for both $d=4$ and $d=2$,
thus it is convenient to plot its anomalous dimension 
$\g_C=\D_C-d+2$, that vanishes at both ends. In Fig. \ref{fig8c}, one sees that
it grows slowly as $4>d>3$ and changes behaviour at $d\sim 2.2$,
where data fluctuate strongly. The behaviour in this region will
be further discussed in Section three.

%-2.2.2-----------------------------------
\subsubsection{Structure constants}

The bootstrap data for the structure constants corresponding to the 
six fields discussed in the previous Section are reported in Table \ref{tab5}.
The exact values at $d=2,4$ are also included for comparison;
for $d=4$ they are obtained from Ref. \cite{ope-4},
\be
f_{\s\s\e}=\sqrt{2}, \qquad f_{\s\s C}=\frac{1}{\sqrt{35}},
\qquad f_{\s\s\e'}=f_{\s\s T'}=f_{\s\s C'}=0,
\ee
while for $d=2$ they are given in Appendix B of Ref. \cite{rast-b},
\be
f_{\s\s\e}=\frac{1}{2}, \quad f_{\s\s\e'}= \frac{1}{64}, \quad
f_{\s\s T'}=\frac{3}{8^3\sqrt{10}},\quad 
f_{\s\s C}=8f_{\s\s T'}, \qquad f_{\s\s C'}= \frac{1}{2^8} .
\ee

\begin{table}
\centering
\be 
\resizebox{1\hsize}{!}{$
\begin{array}{|c|l|l|l|l|l|l|}
\hline
d & c  & f_{\sigma\sigma \epsilon}  & f_{\sigma\sigma\epsilon '} & 
f_{\sigma\sigma T'} & f_{\sigma\sigma C}  & f_{\sigma\sigma C'} \\ 
\hline
\mathbf{4} & \mathbf{1} & \mathbf{ 1.4142136} & \mathbf{0}  & \mathbf{0} 
& \mathbf{0.169031} & \mathbf{0} \\ 
3.75 & 0.998645(45) & 1.34595(20) & 0.02725(25) & 0.00565(5) & 0.139775(25) 
& 0.00140(2) \\ 
3.5 & 0.992277(17) & 1.26142(18) & 0.04432(10) & 0.0091(1) & 0.112(2) 
& 0.0018(3) \\ 
3.25 & 0.976872(16) & 1.16283(7) & 0.05226(7) & 0.01061(17) & 0.086(4) 
& 0.0019(2)  \\ 
3 &  0.946535(15) & 1.051835(35) &  0.05300(5) &  0.010575(15) & 0.065(5)
&  0.0020(5) \\ 
2.75 & 0.893275(15) & 0.929385(35) & 0.04794(8)  & 0.00901(6) & 0.048(4) 
& 0.00235(15) \\ 
2.5 & 0.80711(1) & 0.7963025(45) & 0.03885(2)& 0.006675(25) & 0.0325(25) 
& 0.00285(25) \\
2.25 & 0.677724(2) & 0.653111(16) & 0.027375(35) & 0.00394(14) & 0.0195(15) 
& 0.0035(2) \\
2.2 & 0.64609(7) & 0.62333(6) & 0.0245(5) & 0.00352(7) & 0.0185(35) 
& 0.00375(25) \\
2.15 & 0.61243(8) & 0.59313(8) & 0.0225(5) & 0.0025(5) & 0.017(3) 
& 0.00385(15) \\
2.1 & 0.5768(1) & 0.56249(7) & 0.02018(8) & 0.00265(5) & 0.0155(25) 
& 0.00395(15) \\
2.05 & 0.53935(15) & 0.53143(8) & 0.01785(5) & 0.0023(1) & 0.0135(25) 
& 0.0039(1) \\
2.01 & 0.5082(3) & 0.5058(6) & 0.01605(5) & 0.001925(25) & 0.0155(1) 
& 0.00392(1) \\
2.00001 & 0.500015(15) & 0.4999975(45) & 0.0156225(35) & 0.0018520(5) 
& 0.0148235(15) & 0.003904(1) \\
\mathbf{2} & \mathbf{0.5} & \mathbf{0.5} & \mathbf{0.0156250}  
& \mathbf{0.00185290} & \mathbf{0.0148232} & \mathbf{0.003906} \\ 
\hline
\end{array}
\nonumber
$}
\ee
\caption{Structure constants of six low-lying states for $4>d>2$.
The exact values for $d=2,4$ are given in bold. The normalization
conventions are as in Ref. \cite{sd} (cf. Eq.(\ref{convent}). }
\label{tab5}
\end{table}

\begin{figure}
\centering
\includegraphics[scale=0.35]{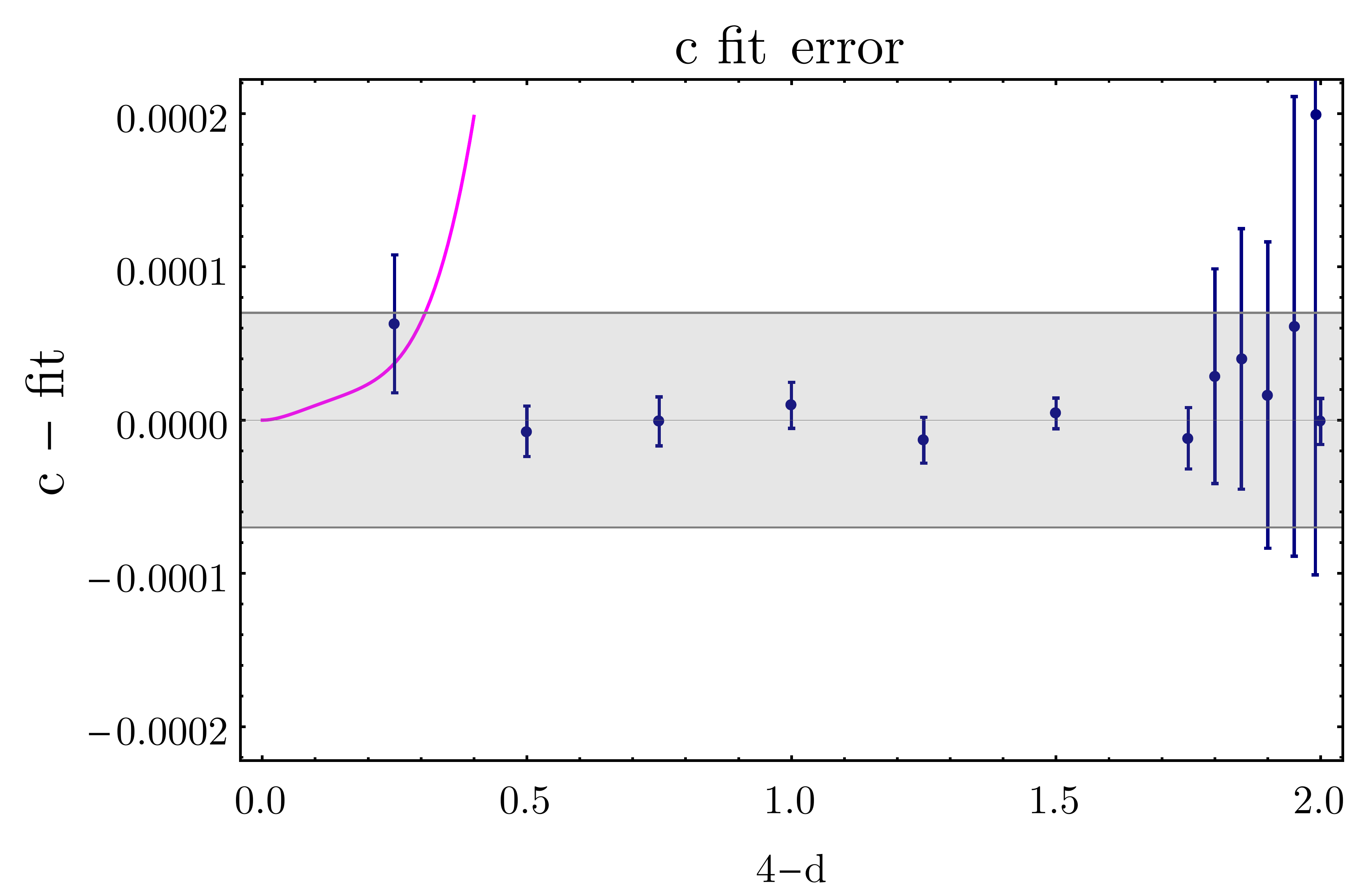}
\caption{Fit of central charge $c$  bootstrap data (blue) with overall error
(gray) and epsilon expansion (magenta) \cite{gopa}.}
\label{fig9}
\end{figure}

The $d$-dependent polynomial fits for the structure constants
are obtained as follows.
The central charge is described by the polynomial:
\ba
c(y) &=&1 - 0.0173616 y^2 - 0.0133068 y^3 - 0.0385653 y^4 + 0.0310843 y^5 
\nl
&&- 0.0196858 y^6 + 0.00436051 y^7 ,
\qquad \qquad\qquad {\rm Err\ }(c) <0.0007 ,
\ea
and the fit error is shown in Fig. \ref{fig9}.
This quantity is very precise with an error less than $10^{-3}$.
The high-order polynomial is necessary to obtain a
good chi-square value: lower-order polynomials would let the data oscillate
out of the fit as $d$ varies. 
A remnant of this behavior is still present in Fig. \ref{fig9},
and was also observed in early bootstrap data \cite{slava-d}.
In order to comply with the fit, we slightly enlarged the error estimate of
$c$ for some $d$ values, as shown in Fig. \ref{fig1b}, for example.

The epsilon expansion of this quantity is known to fourth order \cite{gopa},
\be
c(y)= 1 - 0.0154321 y^2 - 0.0266347 y^3 -0.0039608 y^4,
\qquad {\rm (Epsilon\ expansion)},
\ee
and its relation with the data is also shown in Fig. \ref{fig9}.

The structure constant $f_{\s\s\e}$ is another precise result of 
the bootstrap. Its polynomial fit is:
\ba
\!\!\!\!\!\!\!\!
f_{\s\s\e}(y) &=& 1.41421- 0.235735 y - 0.164305 y^2 + 0.0631842 y^3 
- 0.0371191 y^4
\nl
&&  + 0.0137454 y^5 - 0.00214024 y^6,
\qquad \qquad\quad  {\rm Err\ }\left(f_{\s\s\e}\right) <0.0002 ,
\ea
while the known epsilon expansion is \cite{gopa},
\ba
f_{\s\s\e}(y) &=& 1.41421 - 0.235702 y - 0.168047 y^2 + 0.100996 y^3,
\nl
&&\qquad \qquad\qquad\qquad\qquad{\rm (Epsilon\ expansion)} .
\ea
Note again the close values of the first few coefficients of the
two series.

\begin{figure}
\centering
\includegraphics[scale=0.35]{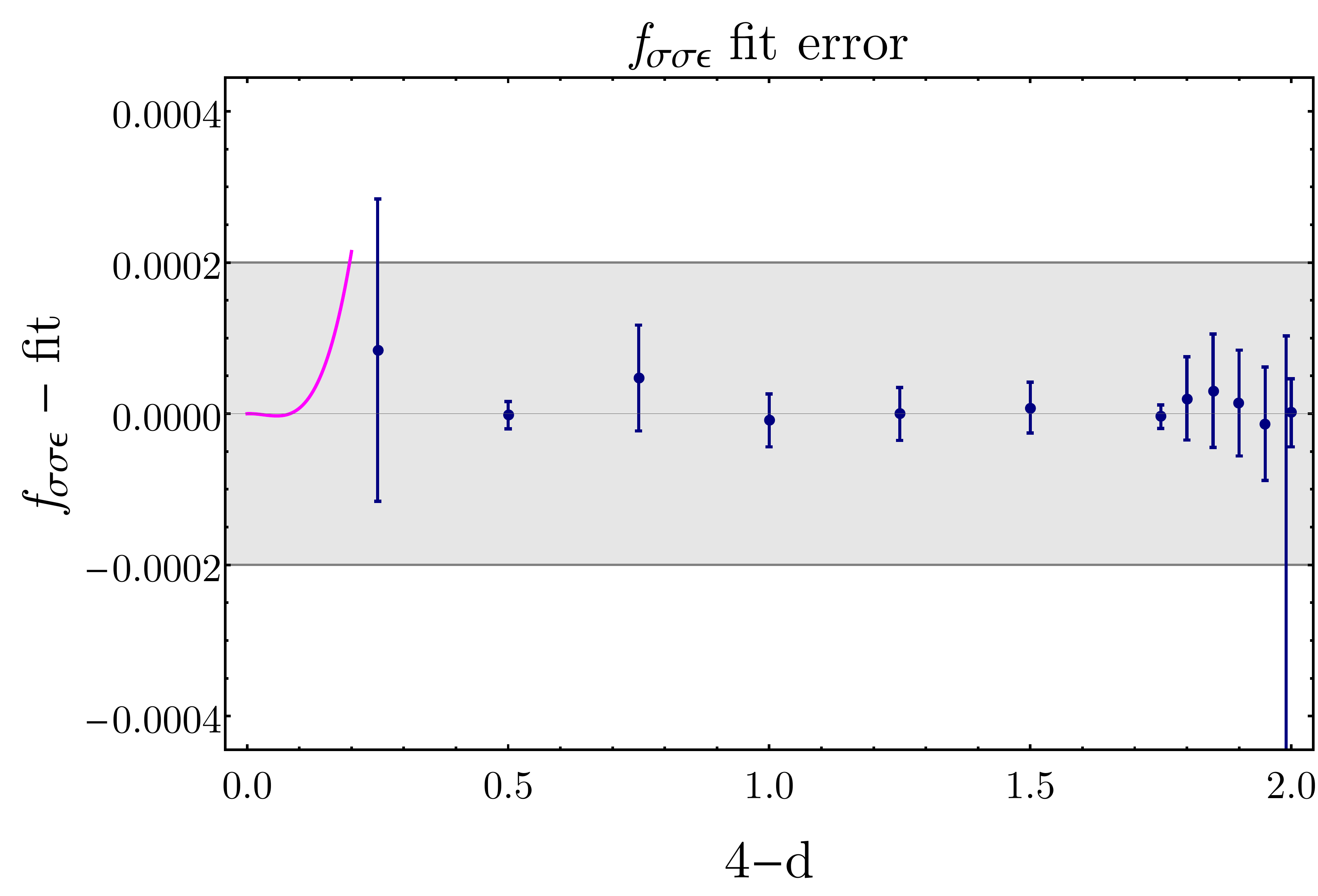}
\caption{Fit of structure constant $f_{\s\s\e}$  with 
bootstrap data (blue) with overall error (gray)
and epsilon expansion (magenta) \cite{gopa}.}
\label{fig10}
\end{figure}

\begin{figure}[h]
\centering
\subfigure[\label{fig11a}]{\includegraphics[scale=0.21]{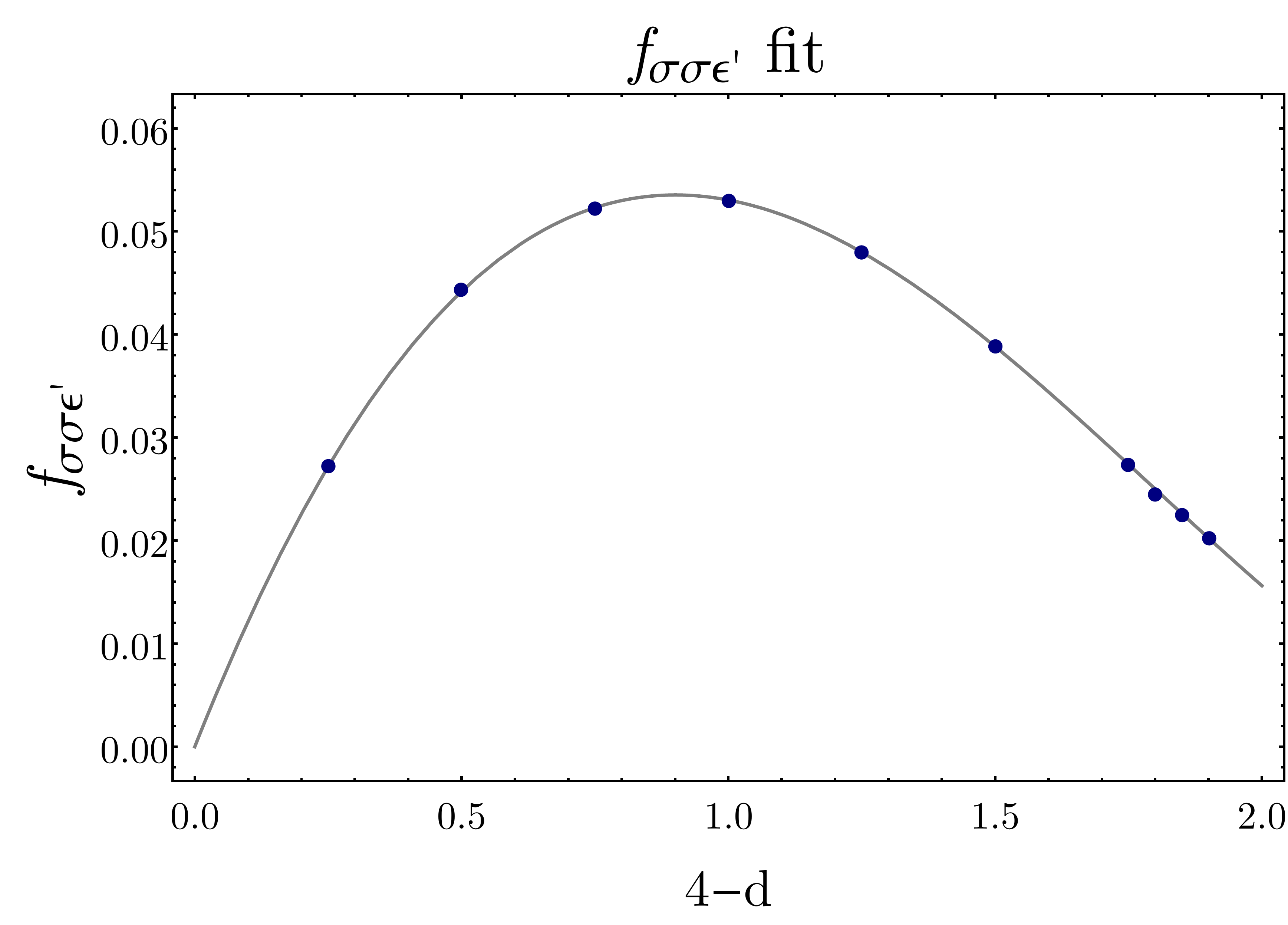}}
\hspace{.1cm}
\subfigure[\label{fig11b}]{\includegraphics[scale=0.21]{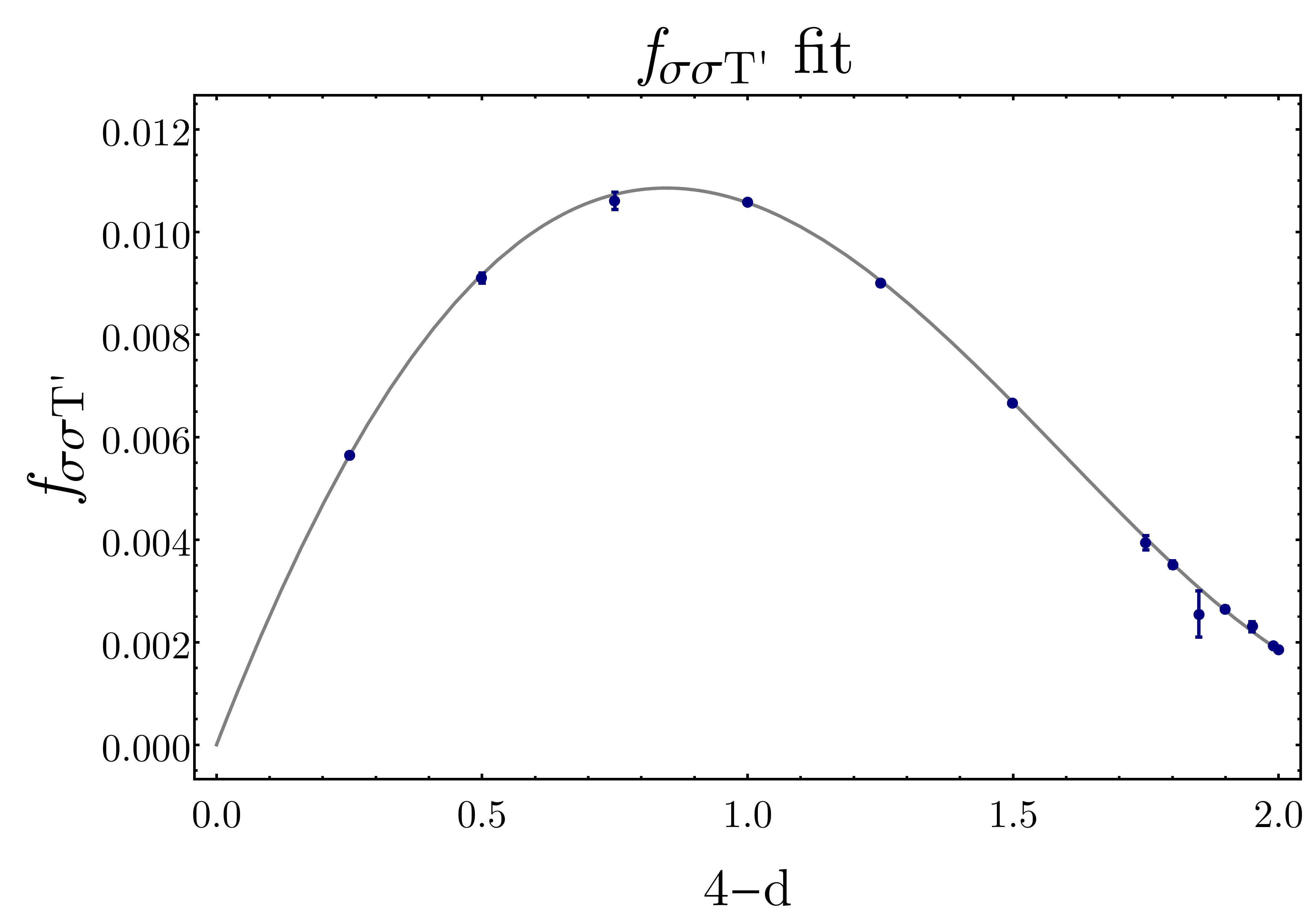}}
\subfigure[\label{fig11c}]{\includegraphics[scale=0.21]{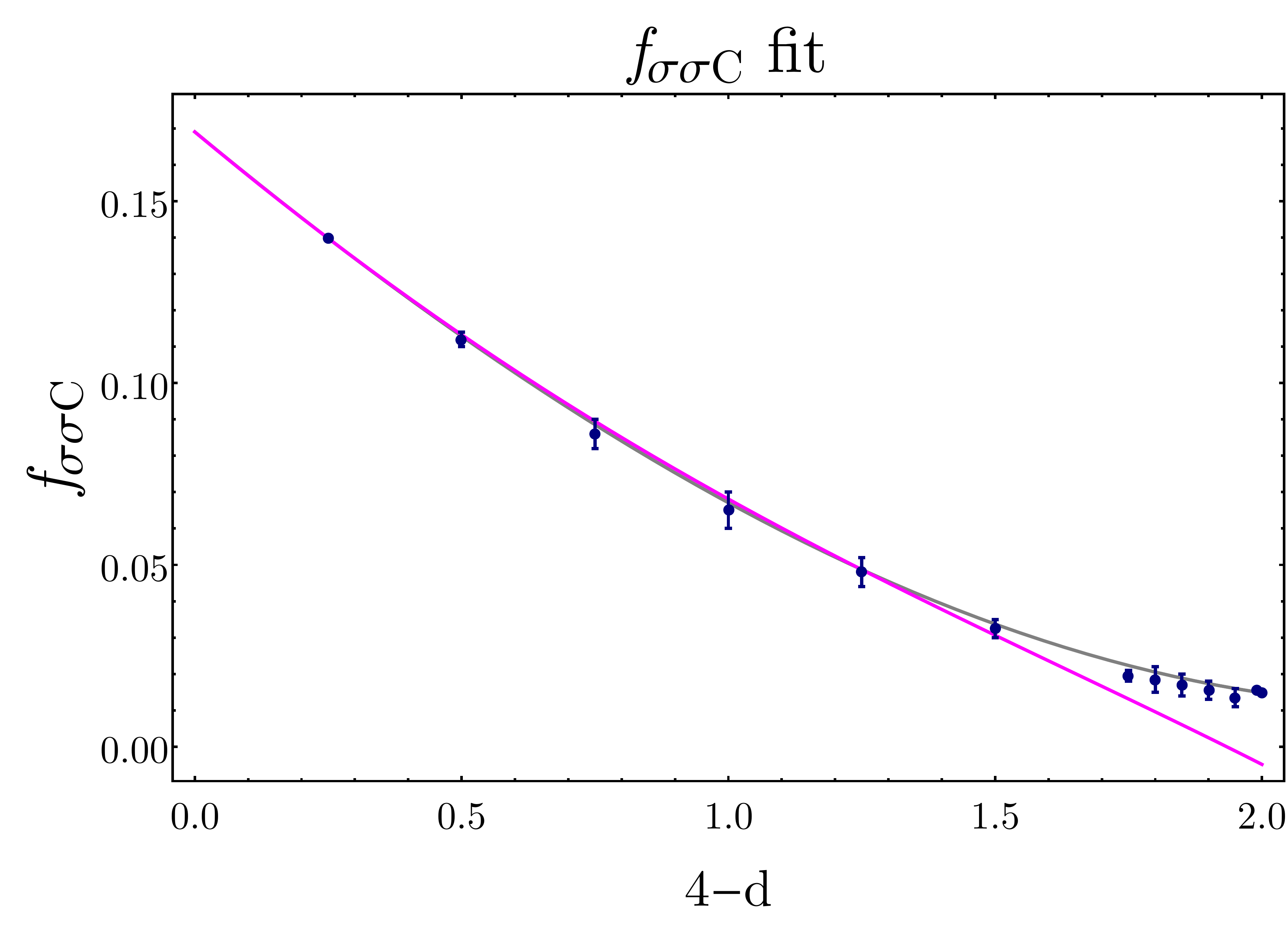}}
\hspace{.1cm}
\subfigure[\label{fig11d}]{\includegraphics[scale=0.21]{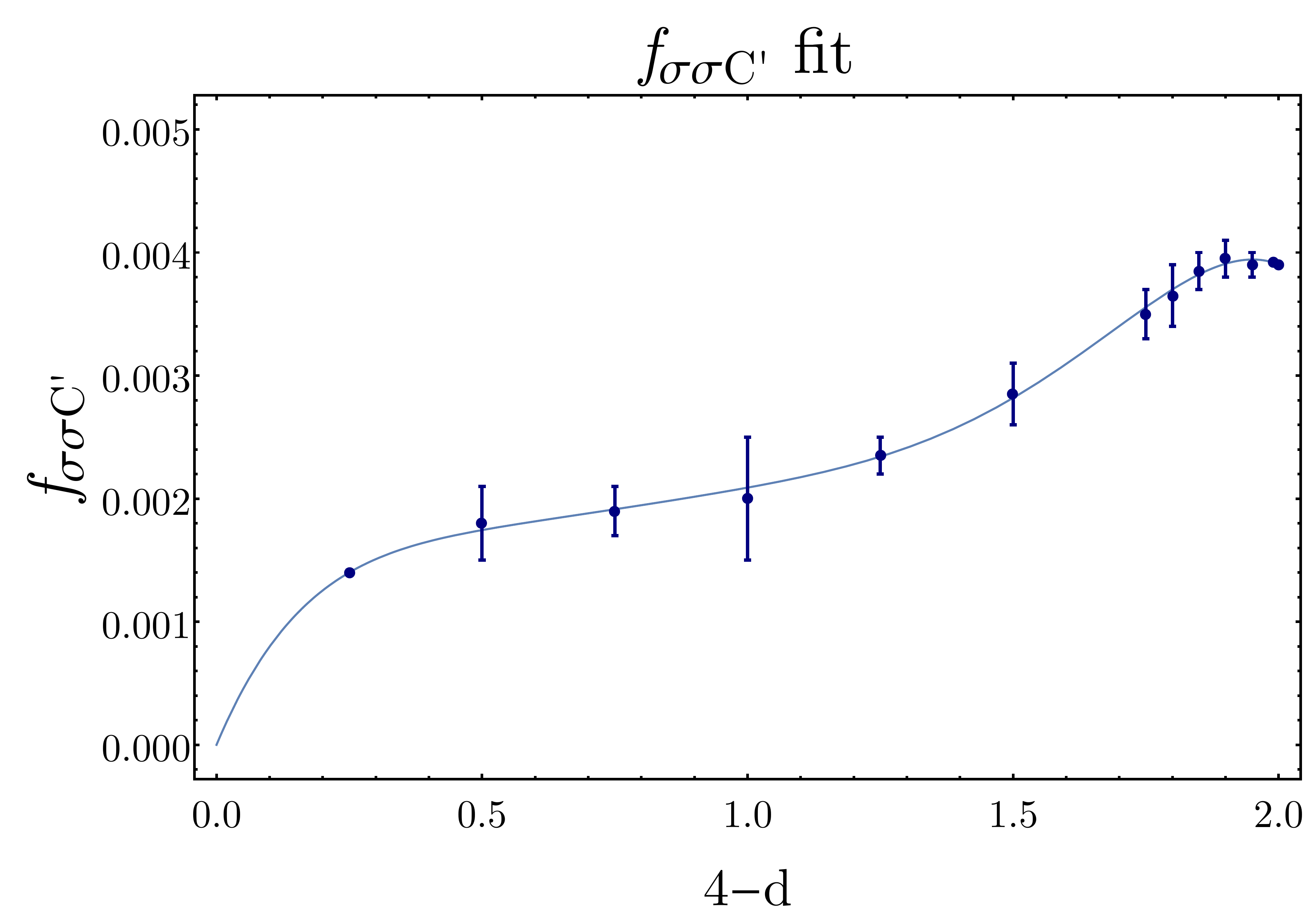}}
\caption{Polynomial fit of $f_{\s\s\e'},f_{\s\s T'},f_{\s\s C},f_{\s\s C'},$
as functions of $y=4-d$. Fig. (c) includes the known epsilon
expansion result \cite{gopa}.}
\label{fig11}
\end{figure}

The other four structure constants are determined with lower 
but yet good precision. Their polynomial fits are:
\ba
f_{\s\s\e'}(y)&=& 0.130874 y - 0.0916396 y^2 + 0.0125655 y^3 + 0.00124417 y^4,
\\
f_{\s\s T'}(y)&=&0.0268204 y - 0.0170251 y^2 - 0.00048868 y^3 + 0.00126381 y^4,
\\
f_{\s\s C}(y)&=&0.1690309 - 0.121369 y + 0.0166922 y^2 + 0.0027202 y^3,
\\
f_{\s\s C'}(y)&=&0.0103034 y - 0.0268178 y^2 + 0.0391145 y^3 - 0.0315244 y^4 
\nl
&& + 0.0132141 y^5 - 0.00220012 y^6,
\ea
and the corresponding errors are:
\ba
&&\!\!\!
\begin{array}{|c|c|c|c|}
\hline
 &4>d>2.2 & 2.2\ge d\ge 2.1 & 2.1>d>\ge 2 \\
\hline
{\rm Err}\left(f_{\s\s\e'} \right) & < 0.003 &\sim 0.007 & < 0.003\\
{\rm Err}\left(f_{\s\s T'} \right) & < 0.0003 &\sim 0.001  
& < 0.0003\\
\hline
\end{array}
\\
&&\nl
&& {\rm Err}\left(f_{\s\s C} \right) < 0.007,
\qquad \qquad
{\rm Err}\left(f_{\s\s\C'} \right) < 0.0005.
\ea
The epsilon expansion is available for the $\ell=4$ leading-twist
field $C$ \cite{gopa}\cite{4th}:
\ba
f_{\s\s C}(y)&=& 0.16903085 - 0.12244675 y + 0.02131741 y^2 +
0.002168567 y^3 
\nl
&& - 0.0019760553 y^4,
\qquad \qquad\qquad {\rm (Epsilon\ expansion)} .
\ea
The comparison with the bootstrap results is surprisingly accurate, as shown
in Fig. \ref{fig11c}. The plots of the other three structure constants
are also shown in this Figure.

%-2.3----------------------------------------
\subsection{Analysis of data and applications}

The numerical bootstrap results in various dimensions obtained in
this work have been recast in the form of simple polynomials of the variable
$y=4-d$, spanning the range $4 \ge d \ge 2$. In the most precise cases,
namely for the dimensions $\D_\s, \D_\e, \D_{\e'}$, central
charge $c$ and structure constant $f_{\s\s\e}$, the error
of the fits is rather small and bounded by a $d$-independent value.
This has been chosen to be larger than the uncertainty of individual
data points, so as to leave some space for potential unknown systematic
errors of the bootstrap approach. 
All data show larger fluctuations in the region $2.3>d>2$, where
the conformal theory rearrange itself to match the $d=2$ limit --
a topic to be discussed in the next Section.

The agreement with the existing results in the literature, often
developed over decades of investigations, are very good:
the bootstrap data, while being superior, are never inconsistent
with them. In particular, the polynomials obtained here 
and those of the (unresummed) epsilon expansion are 
extremely close in their leading terms, allowing for precise
matches in the region $4>d>3.8$.

The physical applications of these results are likely to be numerous:
here, we shall mention two examples.
The first one is the study of convergence for the epsilon expansion. 
This approach is straightforwardly and universally applied 
in field theory descriptions of critical phenomena,
but it requires resummation techniques of the badly convergent 
perturbative series, such as the use of Borel transforms \cite{zinn}.
The precise data of our work can provide a clean case for
checking and refining the optimization methods involved in the
resummations \cite{panzer}\cite{serone}.

The second application is the study of the critical point of the Ising
model with long-range interactions and its relation with the present,
short-range case. This problem also has a long history and has been
recently discussed in Refs. \cite{slava-long}\cite{tromb}.
In the definition of the model, the power-law
decay of the spin-spin interaction $\a=d+s $ involves explicitly the 
space dimension and it is natural to discuss the 
equivalence/inequivalence of long-range and short-range universality classes 
by varying $d$ continuously.
The precise data presented here can help testing the existing conjectures
on the phase diagram \cite{slava-long}\cite{tromb}.

%-2.3.1--------------------------------------------- 
\subsubsection{The issue of unitarity}

In the recent work \cite{slava-uni}, the unitarity of the scalar field
theory in non-integer dimension has been investigated. It is
known that the free theory contain additional fields, called
evanescent operators, whose correlators vanish algebraically at
integer $d$, but involve negative-norm states for non-integer $d$. 
They occur for dimensions $\D\ge 15$ in $d=4$.
After including the $\l\phi^4$ interaction, an example was found
of an evanescent operator with $\D = 23$ acquiring complex anomalous 
dimension to leading order in the epsilon expansion.
These non-unitary states could be present in the critical Ising model 
for non-integer $d<4$.

Non-unitary states are problematic for the numerical
bootstrap approach that assumes real $\D$ values 
and an expansion in conformal partial waves with
positive coefficients $f_{\s\s\cal O}^2$.
This issue has already been discussed in Refs. \cite{slava-uni}\cite{behan},
leading to the conclusion that if non-unitary states appear in the
partial-wave expansion of the four-spin correlator, 
they should be very high up in the spectrum and 
give negligible exponentially small contributions.
The result of the three-correlator conformal bootstrap approach \cite{behan}
at fractional dimensions $d=3.75,3.5,3.25$ is particularly interesting
in this context, because it includes a larger set of partial waves 
and thus can better tests non-unitarity for non-integer dimension.
For these fractional $d$ values, it was found that the unitarity region shrinks 
to a small island in the $(\D_\s,\D_\e)$ plane around the Ising point,
but does not vanish within the precision attained.

In our work, we do not find instabilities of the numerical
routines that could be interpreted as signs of non-unitarity,
while reaching a precision that is superior to that of the mentioned works
\cite{slava-d}\cite{behan}. As a matter of fact, we must conclude again that
the evanescent operators do not contribute
significantly to the partial-wave expansion of the correlator
$\langle \s \s \s \s \rangle$.
We also note the good agreement for $4>d>3.8$  
with the epsilon expansion results that do not rely on the unitarity
of the theory. 
Let us finally quote the bootstrap approach of Ref. \cite{gliozzi1}
that does not require unitarity of the theory.

%-3--------------------------------------------- 
\section{Leading twists and the $d=2$ limit}

In this Section we describe the conformal fields with lowest dimension  for
each spin value.  In conformal
theories with $d>2$, their anomalous dimensions $\g_\ell$, defined by
$ \D_\ell =d-2+\ell+\g_\ell $, are small numbers, with asymptotic
large-$\ell$ value:
\be
\lim_{\ell\to\infty} \g_\ell =2\g_\s .
\label{ell-asym}
\ee
Moreover, as a function of $\ell$, the $\g_\ell$ are fitted by a 
monotonically increasing and convex curve (Nachtmann theorem) \cite{nacht}.
Our analysis of the numerical bootstrap data will follow the same
strategy of the previous Section: first compare the $d=3$ data with the
better three-correlator bootstrap results \cite{sd}; 
then, if this check is passed, extend the analysis to other $d$ values.

\begin{figure}
\centering
\subfigure[\label{fig12a}]{\includegraphics[scale=0.4]{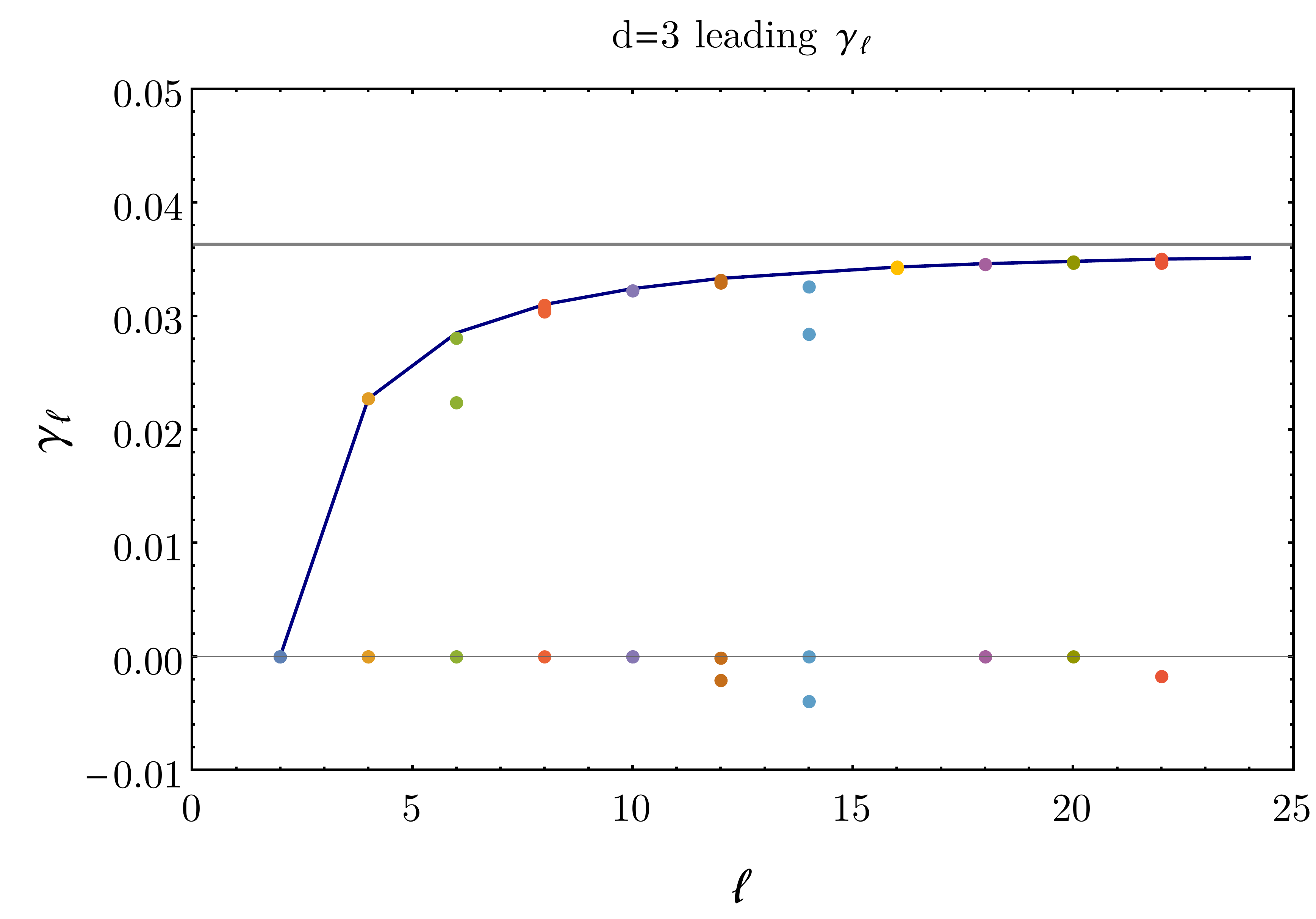}}
\subfigure[\label{fig12b}]{\includegraphics[scale=0.4]{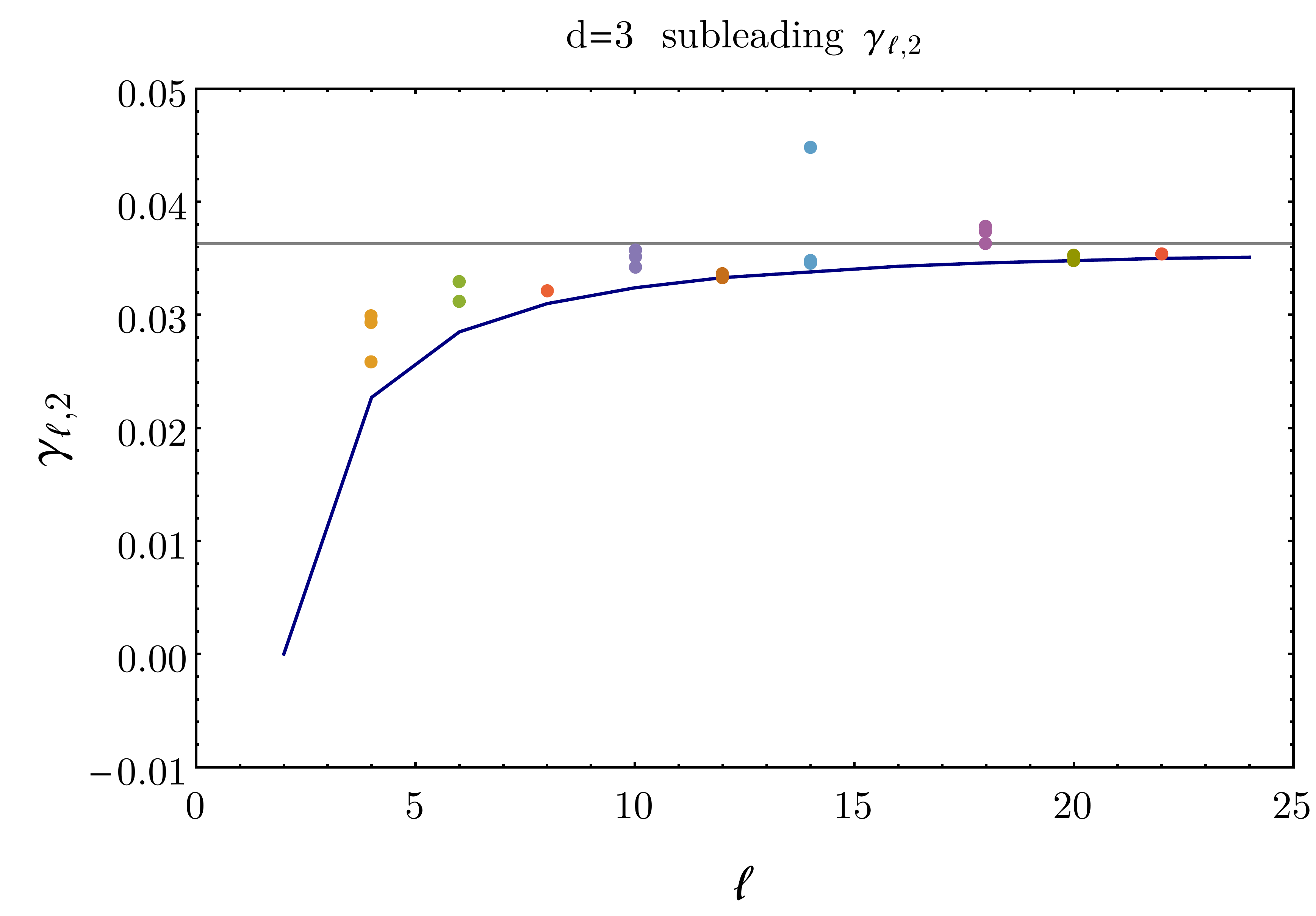}}
\caption{Anomalous dimension of leading (a) and subleading (b) twists with 
$2\le\ell\le 22$ in $d=3$ compared with three-correlator results (blue curve)
and $\ell\to\infty$ asymptotic value (gray line).}
\label{fig12}
\end{figure}

The data are again expressed by curves
$\g_\ell(\D_\s)$ that fluctuate within the range of $\Delta_\s$
identified as the Ising theory, $0.51814\le \D_\s \le 0.51817$
(see Fig. \ref{fig1}).
Since this interval correspond to four data takes, we obtain four
values of $\g_\ell$ for each $\ell$ value, that are represented as
points in Fig. \ref{fig12a} for $\ell=2,4,\dots,22$. 
In this figure, they are compared with the precise results
of the three-correlator bootstrap (joined by the blue curve) and the
asymptotic value (\ref{ell-asym}) (grey line). One sees that the bootstrap
solution oscillates between vanishing anomalous dimensions $\g_\ell=0$,
corresponding to higher-spin conserved currents (free-theory), 
and non-vanishing
values $\g_\ell\neq 0$. The presence of spurious free-theory solutions
is a feature of the Extremal Functional Method already observed in
\cite{sd}.

We remark that the non-vanishing $\g_\ell$ values match rather well the blue
curve for most spin values, namely the expected behavior is reproduced. 
Regarding the errors, in the previous Section it was given by the
fluctuation among the four data points; however, in the present case,
this would be as large as the data value, corresponding to no predictions
at all.  Therefore, we shall consider an alternative point of view:
we discard the free-theory $\g_\ell=0$ solutions and keep the non-vanishing
ones, to which we do not assign a definite error.
This choice is justified for the qualitative analyses carried out
in this Section.

It is also interesting to  study the higher levels, i.e. the
subleading twists, $\g_{\ell,2}> \g_{\ell,1}\equiv \g_\ell$: 
their values span a large range up to $\g_{\ell,2}\sim 2.5$ for
$\ell=22$ and show wide fluctuations.
Nonetheless, some of the points lie slightly
above the blue curve, as shown in Fig. \ref{fig11b}: by inspection, one
find that these $\g_{\ell,2}$ values occur in combination with trivial
first solutions $\g_{\ell}=0$ and thus are other would-be measures of
leading-twist dimensions.  However, their values are slightly overestimated,
because they are pushed up by the level repulsion with $\g_\ell=0$.

In conclusion, the comparison with the precise three-correlator
results \cite{sd} shows that the expected behaviour at $d=3$ is well
reproduced in our data by the leading states for any $\ell=2,4,\dots,22$,
once the trivial cases $\g_\ell=0$ are discarded.

\begin{figure}
\centering
\subfigure[\label{fig13a}]{\includegraphics[scale=0.5]{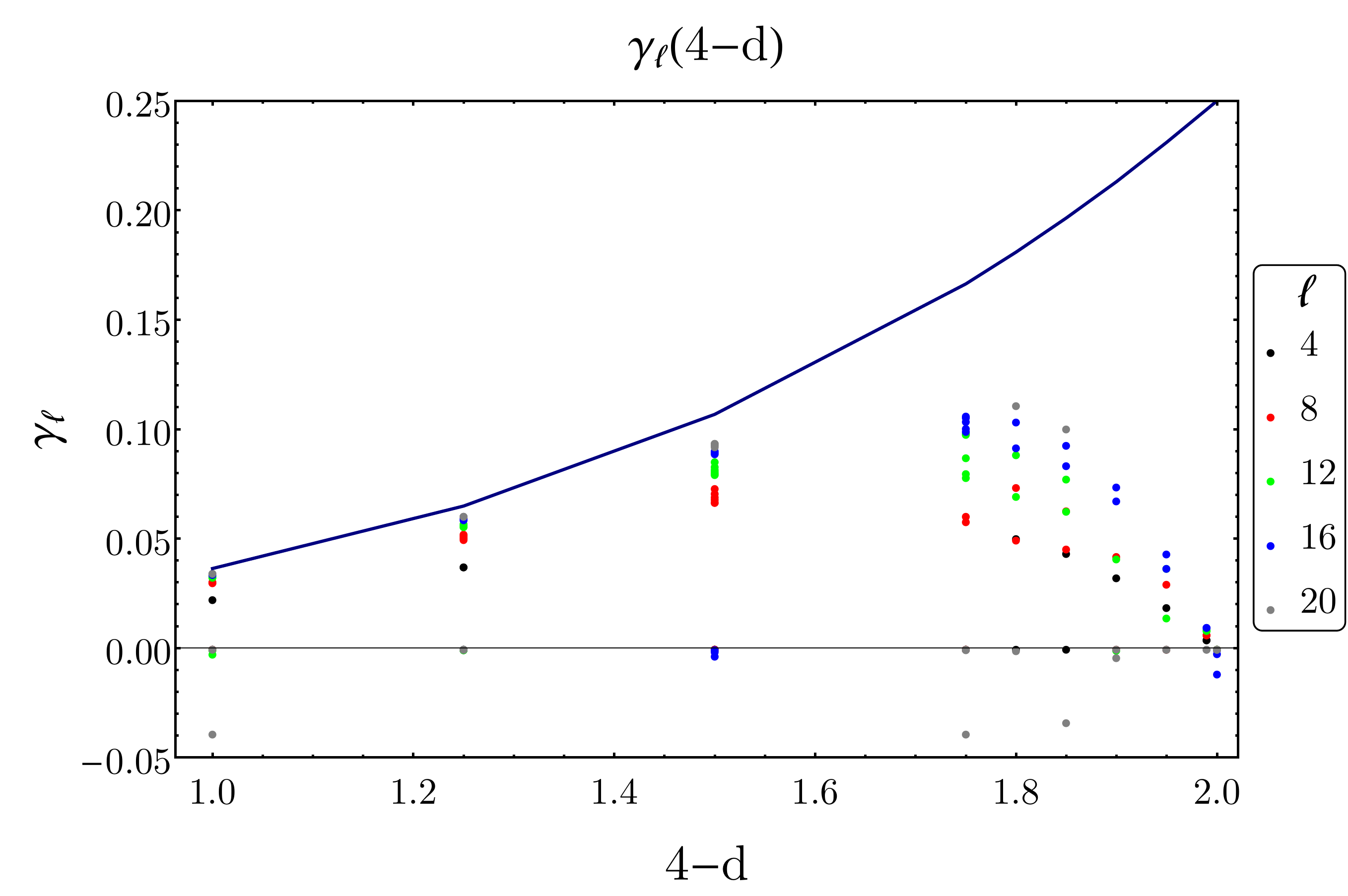}}
\subfigure[\label{fig13b}]{\includegraphics[scale=0.5]{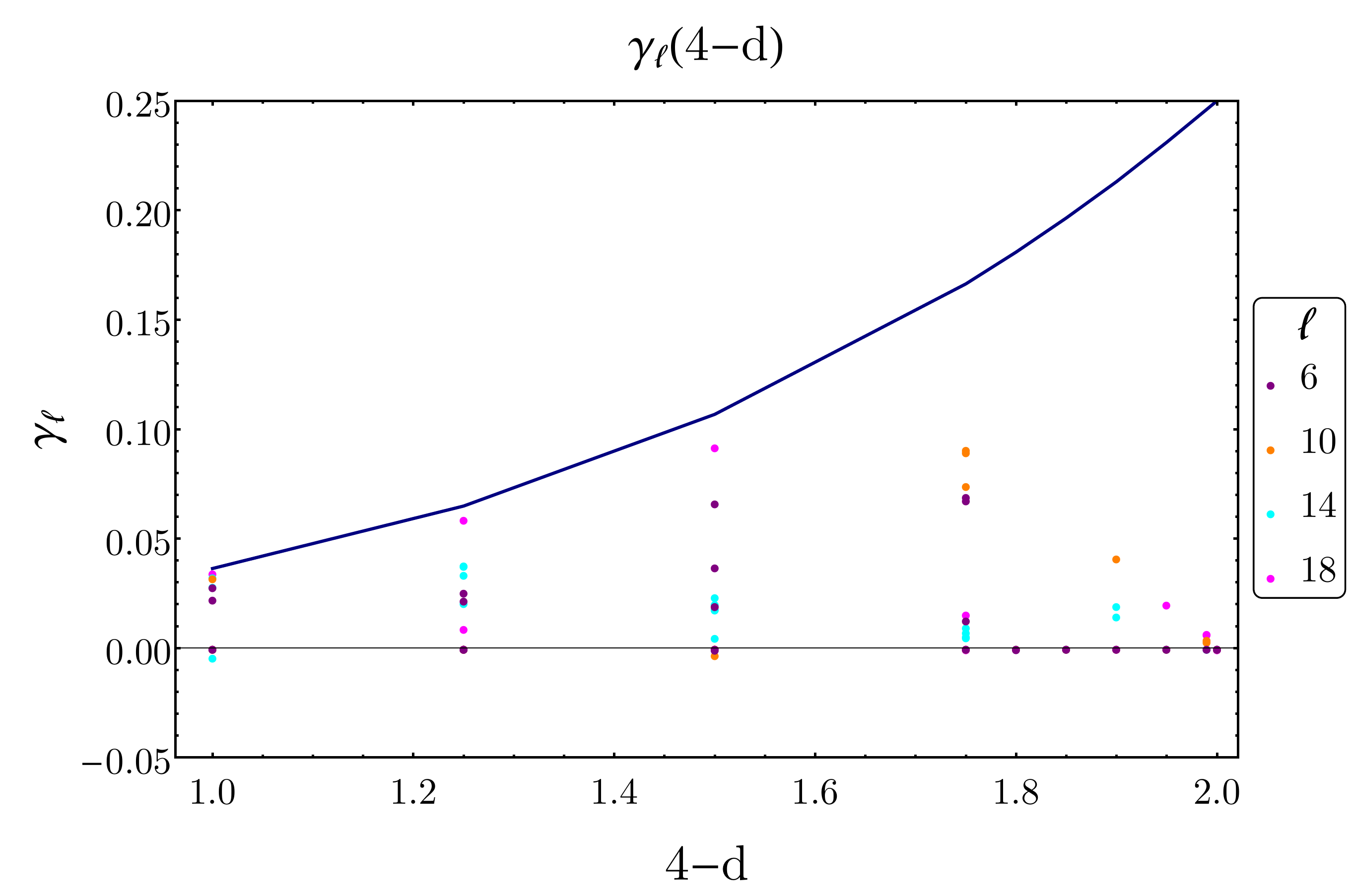}}
\caption{Dimension dependence of anomalous dimensions $\g_\ell(4-d)$:
for the $\ell$ values divided in two sets as given by the color tables.
The blue curve shows the large $\ell$ asymptotic value 
equal to $2\g_\s(4-d)$.  }
\label{fig13}
\end{figure}

We now consider the results for other dimensions $3\ge d \ge 2$.
In Fig. \ref{fig13a},\ref{fig13b}, 
we plot the $d$-dependence of the $\g_\ell(d)$ data, 
represented as points with colors associated to spin values,
dividend in two sets corresponding to $\ell=4k$ and  $\ell=4k+2$,
$k=1,2,\dots$, respectively.
Looking at the first set, we see that
the $\ell$ sequences reach the asymptotic values $2\g_\s(d)$
(blue curve) fairly well for $d=3,2.75,2.5$.
However, starting from $d=2.25$, their behavior is different and the
$\g_\ell$ gradually go to zero towards $d=2$, while approximatively keeping
their monotonicity in $\ell$, i.e. $\g_\ell(d)<\g_{\ell+4}(d)$. In the case
of the $\ell=4$ field $C$, this $d$-dependence was already present in
Fig. \ref{fig8c}, but we now see that it holds for all $\ell$ values.
At $d=2$, the leading twists become higher-spin conserved currents 
belonging to the Virasoro conformal block of the Identity field, as expected.
The behaviour of the second set of $\ell$ values in Fig. \ref{fig13b}
is analogous, but there are stronger fluctuations and the monotonicity
in $\ell$ is not always respected.

The observed $d$-dependence of the leading-twist dimensions $\g_\ell$ 
establishes that the asymptotic limit (\ref{ell-asym})
is violated or, at least, is far fetched, for
$d\le 2.2$, while the Nachtmann theorem becomes void; also,
the $\ell\to\infty$ and $d\to 2$ limits do not commute.
We conclude that at $d=2.2$ the towers of conformal fields begin to
rearrange themselves to comply with the Virasoro representations
at $d=2$. 

\begin{figure}
\centering
\includegraphics[scale=0.48]{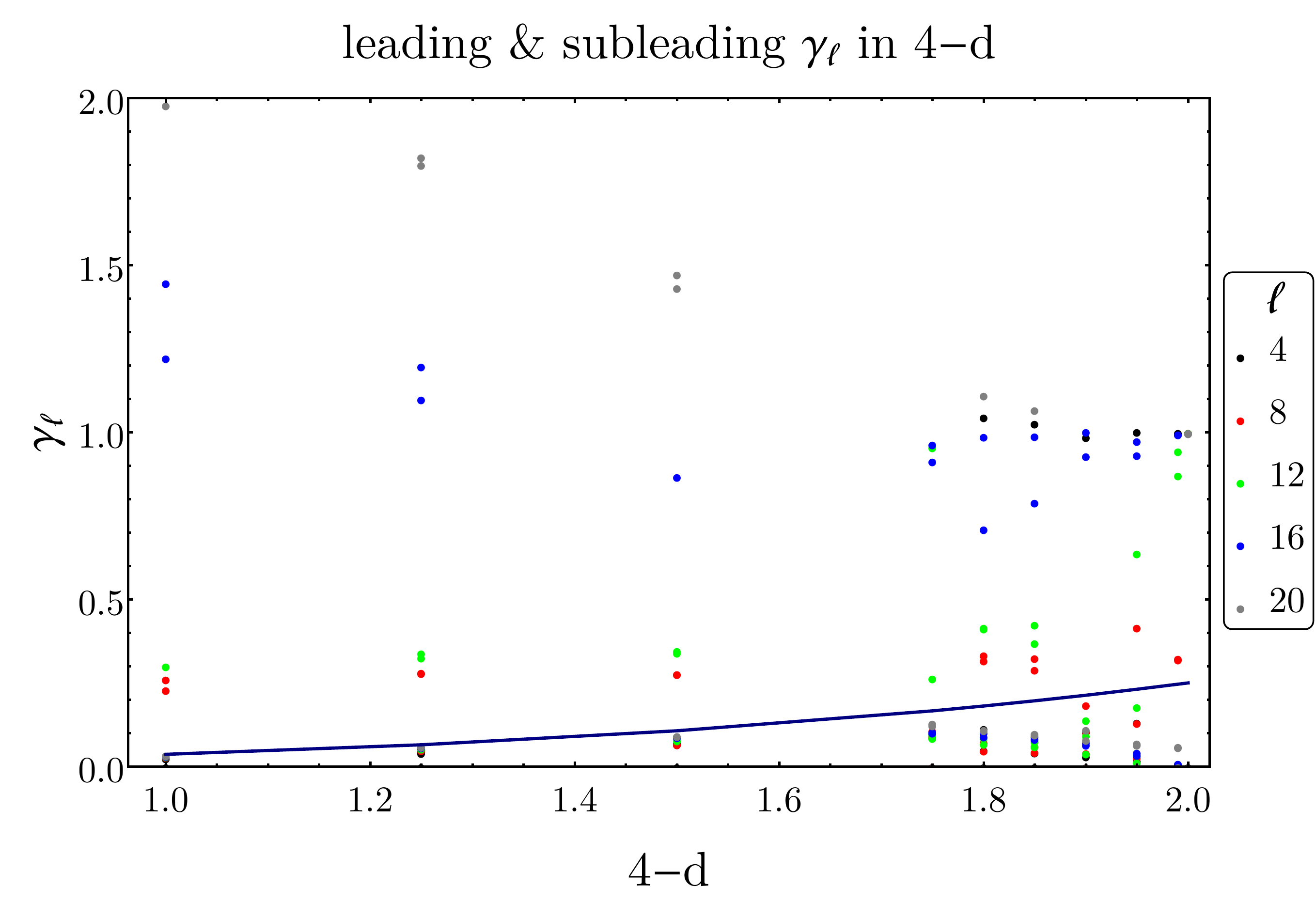}
\caption{Dimension dependence of anomalous dimensions $\g_\ell(4-d)$
for the subleading twists (above the curve) for some $\ell$ values.
The blue curve shows the large $\ell$ asymptotic value 
equal to $2\g_\s(4-d)$. }
\label{fig14}
\end{figure}

The rearrangement of states is also visible for
the subleading twists $\g_{\ell,2}$, as shown in Fig. \ref{fig14} for
some selected $\ell$ values. The subleading states are above the blue
curve, while the leading ones are below it, as in Fig. \ref{fig13a}.
We see that the $\g_{\ell,2}$ anomalous dimensions are rather big for
large $\ell$ values, but as $d$ is decreased they gradually converge to
the common $d=2$ limit $\g_{\ell,2}=1$ corresponding to the Virasoro
conformal block of the energy field $\e$. This behavior is clearer for
the large $\ell$ values, since bigger $\g_{\ell,2}$ values fluctuate
less.

In conclusion, in this Section we have shown that the leading twists
fields in our data match rather well the expected behavior (\ref{ell-asym})
as a function of $\ell$ \cite{nacht} for any dimension $3\ge d > 2.25$
and are also quantitatively good at $d=3$
(if trivial solutions $\g_\ell=0$ are disregarded).
For $d\le 2.25$, this behavior is abandoned: the anomalous dimensions
converge to zero in order to fit in the 
$d=2$ Virasoro conformal block of the identity field. The
subleading twists similarly approach the conformal block of the energy.
The transition region $2.25\ge d> 2$ is also characterized by the
lower precision in the determination of conformal dimensions and
structure constants as discussed in Section 2.2. After the transition,
the $d=2$ numerical data are again very good, even superior to 
the $d=3$ case.

This analysis yields a glimpse of the differences between conformal
theories in two and higher dimensions we were alluding to in the
Introduction. Although the low-lying states discussed in Section 2
depends smoothly on dimension, the states higher up in the spectrum
change considerably and the Virasoro towers split up, with
anomalous dimensions monotonically increasing with $\ell$.
It would be interesting to study these phenomena analytically by 
developing an expansion in $d=2+\eps$ (and possibly large $\ell$).

%-4--------------------------------------------- 
\section{State decouplings at the Ising point}

In this Section we analyze the bootstrap data on the unitarity boundary
in the vicinity of the Ising point and discuss the decoupling of
states (vanishing structure constant) that take place while 
approaching this theory. In two dimensions, the Ising
conformal bootstrap is obeyed by a highly reduced set of states,
owing to the decoupling of Virasoro null  (zero-norm) vectors.
In higher dimensions, we generically expect a reduction of states 
for theories living on the unitarity boundary, where some
squared structure constants parametrically change from positive to
negative values. We should then search for specific decouplings as the
Ising point is approached along the unitarity boundary, by tuning
$\D_\s$.
Our analysis will start by determining the structure constants that
vanish in the $d=2$ model, then check those that are better identified 
numerically and follow their evolution for $2\le d \le 3$.

This analysis is motivated by the observation that some structure
constants for $\ell=0,2$ states do appear to vanish as approaching the
$d=3$ Ising model \cite{slava-IM}.  In this Reference, the connection with the
$d=2$ theory was already discussed: here we further analyze this issue
using the high-quality data for $2\le d\le 3$.

%-4.1--------------------------------------------- 

\subsection{Counting of quasi-primary states on the $d=2$ unitarity
boundary and Virasoro null vectors}

The conformal theory obeying the $d=2 $ bootstrap equations on the
unitarity boundary has been previously identified as the following non-unitary
interpolation of the $c<1$ Virasoro minimal models.
Their central charge and  conformal dimensions are
parameterized by the integer $m$ and read (Kac table) \cite{cft}:
\ba
c&=& 1-\frac{6}{m(m+1)}, \quad\qquad m=3,4,\dots,
\nl
\!\!\!\! \!\!
h_{rs}&=& h_{m-r,m+1-s}=\frac{\left((m+1)r-ms\right)^2-1}{4m(m+1)},
\quad 1\le r \le m-1,\quad 1\le s\le m .
\label{kac}
\ea
The conformal fields corresponding to spin and energy of the Ising model
 are identified as follows:
\ba
\s&=&\f_{12}, \qquad\quad \D_\s=2h_{12}= \frac{1}{2}- \frac{3}{2(m+1)},
\qquad \frac{1}{8} \le \D_\s \le \frac{1}{2},
\nl
\e&=&\f_{13}, \qquad\quad \D_\e=2h_{13}=2-\frac{4}{m+1},
\qquad\quad\ \  1 \le \D_\e \le 2.
\label{ds-de}
\ea
The bootstrap equations for the $\langle \s\s\s\s\rangle$ correlator
involve the conformal partial waves 
determined by the operator product expansion,
\be
\f_{12}\cdot \f_{12}= \f_{11} +\f_{13},
\label{ope}
\ee
that holds for any $m$ value, and is identified here as  $\s\cdot\s=I +\e$.

The interpolating conformal theory is obtained by assuming real $m\ge 3$ values.
Eliminating this parameter from $\D_\s,\D_\e$ in (\ref{ds-de}),
one obtains the curve:
\be
\Delta_\e=\frac{2}{3}\left(1+4\D_\s \right) ,
\ee 
that matches the unitarity boundary found numerically at $d=2$ to the
right of the Ising point, i.e of the kink in Fig. \ref{fig3c} 
\cite{slava-d}. Note
also that the central charge, $c= \D_\s(5-4\D_\s)/(1+\D_\s)$,
correspondingly interpolates between the values $1/2\le c\le 1$,
i.e. over all minimal models. The inverse relation $\D_\s=[5-c
  -\sqrt{(c-1)(c-25)}]/8$ identifies the field $h_{12}$ as one of the
two Virasoro representations with null vector at level $2$, the other
being $h_{21}$.  The next minimal model encountered on the unitarity
line on the right of the Ising model is the Tricritical model for
$m=4$ with $\D_\s\equiv 2h_{12}=1/5$ and $\D_\e\equiv 2h_{13}=6/5$. 
Regarding the boundary on the left of the Ising point,
the corresponding theories are believed not to correspond to any
conformal theory since basic Virasoro descendants are 
missing in the spectrum \cite{slava-IM}.

In the following, we count the number of (quasi)-primary states that
occur in the bootstrap partial waves, i.e. are generated by the OPE
(\ref{ope}), in the cases of the Ising and Tricritical Ising models. 
From this counting we shall obtain the list of low-lying states that
decouples as approaching the Ising model from the right, i.e. for
$\D_\s\to (1/8)^+$, along the unitarity boundary, namely the states that
are proper null vectors of this model. We shall adapt a counting
argument due to A. Zamolodchikov \cite{zam}.

The characters of Virasoro representations $\c^{\rm Vir}_h(q)$ 
with weight $h$ and central charge $c$ are generating functions for the 
multiplicities  $d_N(h,c)$ of descendant states, as follows \cite{cft}:
\be
\wt\c_h (q)=\sum_{N=0}^\infty d_N(h,c)\, q^N= q^{-h+c/24}\c^{\rm Vir}_h(q) .
\label{chi}
\ee
Among the Virasoro descendants, the quasi-primary fields cannot
be written as derivatives of other fields: in terms
of states, $|h, QP\rangle \neq L_{-1} |h, {\rm desc.}\rangle$.
The corresponding multiplicities $d^{QP}_N$ are then given by 
$d^{QP}_N=d_N-d_{N-1}$ (with $d_{-1}=0$). 
It follows that the generating function 
for quasi-primary states is given by:
\be
\wh\c_h(q)=\sum_{N=0}^\infty d^{QP}_N q^N = (1-q) \wt\c_h(q) .
\label{chi-qp}
\ee
In the case of the Identity representation, one should take into account
that the vacuum already obeys $L_{-1}|0\rangle=0$, thus the previous
expression is modified as follows,
\be
\wh\c_0(q)=\sum_{N=0}^\infty d^{QP}_N q^N = (1-q) \wt\c_0 (q) +q.
\label{chi-vac}
\ee
For non-degenerate unitary representations of the Virasoro algebra
with $c \ge 1$, the number of descendants at level $N$ is equal to the 
number of partitions of $N$ \cite{cft}; one readily finds the following 
generating functions:
\be
\wh \c_h(q)=\left(\prod_{k=2}^\infty (1-q^k)\right)^{-1}, \qquad
\wh \c_0(q)=(1-q)\left(\prod_{k=2}^\infty (1-q^k)\right)^{-1}+q .
\label{chi-h}
\ee
In the case of the critical and tricritical Ising models, one
should insert the known form of the Virasoro characters for the
representation $(r,s)$ of the $m$-th 
minimal model \cite{cft} into Eq. (\ref{chi}),(\ref{chi-qp}).
The result is:
\be
\wh\c_{rs}(q)=\left(\prod_{k=2}^\infty (1-q^k)\right)^{-1}
\sum_{n=-\infty}^\infty q^{n^2m(m+1)}
\left(q^{n((m+1)r-ms)}-q^{rs+n((m+1)r+ms)} \right).
\label{chi-rs}
\ee
This formula is modified by adding $(+q)$ for $r=s=1$ and 
is going to be used for the values $m=3,4$.

In order to obtain the generating function ${\cal N}$ of quasi-primary
multiplicities entering the $\langle \s\s\s\s\rangle$ bootstrap,
i.e. of conformal partial waves, we consider the Virasoro
representations in the r.h.s. of the operator product expansion
(\ref{ope}) and write the expression:
\ba
{\cal N}_m(q,\bar q)&=&\vert\wh\c_{11}(q)\vert^2+
\vert q^{h_{13}}\wh \c_{13}(q)\vert^2
\nl
&=& \sum_{\D=0}^\infty\sum_{|\ell|\le \D} d^{QP}(\D,\ell)\, x^\D\, y^\ell,
\qquad q=xy,\ \ \ \bar q =xy^{-1}.
\label{mult}
\ea
In this formula, the generating sum is rewritten in terms of
conformal dimensions $\D=N+\ov N$ and spins $\ell=N-\ov N$ of
descendant states.

We now discuss the counting of quasi-primaries in the interpolating
theory on the unitarity boundary between IM and TIM, i.e. for 
$1/8 < \D_\s<1/5$. The formula ${\cal N}_m$ (\ref{mult}) with $m=4$ describes
the TIM case and this counting remains valid for lower $\D_\s$ values
approaching the IM from the right, only the value $h_{13}=\D_\s/2$
changes.  Such extrapolation of ${\cal N}_4$ to $\D_\s =1/8^+$ should
be compared with the expression ${\cal N}_3$ that holds at the IM
point $\D_\s =1/8$.

The multiplicities $d^{QP}(\D,\ell)$ obtained by these two formulas
are given in Table \ref{tab6} for the low-lying states (only even
$\ell$ values are relevant).  The counting for a generic
$c>1$ theory is also shown for comparison; this is obtained by using the
expressions (\ref{chi-h}) for the characters in (\ref{mult}) (in all cases
$h_{13}=1/2$).

\begin{table}
\centering
\be
\begin{array}{r|rrrrrrrrrrrr}
\Bell&&  &  & 
\multicolumn{6}{c}{\underline{\bf c>1 \rm\bf\ \  theory}}  &  &  &\\
10&  &  &  &  &  &  &  &  &  & 4&12&\\
8 &  &  &  &  &  &  &  & 3& 7& 0& 0&\\
6 &  &  &  &  &  & 2& 4& 0& 0& 3& 7&\\
4 &  &  &  & 1& 2& 0& 0& 2& 4& 0& 4&\\
2 &  & 1& 1& 0& 0& 1& 2& 0& 2& 2& 8&\\
0 &\ 1&\ 0&\ 0&\ 1&\ 1&\ 0&\ 1&\ 1&\ 4&\ 0&\ 4&\\
\hline
  & 1& 2& 3& 4& 5& 6& 7& 8& 9&10&11&{\bf \D}\\
\end{array}
\nonumber
\ee
\be
\ \ \ \begin{array}{r|rrrrrrrrrrrr}
\Bell&&  &  & 
\multicolumn{6}{c}{\underline{\bf Tricritical\ Ising }}   &  &  &\\
10&  &  &  &  &  &  &  &  &  & 4&{\CC\bf 6}&\\
8 &  &  &  &  &  &  &  & 3&{\CC\bf 4}& 0& 0&\\
6 &  &  &  &  &  & 2&{\CC\bf 2}& 0& 0&{\CC\bf 3}&\boxed{\CC\bf 4}&\\
4 &  &  &  & 1& 2& 0& 0& 2&\boxed{\CR\bf 2}& 0& 0&\\
2 &  & 1&\boxed{1}& 0& 0& 1&\boxed{\CR\bf 2}& 0& 0& 2& 4&\\
0 &\ 1& 0& 0&\ 1&\boxed{\CC\bf 1}&\ 0&0 &\ 1&{\CR\bf 4}& 0&{\CG\bf 1}&\\
\hline
  & 1& 2& 3& 4& 5& 6& 7& 8& 9&10&11&{\bf \D}\\
\end{array}\nonumber
\ee
\be
\begin{array}{r|rrrrrrrrrrrr}
\Bell&&  &  &  \multicolumn{6}{c}{\underline{\bf Ising }}
  &  &  &\\
10&  &  &  &  &  &  &  &  &  & 2& 2&\\
8 &  &  &  &  &  &  &  & 2& 1& 0& 0&\\
6 &  &  &  &  &  & 1& 1& 0& 0& 2& 0&\\
4 &  &  &  & 1& 1& 0& 0& 1& 0& 0& 0&\\
2 &  & 1& 0& 0& 0& 1& 0& 0& 0& 1& 1&\\
0 &\ 1&\ 0&\ 0&\ 1&\ 0&\ 0&\ 0&\ 1&\ 1&\ 0&\ 0&\\
\hline
  & 1& 2& 3& 4& 5& 6& 7& 8& 9&10&11&{\bf \D}\\
\end{array}\nonumber
\ee
\caption{Number of quasi-primary states in the $d=2$
  $\langle\s\s\s\s\rangle$ bootstrap as a function of conformal dimension and
  spin, for a generic $c>1$ theory, the tricritical and critical Ising
  Model, respectively. Boxed numbers indicate TIM channels that completely
  disappear in IM; cyan (red) numbers correspond to numerically uncertain
  (unseen) states; the green channel $(\D,\ell)=(11,0)$ does not appear in
  the four-spin correlator.}
\label{tab6}
\end{table}

The analysis of the three tables leads to the following observations:
\begin{itemize}
\item
The generic theory contains many more quasi-primaries than any minimal
model, thus confirming the reduction of states on the unitarity boundary. 
More importantly, the IM has less states than the TIM. 
The latter fact follows from the properties of Virasoro representations.
Any primary field $\f_{rs}$ has a first null vector at level $N=rs$, 
that projects out a sub-tower of states: in the case of 
the energy field $\e=\f_{13}$ appearing in
the operator product expansion (\ref{ope}), this occurs
at level $N=3$. At the IM point, the reflection symmetry of the
$m=3$ Kac table (\ref{kac}) implies the following field identification:
\be
\f_{13}\equiv\f_{21} .
\label{f-12}
\ee
Thus, an additional degeneracy occur at level $N=2$, implying
a further projection of states. Other projections due to higher-$N$ 
null vectors are taken into account in the expressions of the
characters (\ref{chi-rs}).
\item
In the case of a given correlator, such as $\langle \s\s\s\s\rangle$, the
$d^{QP}(\D,\ell)$ quasi-primary states for a given $(\D,\ell)$ pair
combine in a single amplitude and a unique structure constant. Some
amplitudes could be vanishing for this particular observable, thus the
counting argument gives an upper bound of the possible bootstrap
channels, in general.
Since the expression of the $\langle \s\s\s\s \rangle$
correlator is known exactly for the interpolating theory, 
the structure constants as a function of
$\D_\s$ have been obtained in Refs. \cite{rast-b}\cite{slava-IM}.  
By inspection, one finds that the channel $(\D,\ell)=(11,0)$
is absent in the TIM (green digit in Table \ref{tab6}), while all the
other ones in the Table are present. 
\item
The comparison of the TIM and IM tables shows the following interesting cases:
\ba
&& d^{QP}_{TIM}(\D,\ell) \ge  d^{QP}_{IM}(\D,\ell) >0,
\label{d-ok}\\
&& d^{QP}_{TIM}(\D,\ell) > 0\ \ {\rm and}\ \ d^{QP}_{IM}(\D,\ell) =0 .
\label{m-con}
\ea
While moving from the TIM to the IM on the unitarity boundary,
the first case gives a channel that remains open in the limit to the
IM and therefore the corresponding state/field 
${\cal O}(\D,\ell)$ exists at $\D_\s=1/8$. 
In the second case, the channel is open in the TIM but 
should close reaching the IM, thus the following decoupling 
should be observed: 
\be
\lim_{\D_\s\to \left(\frac{1}{8}\right)^+} f_{\s\s{\cal O}(\D,\ell)}
\left(\D_\s \right)=0 .
\ee
\end{itemize}

In conclusion, the occurrence of pairs of multiplicities of the type
(\ref{m-con}), i.e. non-vanishing in TIM and vanishing in IM, gives a
necessary condition for the decoupling of the corresponding state.
Such occurrences within the low-lying spectrum are represented by
boxed numbers in Table \ref{tab6}.  The vanishing at
$\D_\s=1/8$ of the corresponding
structure constants has been checked by plotting the explicit expressions
for $f_{\s\s\cal O}(\D_\s)$ obtained from the exact form of $\langle
\s\s\s\s\rangle$ \cite{rast-b}.
We remark that the state decouplings discussed here could also have
been obtained from the direct analysis of these structure constants,
but we find that the counting argument is more instructive and general.
Note also that all structure constants are real positive 
numbers in spite of the non-unitarity of the interpolating 
theory \cite{slava-uni}.

%-4.2--------------------------------------------- 
\subsection{Numerical spectrum near the Ising point and decoupling 
of states for $2\le d\le 3$}

The quality of the bootstrap data at $d=2.00001$ is also
reported in Table \ref{tab6} using the following color code:
\begin{itemize}
\item
Black digits represent states that are well observed and are close to the
exact solution for $\D_\s\ge 1/8$.
\item
Red digits correspond to states not observed, owing to the large errors
of subleading fields.
\item
Cyan digits represent observed states whose dimensions are not very close 
to the exact solution.
\end{itemize}
In conclusion, the decoupling of states at the Ising model can be
described in our numerical setting for the well-seen
spin-two state $(\D,\ell)=(3,2)$ and the less precise spin-zero state
$(\D,\ell)=(5,0)$. These two decouplings are actually due to the
$N=rs=2$ null vector of the energy field $\f_{21}$ in the IM, Eq. (\ref{f-12}),
in the chiral and scalar channels, respectively.

\begin{figure}[h]
\centering
\includegraphics[scale=0.4]{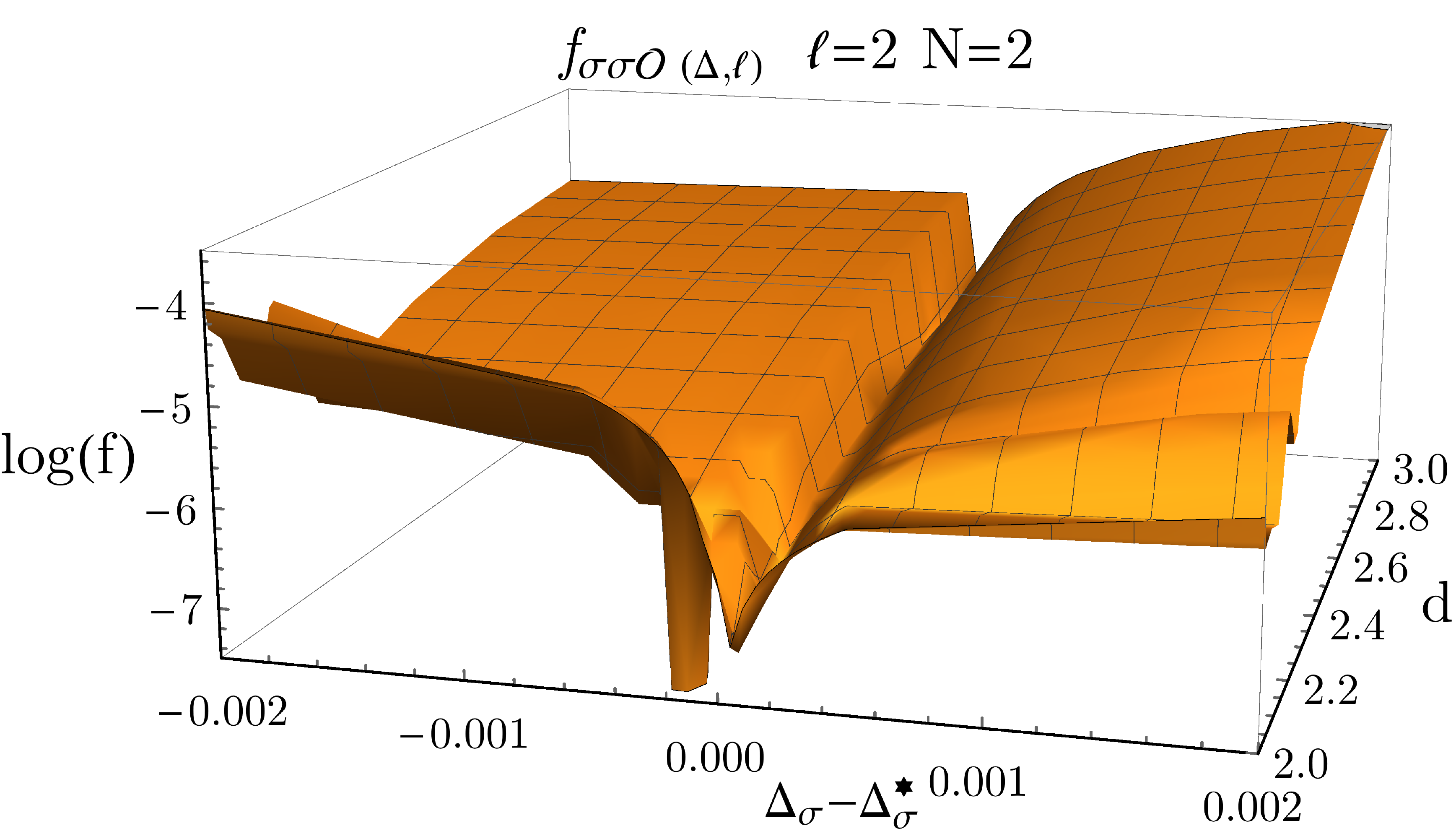}
\caption{Evolution of the structure constant $f_{\s\s{\cal O}}$ for the
$\ell=2$ null state as a function of dimension $d$ and displacement 
from the Ising point $\D=\D_\s^*$.
The fluctuations observed to the right of $\D=\D_\s^*$ for 
$2\le d\le 2.3$ are numerical uncertainties due to coarse sampling 
of the unitarity boundary. The big dip and the rise observed on
the left of $\D=\D_\s^*$ for  $d\ge 2$ are due to the decoupling 
of a left state that disappear at $d\sim 2.3$.}
\label{fig15}
\end{figure}

We now extend the analysis to $d>2$ by studying 
the behaviour of the corresponding structure constants
on the unitarity boundary to the right of the Ising points, $\D_\s \ge \D_\s^*$.
The numerical results for the spin-two state are shown in Fig. \ref{fig15}.
The vertical axis reports the logarithm of its structure constant
$f_{\s\s{\cal O}(3,2)}$ as a function of $d$ and the 
displacement w.r.t. the Ising point $\D_\s - \D_\s^*$ (recall that this
point is determined with error $<0.0001$ over the entire range).
We find that the vanishing of the structure constant
at $\D_\s=\D_\s^*$ continues above two dimensions, and smoothly
connects to $d=3$.  Therefore, the simplest level-two Virasoro 
null-vector condition at $d=2$ is found to correspond to
the decoupling of the $\ell=2$ state observed earlier at $d=3$ \cite{slava-IM}.

This result gives remarkable support to the existence of 
specific state projections identifying the Ising conformal theory 
in three dimensions, one of the questions addressed in the Introduction.
Let us postpone the discussion on $d>2$
null vectors to the Conclusions and analyze now the numerical precision 
of the result in Fig. \ref{fig15}.

In general, it is difficult to make precise quantitative comparisons
of structure constants in the $d=2$ and $d=3$ theories, 
both in practice and in principle. Let us nonetheless report
the numerical values for $f_{\s\s{\cal O}(3,2)} $ in the two cases, 
respectively at the critical point, the right edge of Fig. \ref{fig15}
and far off the Ising point:
\be
\begin{array}{|c|ccc|c|}
\hline
\D_\s & \D_\s^* & \D_\s^*+0.002 &  \D_\s^*+0.06 &{\bf d} \\
\hline
f_{\s\s{\cal O}(3,2)}& 0.0003 & 0.006 & 0.04 & {\bf 2} \\
f_{\s\s{\cal O}(3,2)}& 0.003  & 0.03  & 0.2  &{\bf 3} \\
\hline
\end{array}
\label{null-val}
\ee
We see that the structure constant roughly drops one order of magnitude
in the vicinity of the critical point and two orders w.r.t. its far-off
value, both in the $d=2$ and $d=3$ cases.

Next, we give the values of other non-vanishing structure constants.
In the case of the stress tensor, the 
leading $\ell=2$ state, $f_{\s\s T}$ is practically 
constant over the  $\D_\s$ intervals, and read:
\be
f_{\s\s T} = 0.125\ (d=2),\qquad 0.326\ (d=3) .
\ee
These values are determined with $O(10^{-3})$ relative error. 
The structure constant of the subleading field $T'$, i.e. the next
non-vanishing one in the $\ell=2$ channel, takes the following values 
at the Ising point $\D_\s^*$:
\be
f_{\s\s T'} \sim 0.002 \ (d=2), \qquad 0.01\ (d=3) ,
\ee
and is determined with relative error $O(10^{-1}- 10^{-3})$. 

The comparison of $f_{\s\s T}$, $f_{\s\s{\cal O}(3,2)}$ and $f_{\s\s T'}$
for both dimensions shows that their off-Ising values differ
by roughly one order of magnitude relative to each other, i.e. follow
the typical ordering of subleading states.
However, at the Ising point the structure constant $f_{\s\s{\cal O}(3,2)}$
is one order of magnitude smaller than $f_{\s\s T'}$ and 
comparable with the error of the latter.
This behavior let us conclude that the minimal values taken 
by $f_{\s\s{\cal O}(3,2)} $ are
consistent with zero within numerical precision.

The corresponding analysis of the structure constant for the
$(\D,\ell)=(5,0)$ state shows that it goes to zero at the Ising point for 
$2\le d < 2.2$, but stays constant above this dimension,
namely this null state does not extends to $d=3$. The dimension at
which the behaviour changes is approximatively the same where
the towers of leading-twist fields enter the higher-dimension regime,
as discussed in Sections three.

%-5---------------------------------------------

\section{Conclusions}

In this work, we have analyzed the critical properties of the Ising
model in continuous dimension $4> d \ge 2$ by using the numerical
conformal bootstrap.  High quality data for the critical exponents and
other low-lying conformal dimensions have been obtained together with
the corresponding structure constants and have been fitted with simple
polynomials in $d$. The comparison with other methods has
shown good consistency and improvement.

Two qualitative results show the interplay between conformal field
theories in $d=2$ and $d>2$. The first one is the behaviour of leading
and subleading twists fields for spin $4 \le \ell \le 20$, that
gradually acquire independent anomalous dimensions for $2<d< 2.2$ and
then enter the expected higher-dimensional regime for $d>2.2$.  
The second observation is the persistence above two dimensions of the
decoupling of one state corresponding to the simplest $\ell=2$
null-vector of the Ising energy field $\e=\f_{21}$ in two dimensions.
As discussed in the Introduction, this result is very intriguing
and encouraging deeper analyses. For example, it would be interesting
to develop analytic approaches based on 
$d=2+\eps$ and/or large $\ell$ 
perturbative expansions \cite{alday}.

Let us add some remarks concerning the possibility of specific
state decouplings for the Ising model in $d>2$, i.e. of a consistent
solution of the conformal bootstrap on a 'smaller' set of states.
As emphasized in Ref. \cite{slava-IM}, the presence of a kink, i.e.
a singularity, on the unitarity
boundary at the Ising point indicates that this model cannot be continuously
deformed (even allowing some non-unitarity): this rigidity 
is already an indication of a smaller bootstrap.
The main question is, in our opinion, how to formulate
a consistent projection of states. Algebraic conditions directly involving
critical exponents and other conformal data are not expected for $d>2$
 -- at least, a naive attempt to generalizing the level-two 
null-vector condition to $d>2$ has
failed. Let us finally quote the following works analyzing the  
null-vectors of the $d>2$ conformal representations \cite{null-ref}.

\bigskip\bigskip

{\bf Acknowledgments}

The authors would like to thank D. Bernard, F. Gliozzi, R. Guida,
Z. Komargodski, L. Rastelli, M. Serone and A. Trombettoni for
interesting scientific exchanges, C. Behan and J. Henriksson for
communicating their results \cite{behan}\cite{4th}, and in particular
S. Rychkov for discussions, data \cite{slava-d} and comments on the
manuscript.  A. C. acknowledge the hospitality and support by the
\'Ecole Normale Sup\'erieure, Paris, and all authors thanks the
G. Galilei Institute for Theoretical Physics, Arcetri, where part of
this work has been done.

\appendix

%-app A---------------------------------------------------

\section{Numerical methods}

The conformal bootstrap for the four-point correlator
$\langle\sigma(x_1)\sigma(x_2)\sigma(x_3)\sigma(x_4)\rangle$ of the
scalar primary field $\s(x)$ with given conformal dimension
$\Delta_\sigma$ is based on expanding the correlator in conformal
partial waves (conformal blocks) and imposing the crossing
symmetry. The resulting functional equation can be written:
\be
F_{0, 0}(u, v) + \sum_{(\Delta, \ell) \neq (0, 0)} 
p_{\Delta, \ell} \, F_{\Delta,\ell}(u, v) = 0.
\label{crossing-eq}
\ee
In this expression, the summation runs  over the primary fields 
appearing in the $\sigma \cdot \sigma$ operator product expansion,
with the exclusion of the identity field, whose
contribution is singled out in 
$F_{0, 0}(u,v)$; $u$ and $v$ are the cross ratios,
\be
u = z\bar{z} =\frac{ x^2_{12} x^2_{34}}{x^2_{13} x^2_{24}}, \qquad 
v = (1 - z)(1 - \bar{z}) = \frac{x^2_{14} x^2_{23}}{x^2_{13} x^2_{24}},
\ee
the coefficients
$p_{\Delta, \ell} = f_{\sigma\sigma\mathcal{O}(\D,\ell)}^2$ are squared
structure constants and should be positive in an unitary theory. 
The functions $F_{\Delta, \ell}(u, v)$ are defined as:
\be
F_{\Delta, \ell}(u, v) = 
v^{\Delta_{\sigma}} G_{\Delta, \ell}(u, v) - u^{\Delta_{\sigma}} G_{\Delta, \ell}(v, u),
\ee
where $G_{\Delta, \ell}(u, v)$ is a conformal block for
the primary field with dimension $\Delta$ and spin $\ell$.
The overall dependence on $\D_\s$ in all these formulas is left implicit.
 
In Reference \cite{slava-I}, the problem of finding the spectrum of
$(\D,\ell)$ and the coefficients $p_{\Delta, \ell}>0$ satisfying the
function equation (\ref{crossing-eq}) and the unitarity
constraint was reformulated into the following optimization problem.
The functions $F_{\Delta,\ell}(u, v)$ (suitably discretized) are considered as
elements of a vector space and a linear functional $\Lambda$  
is introduced that should obey the following conditions:
\ba
&&\Lambda(F_{0, 0}) = 1, \qquad\qquad {\rm normalization},
\nl
&&\Lambda\left(F_{\Delta,\ell}\right) \ge 0,  \qquad\quad 
\forall\ (\Delta, \ell) \ {\rm in\ the\ spectrum}.
\label{pos-cond}
\ea
If such a functional is found for a given spectrum $\{(\Delta, \ell)\}$,
then the crossing-symmetry equation (\ref{crossing-eq}) cannot
be satisfied for unitary theories and the corresponding conformal theory
is ruled out. 

The form of the functional over the truncated basis that we are going to use 
is:
\ba
\Lambda: F_{\D,\ell}(u, v) 
& \mapsto & \sum_{0 \le m+2n \le 2n_{\max}+1} \lambda_{m, n}\ 
F^{(m,n)}(\D,\ell),
\nl
F^{(m,n)}(\D,\ell) &= &\left.
\partial_a^m\partial_b^n\, F_{\D,\ell}(a, b)\right|_{a = 1, b= 0} ,
\label{f-der}
\ea
where $z = (a + \sqrt{b})/2$ and $\bar{z} = (a - \sqrt{b})/2$
and $\l_{m,n}$ are the coefficients defining the functional.

In our implementation, the calculation of the conformal blocks and
their derivatives with respect to $a$ and $b$ follows the methods of
Ref. \cite{slava-diag}.  The conformal blocks along the diagonal 
$z =\bar{z}$ are expanded in series of $\rho = z/(1 +\sqrt{1-z})^2$:
\begin{eqnarray}
G_{\Delta, \ell}(z, z) = (4\rho)^\Delta \sum_{n=0}^\infty b_n\, \rho^n, 
\qquad b_0 = 1,
\label{g-series}
\end{eqnarray}
where the $b_n$ are determined by recursion relations following from the
quadratic and quartic Casimir equations. 
The series (\ref{g-series}) is approximated by a polynomial with maximal degree
$k_{\max}$, and the needed derivatives with respect to $a$ 
are easily obtained by using the map from $\rho$ to $a$. The derivatives 
 with respect to $b$ are also determined by solving
a recursion relation involving the quadratic Casimir (See
Appendix C of \cite{slava-IM-old}). Moreover,
an additional approximation for the conformal blocks and
their derivatives is introduced by replacing the poles at 
$\Delta_j$ with small residues with the some poles with the 
first largest residue \cite{sd-on} \footnote{
This approximation assumes that
  conformal blocks have simple poles, that is correct for dimensions 
$d\neq 2,4$. For this reason, the two-dimensional theory was
studied at $d=2.00001$.}

As a result, the functions $F^{(m, n)}(\Delta, \ell)$ (\ref{f-der})
are expressed as:
\be
F^{(m, n)}(\Delta, \ell) 
\sim \chi_\ell(\Delta) \ P^{ (m,n)}_\ell(\Delta),
\ee
where the factor
$\chi_\ell(\Delta)$  is positive for all $\Delta$ in unitary theories
and $P_\ell^{(m,n)}(\Delta)$ are polynomials.
After removing the non-polynomial part $\chi_\ell(\Delta)$, 
the conditions (\ref{pos-cond}) become:
\ba
\!\!\!\!\!\!\!\!\!\!\!\!\!\!\!\!\!\!\!\!
&&\sum_{0\le m+2n\le 2n_{\max}+1}\lambda_{m, n} \, F^{(m, n)}(0, 0) = 1,
\nl
\!\!\!\!\!\!\!\!\!\!\!\!\!\!\!\!\!\!\!\!
&&\sum_{0\le m+2n\le 2n_{\max}+1}\lambda_{m, n}\, 
P_\ell^{m,n}(\Delta_{\min}(\ell) + x) \ge 0,
\ \  x \in [0, \infty), \ell = 0, 2, \cdots, \ell_{\max},
\label{pol-cond}
\ea
where $\Delta_{\min}(\ell) = D -2 + \ell$ for $\ell > 0$. 
Equation (\ref{pol-cond}) is the final form that is used for implementing the 
numerical polynomial optimization by means of the SDPB solver \cite{SDPB}.
These results are well established in the bootstrap literature and 
were summarized here for the sake of the presentation.

Next, the SDPB solver is run for various values of the dimension
$\Delta_{\min}(0)$ of the lowest scalar primary field.  Thus, we can
identify the allowed and disallowed regions for the bootstrap, namely
the unitarity boundary as a function  of $\D_\s$ (see e.g. Fig. \ref{fig1a}). 
In our simulations, we used the functional with two values of
$n_{\max}$ in (\ref{f-der}), i.e. $n_{\max} = 16$ and $n_{\max} = 18$, 
corresponding to $153$ and $190$  $F^{(n,m)}$ components, respectively. 
We also chose the maximal values $\ell_{\max} = 50$ and $k_{\max} = 120$ 
for all dimensions $d$. In Table \ref{tab7} we list the parameters 
of the SDPB solver that have been used in this work.

\begin{table}[h]
\centering
\be 
\begin{array}{|l|l|}  
\hline
{\rm Parameter} & {\rm Value}\\   
\hline
{\rm findPrimalFeasible } &  {\rm true} \\
{\rm findDualFeasible } &  {\rm true }\\
{\rm detectPrimalFeasibleJump } &  {\rm false} \\
{\rm detectDualFeasibleJump } &  {\rm false} \\
{\rm precision } &  704  \\
{\rm dualityGapThreshold} & 10^{-30} \\
{\rm primalErrorThreshold} & 10^{-30}  \\
{\rm dualErrorThreshold} & 10^{-30}  \\
{\rm initialMatrixScalePrimal} & 10^{20}  \\
{\rm initialMatrixScaleDual} & 10^{20}  \\
{\rm feasibleCenteringParameter} &  0.1 \\
{\rm infeasibleCenteringParameter} & 0.3 \\
{\rm stepLengthReduction} &  0.7\\
{\rm choleskyStabilizeThreshold} & 10^{-40} \\
{\rm maxComplementarity} & 10^{100} \\
\hline
\end{array}
\nonumber
\ee
\caption{Parameters employed in the SDBP program.}
\label{tab7}
\end{table}

We now briefly summarize the implementation of the Extremal Functional
Method \cite{extr} for finding the spectrum of $(\D,\ell)$ values for
states participating the bootstrap (\ref{crossing-eq}) 
and the corresponding structure
constants $f_{\s\s {\cal O}(\D,\ell)}$ .  The functional
(\ref{pol-cond}) is called extremal when it is evaluated on the
unitarity boundary, that has been identified in previous steps.  The
zeros of the extremal functional for each $\ell$ value identify the
spectrum of dimensions $\D_{\ell,i}$ as functions of the boundary
parameter $\D_\s$, that are the data discussed in the text.

The structure constants are obtained by solving
the truncated crossing symmetry equations:
\be
F^{(m, n)}(0, 0) + \sum_{(\Delta, \ell)\neq (0, 0)} 
p_{\Delta, \ell} F^{(m, n)}(\Delta, \ell) = 0 ,
\label{extr-eq}
\ee
where the integers $(m ,n)$ take all the positive values
with $ m + 2n \le 2n_{\max}+1$.
The sums run over the $(\D,\ell)$ spectrum determined earlier.
Since Eq. (\ref{extr-eq}) is an over-constrained linear system, the
solution doesn't exist in general. Therefore, an
approximated solution is found by considering the goal programming with the
Chebyshev method \cite{goal}.

%-bib--------------------------------------------

\end{document}